\documentclass[aps,pra,twocolumn,10pt,showpacs,superscriptaddress]{revtex4-1}
\usepackage[utf8]{inputenc}
\usepackage[english]{babel}
\babeltags{en = english}
\usepackage{amsmath,amssymb,amsbsy,graphicx,hyperref,braket,mathtools,wrapfig,cleveref,
color,epsfig,import,bm,eufrak,ulem,comment,stackengine,lipsum}
\usepackage{todonotes}
\renewcommand{\v}[1]{\ensuremath{\mathbf{#1}}}

\newcommand{\D}{^{\dag}}
\newcommand{\p}{^{\prime}}

\newcommand{\MD}{^{\mathstrut}}

\providecommand*{\iu}{\ensuremath{\mathrm{i}}}

\DeclarePairedDelimiter\abs{\lvert}{\rvert}

\newcommand{\rom}[1]{\uppercase\expandafter{\romannumeral #1\relax}}
\newcommand{\Id}{1\!\!1}

\newcommand{\beginsupplement}{
    \addto\captionsenglish{\renewcommand{\figurename}{FIG.}}
    \setcounter{figure}{0}
    \setcounter{equation}{0}
    \setcounter{section}{0}
    \renewcommand{\thefigure}{S-\arabic{figure}}%
    \renewcommand{\thesection}{\Alph{section}}
    \renewcommand{\thesubsection}{\roman{subsection}}
    \numberwithin{equation}{section}
}

\newcommand{\dtilde}[1]{\stackon[-.3pt]{$#1$}{\stackon[-.5pt]{$\scriptscriptstyle\sim$}{$\scriptscriptstyle\sim$}}} %

\begin{document}

\title{Topological Phase Transitions in the Disordered Haldane Model}
\author{J. Mildner}
\affiliation{Department of Physics, King's College London, Strand, London WC2R 2LS, United Kingdom}
\author{M. D. Caio}
\altaffiliation{These authors contributed equally to this work}
\affiliation{Department of Physics, King's College London, Strand, London WC2R 2LS, United Kingdom}
\affiliation{Instituut-Lorentz, Universiteit Leiden, P.O. Box 9506, 2300 RA Leiden, The Netherlands}
\author{G. M\"oller}
\altaffiliation{These authors contributed equally to this work}
\affiliation{Physics and Astronomy, Division of Natural Sciences, University of Kent, Ingram Building, Canterbury CT2 7NZ, UK}
\author{N. R. Cooper}
\affiliation{TCM Group, Cavendish Laboratory, University of Cambridge, Cambridge CB3 0HE, United Kingdom}
\author{M. J. Bhaseen}
\affiliation{Department of Physics, King's College London, Strand, London WC2R 2LS, United Kingdom}

\begin{abstract}
We investigate the phases and phase transitions of the disordered
Haldane model in the presence of on-site disorder. We use the
real-space Chern marker and transfer matrices to extract
critical exponents over a broad range of parameters. The disorder-driven transitions are consistent with the plateau transitions in the Integer Quantum
Hall Effect (IQHE), in conformity with recent simulations of disordered Dirac fermions. 
Our numerical findings are compatible with an additional line of mass-driven transitions with a continuously varying correlation length exponent.
The values interpolate between free Dirac fermions and the
IQHE with increasing disorder strength. We also show that the fluctuations of the Chern marker exhibit a power-law divergence in the vicinity of both sets of transitions, yielding another varying exponent.
We discuss the interpretation of these results.
\end{abstract}
\maketitle

\section{Introduction}
A defining characteristic of topological phases of matter is their
resilience to local perturbations and sample defects. A prominent
example is the Integer Quantum Hall
Effect (IQHE) \cite{Ando1975,Klitzing1980} which is robust to variations in
the sample geometry and is manifest in the presence of
disorder. This robustness to local perturbations makes topological
systems ideal candidates for applications, including
metrology and quantum information processing~\cite{Nayak2008}.
Experimental realizations of topological systems have proliferated in
recent years, and now include solid state devices in two- and
three-dimensions, cold atomic gases and optical systems.
For reviews, see for example Refs~\cite{Hasan2010, Dalibard2011, Qi2011, Carusotto2013,Lu2014,Goldman2014, Kim_2020, Lv_2021, Ni_2023}.

From a theoretical perspective, one of the most challenging and
long-standing problems is the characterisation of the plateau
transitions in the IQHE. This has attracted a great deal of attention
over the years, including scaling theory approaches \cite{Khmelnitskii_1983, Pruisken_1988, Kramer1990,Lutken_2007,Obuse_2012, Dresselhaus2022,Slevin_2023},
network models \cite{Chalker1988, Fulga_2011,Gruzberg2017}, and recent conjectures for the
low-energy field theory \cite{Zirnbauer_2019,Zirnbauer_2021}. Amongst these approaches is the
idea that topological phase transitions can be described in
terms of disordered Dirac fermions \cite{Ludwig1994,Chalker_1996, Guruswamy_2000,Bernard_2002}. This
has been the focus of renewed interest due to recent simulations of
continuum Dirac fermions which confirm their relevance to the IQHE \cite{Sbierski_2021}.

In this work, we explore the critical properties of disordered
topological phase transitions using the real-space topological marker
\cite{Bianco2011} and transfer matrices. The absence of translational invariance makes the real-space
approach especially suitable for numerical studies of critical
exponents \cite{Caio2018b,Ul_akar2020,Goldstein_2021,Ornellas_2022}. We focus on the Haldane model~\cite{Haldane1988} in
the presence of on-site disorder, whose low-energy description
corresponds to disordered Dirac fermions. We confirm that the disorder-driven topological transition is in the universality class of the plateau transition for disordered IQH systems, in conformity with recent work on continuum Dirac fermions \cite{Sbierski_2021}. Our numerical results are compatible with an additional line
of mass-driven transitions with a continuously varying correlation length
exponent $\nu$. The results interpolate between those of free Dirac fermions with $\nu =1$
and those of the IQHE with $\nu \sim 5/2$, with increasing disorder strength. We also observe a power-law divergence of the fluctuations of the Chern marker yielding another continuously varying exponent $\kappa$.
We discuss the interpretations of these findings including the possible need for larger system sizes at weak disorder.

The layout of this paper is as follows. We introduce the model in Section \rom{2} and the Chern marker in Section \rom{3}. We discuss the evolution of the phase diagram in Section \rom{4} before turning our attention to the correlation length exponent in Section \rom{5}. In Section \rom{6} we examine the sample-to-sample fluctuations of the Chern marker exposing its power-law divergence in the vicinity of the transitions. In Section \rom{7} we discuss the variation of the correlation length exponent and the fluctuation exponent as a function of the disorder strength. We conclude in Section \rom{8} and provide Supplementary Material.

\section{Disordered Haldane model}
The Haldane model~\cite{Haldane1988} describes spinless fermions hopping on a
honeycomb lattice with nearest and next-nearest neighbor hopping
amplitudes $t_1$ and $t_2$. In the presence of on-site 
disorder the Hamiltonian is given by 
\begin{eqnarray}
 \hat H&=&-t_1\sum_{\langle i,j\rangle}\left(\hat a\D_i \hat 
a\MD_j+\mbox{h.c.}\right)
   - t_2 \sum_{\langle\!\langle i,j\rangle\!\rangle}\left(e^{\iu \varphi_{ij}} 
\hat a\D_i \hat a\MD_j+\mbox{h.c.}\right) \nonumber \\
& &+M \sum_{i\in A} \hat n_i -M \sum_{i\in B} \hat n_i+\sum_i v_i \hat n_i,
\label{eq:HM_realspace}
\end{eqnarray}
where $v_i\in [-W,W]$ is a random variable drawn from a flat
distribution of width $2W$ and $A,B$ label the two sublattices. Throughout this work we set $t_1=1$ and $t_2 = 1/3$. Here, $\hat a_i^\dagger$ and $\hat a_i$
are fermionic creation and annihilation operators obeying
anticommutation relations $\{\hat a_i,\hat a_j^\dagger\}=\delta_{ij}$
and $\hat n_i\equiv\hat a_i^\dagger \hat a_i$. The energy offset $\pm
M$ breaks inversion symmetry between the $A$ and $B$ sublattices
allowing the possibility of a conventional band insulator when
$W=0$. The phase factor $\varphi_{ij}=\pm \varphi$ is positive
(negative) for anticlockwise (clockwise) hopping and breaks
time-reversal symmetry, allowing the possibility of topological
phases. The Haldane model with $W=0$ was realized using cold
atomic gases~\cite{Esslinger2014}, where the phase factor $\varphi$ is
imprinted via the periodic modulation of the optical lattice. The clean
Haldane model with $W=0$ also played a crucial role in the discovery of
topological insulators~\cite{Kane2005,Kane2005a}; for a review see
Ref.~\cite{Qi2011}. More recently, an extension of this model with
spatially anisotropic first neighbor hopping parameters has been
investigated for $W\neq 0$~\cite{Sriluckshmy2018}. This showed the
presence of disorder-driven Lifshitz and Chern transitions. In this
work, we focus on the isotropic Haldane model with $W\neq 0$. 
We extract the disorder induced critical exponents using the real-space Chern marker and transfer matrix calculations.

\section{Real-space Chern Marker}
In the absence of disorder, the phases of the Haldane model
(\ref{eq:HM_realspace}) are distinguished by the global Chern index~\cite{Chern1946,Berry1984,Haldane1988}
\begin{equation}
   C = -\frac{1}{\pi}{\rm Im}\sum_{n=1}^{n_\text{occ}}\int_{BZ}d{\bf k}\braket{\partial_{k_x}u_{n \bf k}|\partial_{k_y}u_{n \bf k}},
  \label{eq:nu}
\end{equation}
where $u_{n\v k}(\v r)=e^{-i \v k\cdot \v r}\psi_{n \v k}(\v r)$ is
the periodic part of the ground state wavefunction and $n_\text{occ}$ is the number of occupied bands. Here, $n$ is the band index
and the integration
is over the first Brillouin zone. The Chern index is quantized and
takes the values $C=\pm 1$ ($C=0$) in the topological
(non-topological) phases. In the presence of disorder (or in
finite-size samples with open boundary conditions) the definition
(\ref{eq:nu}) is not immediately convenient due to the explicit use of
momentum space. One approach to this problem is to impose periodic
boundary conditions on the disordered sample and to use a supercell
formulation instead~\cite{Ceresoli2007}. Alternatively, one can employ the real-space Chern
marker $c({\bf r}_\alpha)$, defined on a unit cell $\alpha$,
introduced by Bianco and Resta~\cite{Bianco2011}. This can be obtained
from the definition (\ref{eq:nu}) and is given by
\begin{equation}\label{eq:ChernMarker}
    c({\bf r}_\alpha ) = -\frac{4\pi }{A_c} {\rm Im} \sum_{s=A,B} \langle {\bf r}_{\alpha_s} \vert \hat{P} \hat{x} \hat{Q} \hat{y} \hat{P} \vert {\bf r}_{\alpha_s} \rangle.
\end{equation}
Here, $\hat P=\frac{A_c}{(2\pi)^2}\sum_{n=1}^{n_\text{occ}}\int_{BZ} d{\bf k}
\Ket{\psi_{n\bf k}}\Bra{\psi_{n \bf k}}$ is the projector onto the
ground state, $\hat Q =
\hat I-\hat P$ is the complementary projector onto the unoccupied
bands, $A_c$ is the area of a unit cell, and the sum is over the two
sublattice sites within the unit cell $\alpha$. For a clean Chern
insulator with $W=0$ and away from topological transitions, the value of $c({\bf r}_\alpha)$ evaluated in
the center of a finite-size sample reproduces the value that the global
Chern index (\ref{eq:nu}) would have in the presence of periodic boundary
conditions. For further details see Ref.~\cite{Bianco2011} and the
Supplementary Material provided here.

\section{Phase Diagram}
\begin{figure} \centering
\includegraphics[width=0.85\columnwidth]{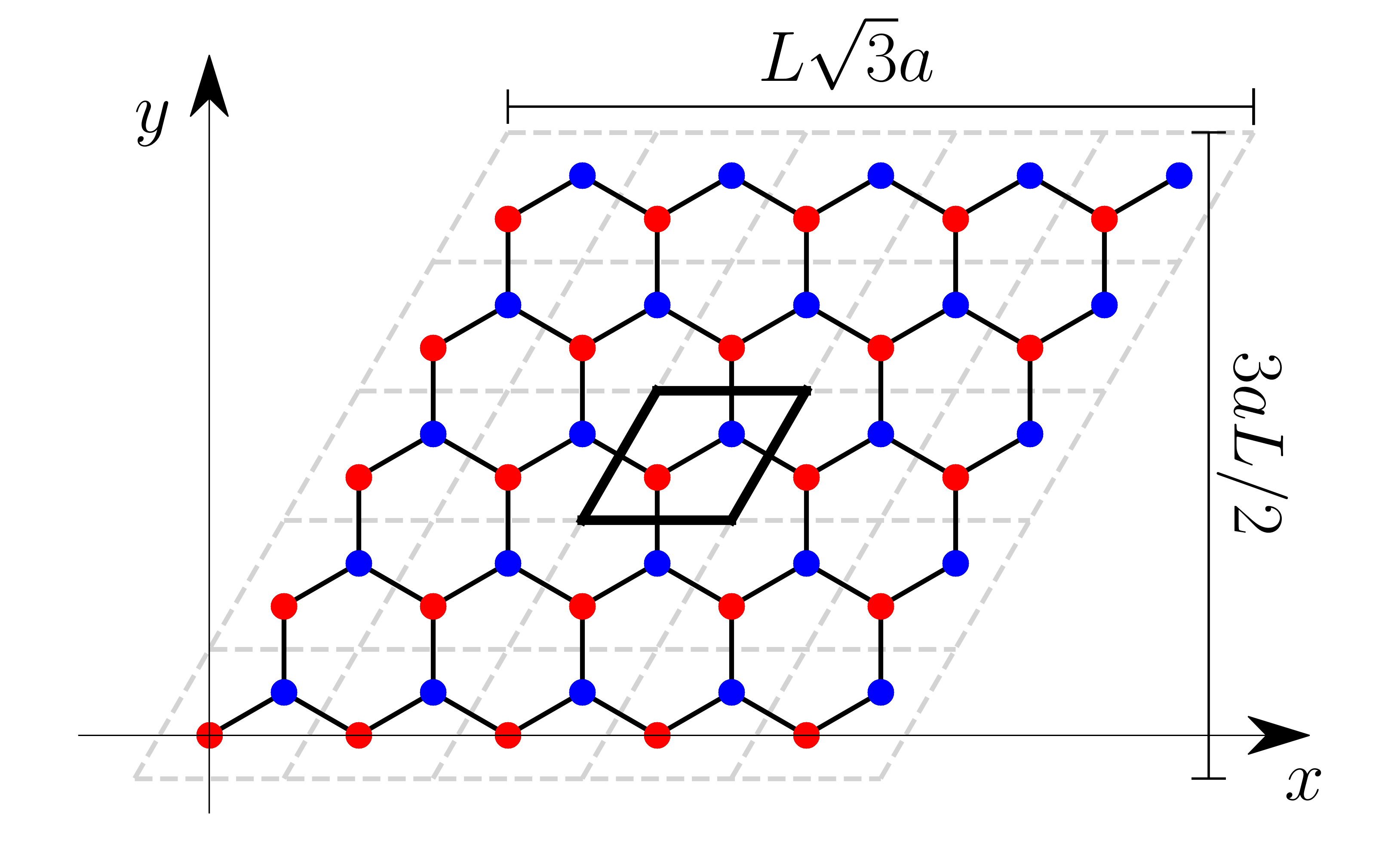}
\caption{Finite-size geometry used in the numerical investigations of
  the Haldane model via the real-space Chern marker. The dashed lines divide the honeycomb lattice
  into unit cells containing two sublattice sites $A$ and $B$, as indicated by the
  red and blue dots. We consider diamond shaped samples with $L$ unit
  cells along each edge and $2L^2$ lattice sites. We
  set the inter-site lattice spacing $a$ to unity. The Chern marker is
  typically evaluated in the center of the sample as indicated by the
  solid lines. }\label{fig:region}
\end{figure}

\begin{figure}[ht]
  \centering  
   \includegraphics[width=.85\columnwidth]{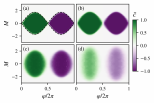}
   \caption{Exact diagonalization results for the disorder averaged
     Chern marker $\bar c$ in the centre of a finite-size sample with
     increasing disorder strength $W$. The results are obtained for an
     $L=17$ sample with $2L^2$ sites, $t_1=1$ and $t_2=1/3$. (a) $W=0$, (b) $W=1$, (c)
     $W=2$, and (d) $W=3$. For $W\neq 0$, we average over $500$
     disorder realizations and for $W=0$ we plot $c$ directly.
     The data in (a) are consistent with the phase
     diagram of the clean Haldane model (dashed
     lines). }\label{fig:MvsP_d}
 \end{figure}

In order to distinguish between the topological and non-topological
phases of the disordered Haldane model (\ref{eq:HM_realspace}) one may
consider the disorder average of the real-space Chern marker $\bar c$, in
the center of a finite-size sample, and averaged over a number of
independent disorder 
realizations~\cite{Bianco2011,Sriluckshmy2018}; see Fig.~\ref{fig:region}.
In Fig.~\ref{fig:MvsP_d} we show the evolution of $\bar c$ with increasing 
disorder strength $W$. 
 \begin{figure}[ht]
  \centering  
   \includegraphics[width=.85\columnwidth]{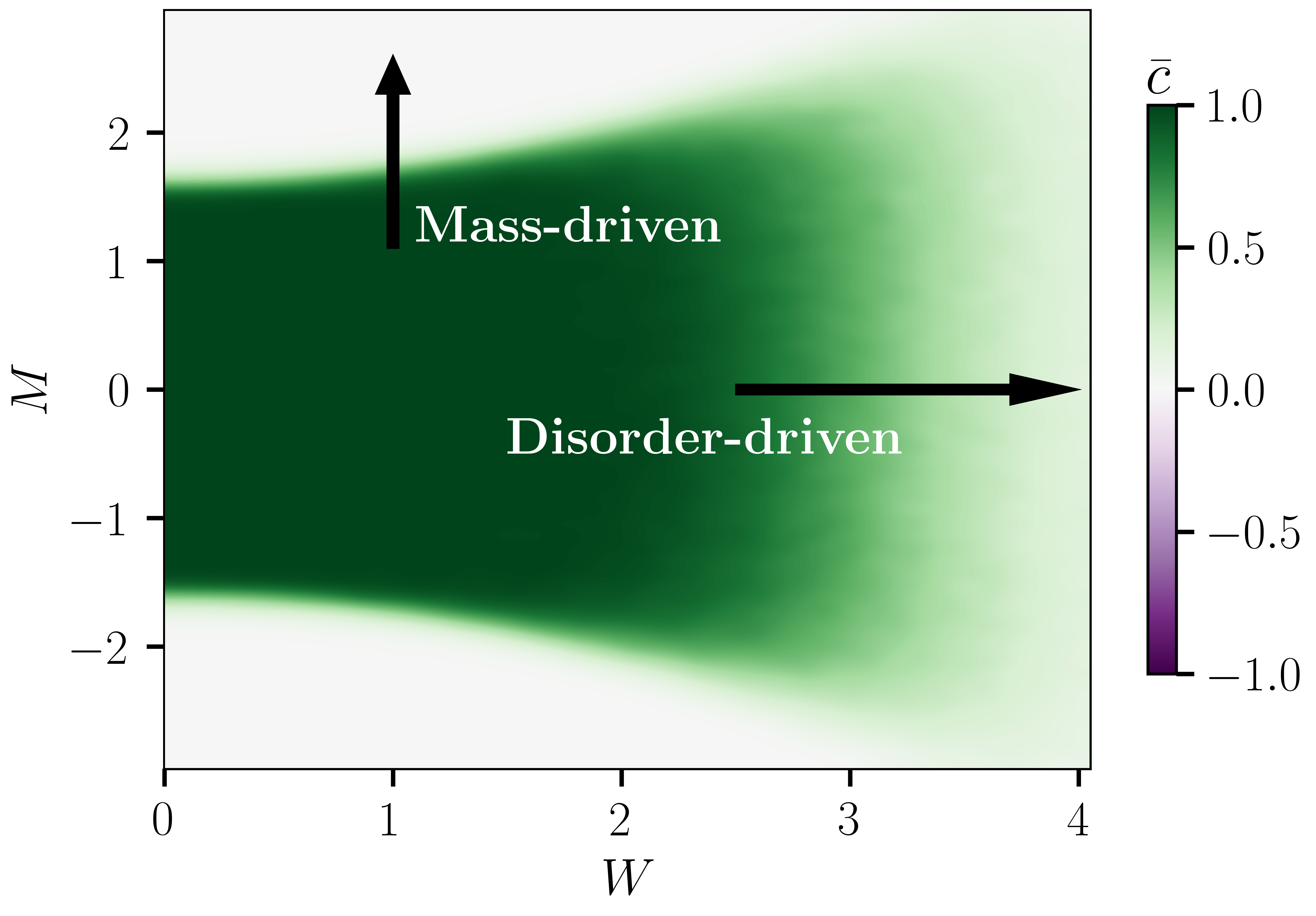}
   \caption{Exact diagonalization results for the disorder averaged
     Chern marker $\bar c$ in the centre of a finite-size sample as a
     function of $M$ and the disorder strength $W$. The results are obtained for an $L=17$ sample with $2L^2$ sites and
     $\varphi=\pi/2$. The data are averaged over $10^3$ disorder
     realizations. We investigate both the disorder-driven transitions (horizontal arrow) and the mass-driven transitions at fixed disorder (vertical arrow).
     }\label{fig:MvsV_phi}
 \end{figure}
  As may be seen from Fig.~\ref{fig:MvsP_d}(a), in the absence of
  disorder the real-space Chern marker $c$ directly reproduces the
  equilibrium phase diagram of the clean Haldane
  model~\cite{Haldane1988}. A vertical slice through
  Fig.~\ref{fig:MvsP_d}(a) shows clear plateaus, with values of
  $c$ close to integers \cite{Bianco2011}. In the vicinity of the transition
  between the topological and non-topological phases, the value of $c$
  is no longer quantized, but smoothly interpolates between the
  plateaus. As shown in our earlier work \cite{Caio2018b}, a finite-size
  scaling analysis of this transition region yields the correlation
  length exponent $\nu=1$. This is in agreement with the low-energy
  description of the clean Haldane model in terms of free Dirac fermions
  \cite{Haldane1988}. As the disorder strength $W$
  increases, the topological regions of the phase diagram expand, in
  agreement with Ref.~\cite{Sriluckshmy2018}; see Figs \ref{fig:MvsP_d}~(b)
  - (d). Strong disorder ultimately destroys the topological phases as illustrated in
  Fig.~\ref{fig:MvsV_phi}. This occurs around $W\sim 3.6$ for $t_1 =1$, $t_2 = 1/3$ and for a range of $\varphi$ and $M$. A notable feature in Fig.~\ref{fig:MvsV_phi} is the
  presence of re-entrant disorder-driven transitions, first from $\bar c=0$
  to $\bar c=1$, and then from $\bar c=1$ to $\bar c=0$ at stronger disorder
  \cite{Sriluckshmy2018}; this may be seen along the line $M=2$ for example. In the next section we use finite-size scaling to extract the critical properties of both the disorder-driven transitions and the mass-driven transitions at fixed disorder strength; see Fig.~\ref{fig:MvsV_phi}.

\section{Finite-Size Scaling}
\begin{figure}[h]
  \centering 
   \includegraphics[width=.95\columnwidth]{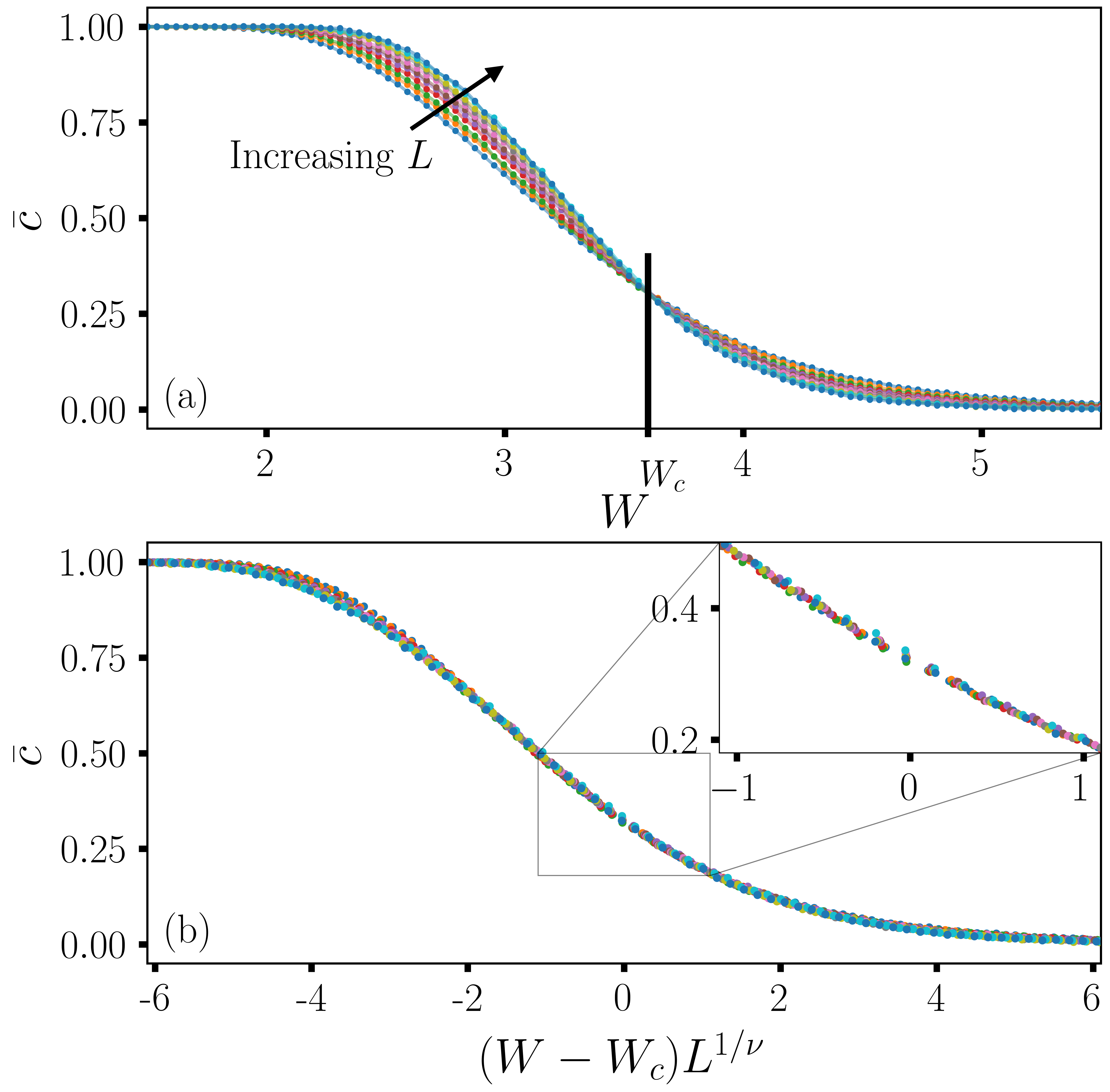}
   \caption{(a) Exact diagonalization results for the disorder
     averaged Chern marker $\bar c$ in the centre of a finite-size
     sample as a function of the disorder strength $W$, with
     $M=0$ and $\varphi=\pi/2$.  The
     results are obtained for $L=15, 17, 19,...,35$, corresponding
     to $2L^2$ site samples, averaged over $3\times 10^4$ disorder
     realizations. At low disorder strength $\bar c$ is pinned at
     unity, and it is zero for strong disorder, corresponding to a
     disorder-driven topological phase transition at $W_c\sim
     3.6$. By maximizing the overlap between $\bar{c}$ when plotted as a function of $(W-W_c)L^{1/\nu}$
     for different system sizes we extract $\nu=2.42 \pm 0.11$ and $W_c =3.58 \pm 0.02$.
     (b) Collapse of the data in (a) for the rescaling
     $(W-W_c)L^{1/\nu}$ of the horizontal
     axis with $\nu = 2.42$ and $W_c =3.58$. Inset: a zoomed-in portion of the data in the vicinity of the critical point. The results are in good agreement with the transfer matrix calculations which yield $\nu=2.47 \pm 0.09$; see Supplementary Material.
     }\label{fig:CMDD}
\end{figure}

The universal features of topological phase transitions can be obtained from a finite-size scaling analysis of the Chern marker, both in the absence \cite{Caio2018b} and the presence of disorder \cite{Varjas_2020,Ul_akar2020,Goldstein_2021}. In Fig.~\ref{fig:CMDD}(a) we plot the variation of the disorder-averaged Chern marker $\bar c$ as a function of $W$, corresponding to a horizontal slice through Fig.~\ref{fig:MvsV_phi} with $M=0$. It is readily seen that $\bar c$ interpolates between plateaus at $\bar c\sim 1$ and $\bar c\sim 0$, over a broad transition region in the vicinity of a critical
  disorder strength $W_c\sim 3.6$. The width of this transition region
  $\Delta W$ narrows with increasing system size $L$. Assuming that
  the correlation length $\xi \sim
  (W-W_c)^{-\nu}$ is of order the system size when the
  departures from quantization occur, one expects that the width scales as $\Delta W\sim
  L^{-1/\nu}$. More generally, we assume a scaling form 
  $\bar{c} \sim f(\xi/L) \sim \tilde f((W-W_c)L^{1/\nu})$. In Fig.~\ref{fig:CMDD}(a) we maximize the overlap between the $\bar{c}$ curves obtained for different system sizes when plotted as a function of $(W-W_c)L^{1/\nu}$. This yields $W_c = 3.58\pm 0.02$ and $\nu=2.42\pm 0.11$. Replotting the data in Fig.~\ref{fig:CMDD}(a) as a function of $(W-W_c)L^{1/\nu}$ with $\nu = 2.42$ the data collapse onto a single curve; see Fig.~\ref{fig:CMDD}(b) and the inset. This value of $\nu$ is close to, but a little lower than, the most recent numerical results for the correlation length exponent pertaining to the plateau transitions in the IQHE where $\nu \sim 2.6$ \cite{Dresselhaus_2022, Sbierski_2021, Slevin_2023}. It is however, compatible with the spread of results shown in Ref.~\cite{Dresselhaus_2021}. We corroborate these findings with transfer matrix calculations on large strips with width up to $59$ unit cells and length up to $10^7$. We find that $\nu = 2.47 \pm 0.09$, which is numerically close to $5/2$, and in agreement with the scaling of the Chern marker; see Supplementary Material. We further confirm our results for the disorder-driven transition with $M=1$. This yields $\nu=2.47\pm 0.08$, in agreement with the results for $M=0$; see Supplementary Material. This suggests that the disorder-driven transitions in Fig.~\ref{fig:MvsV_phi} are in the same universality class, independent of the value of $M$.
  
\begin{figure}[h]
  \centering 
   \includegraphics[width=.95\columnwidth]{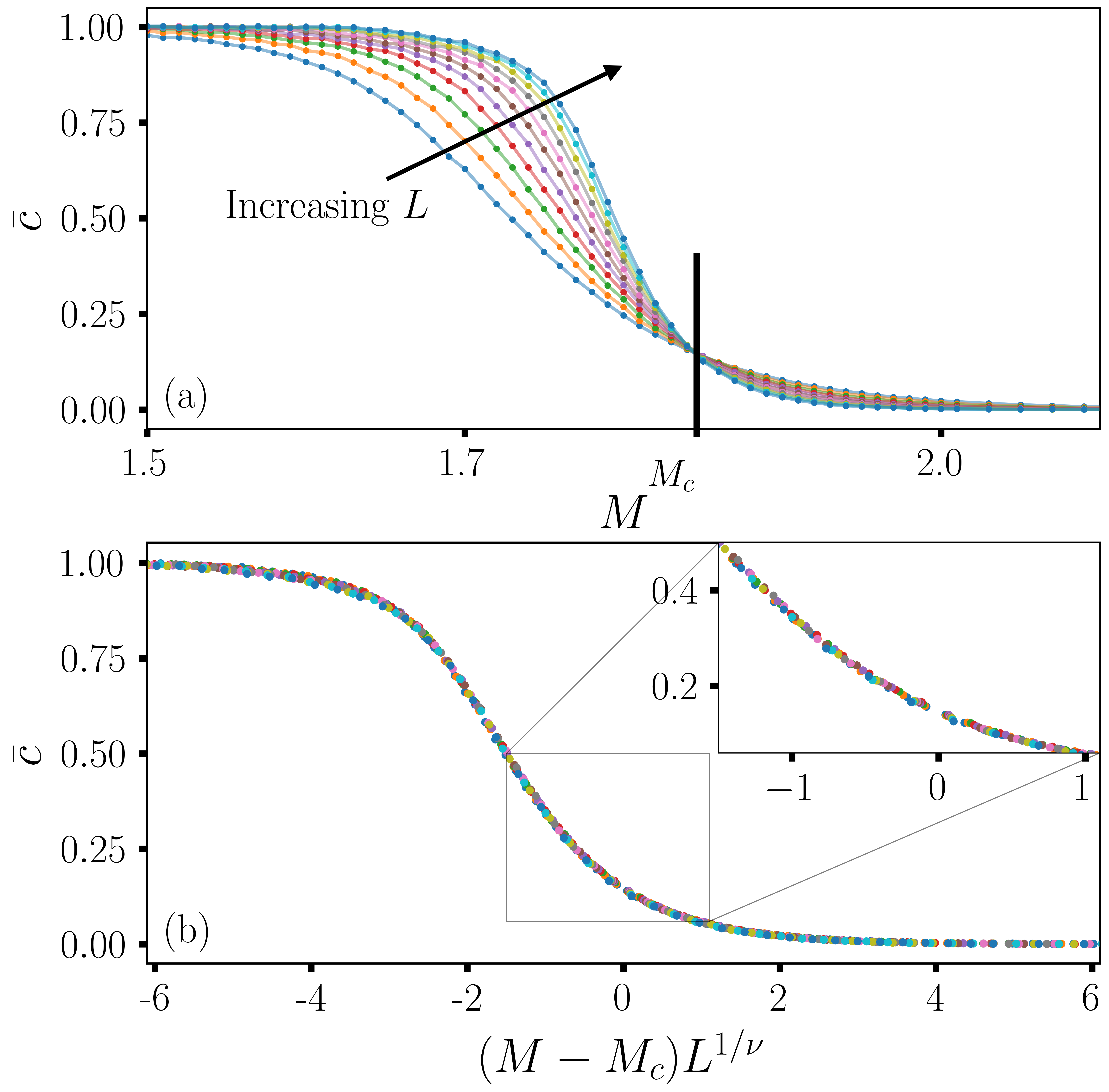}
   \caption{(a) Exact diagonalization results for the disorder
     averaged Chern marker $\bar c$ in the centre of a finite-size
     sample as a function of the inversion-breaking parameter $M$. We set
     $W=1$ and $\varphi=\pi/2$.  The
     results are obtained for $L=13, 15, 17, 19,...,35$, corresponding
     to $2L^2$ site samples, averaged over $10^4$ disorder
     realizations. On transiting from the topological to the non-topological
     phase the topological marker $\bar{c}$ interpolates between
     unity and zero. We extract the value $M_c\sim
     1.85$ from the crossing point. We fit the critical point $M_c = 1.85 \pm 0.01$
     and the critical exponent $\nu=1.06 \pm 0.03$ by maximizing the overlap 
     when plotted as a function of $(M-M_c)L^{1/\nu}$.
     (b) Collapse of the data in (a) for the rescaling
     $(M-M_c)L^{1/\nu}$ of the horizontal
     axis with $M_c=1.85$ and $\nu = 1.06$. This is in agreement with the exponent 
     $\nu=1.05\pm 0.03$ derived via the transfer matrix method.
     Inset: A zoomed-in portion of the data in the vicinity of the critical point illustrating the quality of the collapse.
     }\label{fig:CM_W10}
\end{figure}

  Having examined the disorder-driven transitions in Fig.~\ref{fig:MvsV_phi} we turn our attention to the mass-driven transitions at fixed disorder strength. In Fig.~\ref{fig:CM_W10}(a) we plot the evolution of the Chern marker on transiting from the topological to the non-topological phase with $W=1$ held fixed. The data show a clear crossing point at $M_c \sim 1.85$ in conformity with Fig.~\ref{fig:MvsV_phi}. Re-plotting the data as a function of $(M-M_c)L^{1/\nu}$ with $\nu=1.06\pm 0.03$, the data collapse onto a single curve; see Fig.~\ref{fig:CM_W10}(b). This result is in agreement with transfer matrix calculations which yield $\nu=1.05 \pm 0.03$; see Supplementary Material. This value is close to, but distinct from that of free Dirac fermions with $\nu=1$. We will consider further instances of such departures in Sections~\rom{6} and \rom{7}.

\section{Fluctuations}
\begin{figure}[h]
    \centering  
    \includegraphics[width=.95\columnwidth]{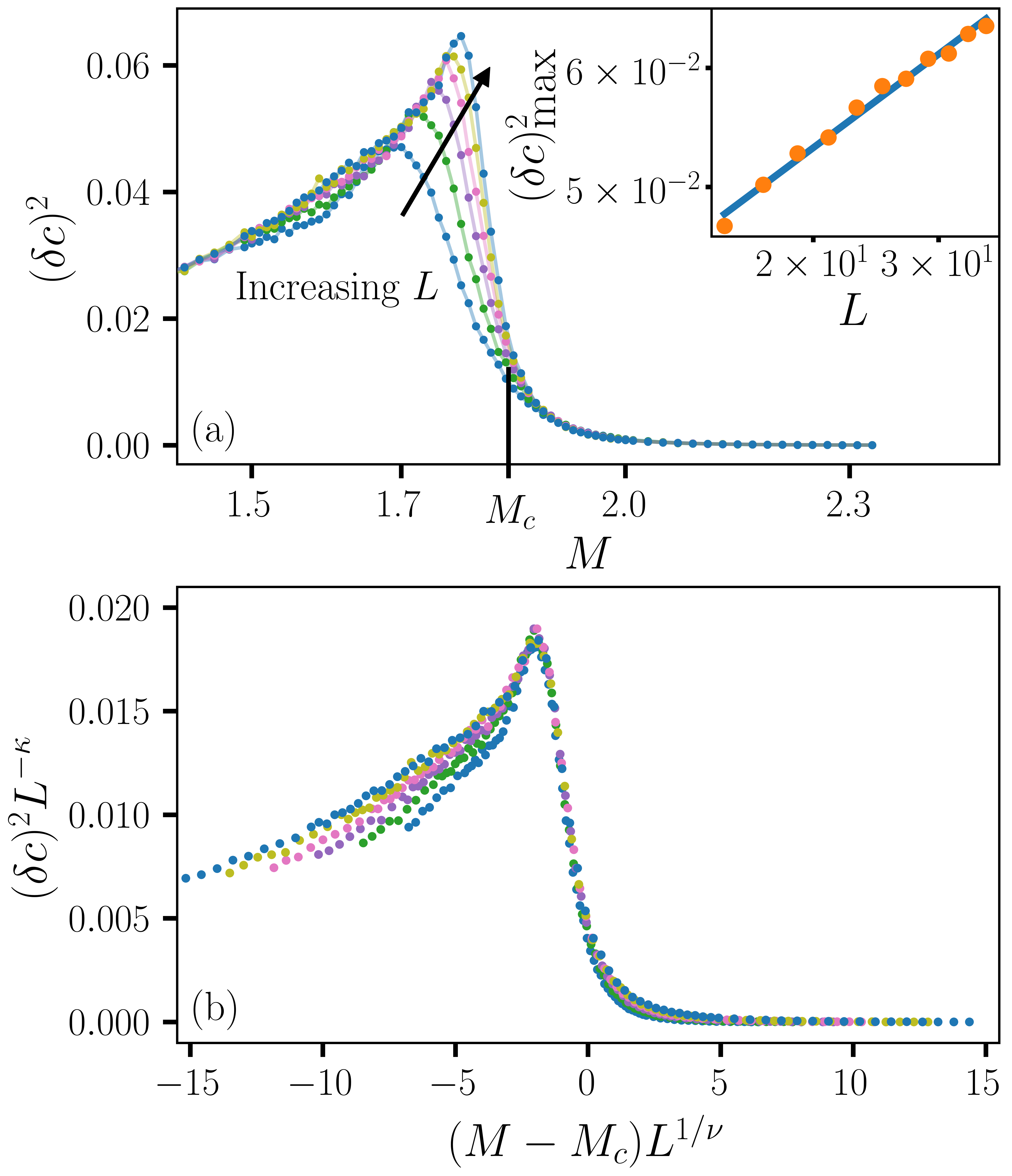}
    \caption{(a) Evolution of the Chern marker fluctuations $(\delta c)^2$ on transiting from the topological to the non-topological phase with $W=1$ held fixed. We set $t_1=3t_2=1$, $\varphi = \pi/2$ and average over $10^4$ disorder realizations with $L=15,19,...,35$. The data show a clear peak on approaching the critical point at $M_c\sim 1.84$. Inset: Evolution of the peak value $(\delta c)^2_\textrm{max}$ with increasing system size. The data corresponds to a power-law divergence $(\delta c)^2_\text{max} \sim L^\kappa$ with $\kappa = 0.36 \pm 0.02$. (b) Scaling collapse of the data shown in panel (a) with $M_c = 1.84$, $\nu = 1.05$ and $\kappa = 0.36$.}
    \label{fig:Fluct_W10}
\end{figure}

Having established the scaling of the disorder averaged Chern marker $\bar{c}$, we now turn our attention to its fluctuations. We examine the sample-to-sample fluctuations of the Chern marker $(\delta c)^2 = \overline{(c - \bar{c})^2}$ in the middle of the system, where the overbar indicates disorder averaging. In Fig.~\ref{fig:Fluct_W10}(a) we plot the evolution of $(\delta c)^2$ on transiting from the topological to the non-topological phase with $W=1$ held fixed. The fluctuations show a clear peak on approaching the critical point at $M_c \sim 1.84$. The value of $(\delta c)^2$ at the peak grows with increasing system size and is consistent with a power-law divergence $(\delta c)^2_\textrm{max} \sim L^\kappa$ with $\kappa = 0.36 \pm 0.02$; see inset of Fig.~\ref{fig:Fluct_W10}(a). We therefore consider a scaling form $(\delta c)^2 \sim L^\kappa g(\xi/L) \sim L^\kappa \tilde{g}((M-M_c)L^{1/\nu})$ in the vicinity of the transition. In Fig.~\ref{fig:Fluct_W10}(b) we plot $(\delta c)^2 L^{-\kappa}$ as a function of $(M-M_c)L^{1/\nu}$ where the values of $M_c = 1.84$ and $\nu = 1.05$ are obtained from the scaling of the Chern marker. The data collapse in the vicinity of the transition highlighting the consistency with our Section~\rom{5} results for $\nu$. In Section \rom{7} we explore the variation of the exponents $\nu$ and $\kappa$ with increasing disorder strength.

\section{Variation of the exponents}
\begin{figure}[h]
  \centering 
   \includegraphics[width=.95\columnwidth]{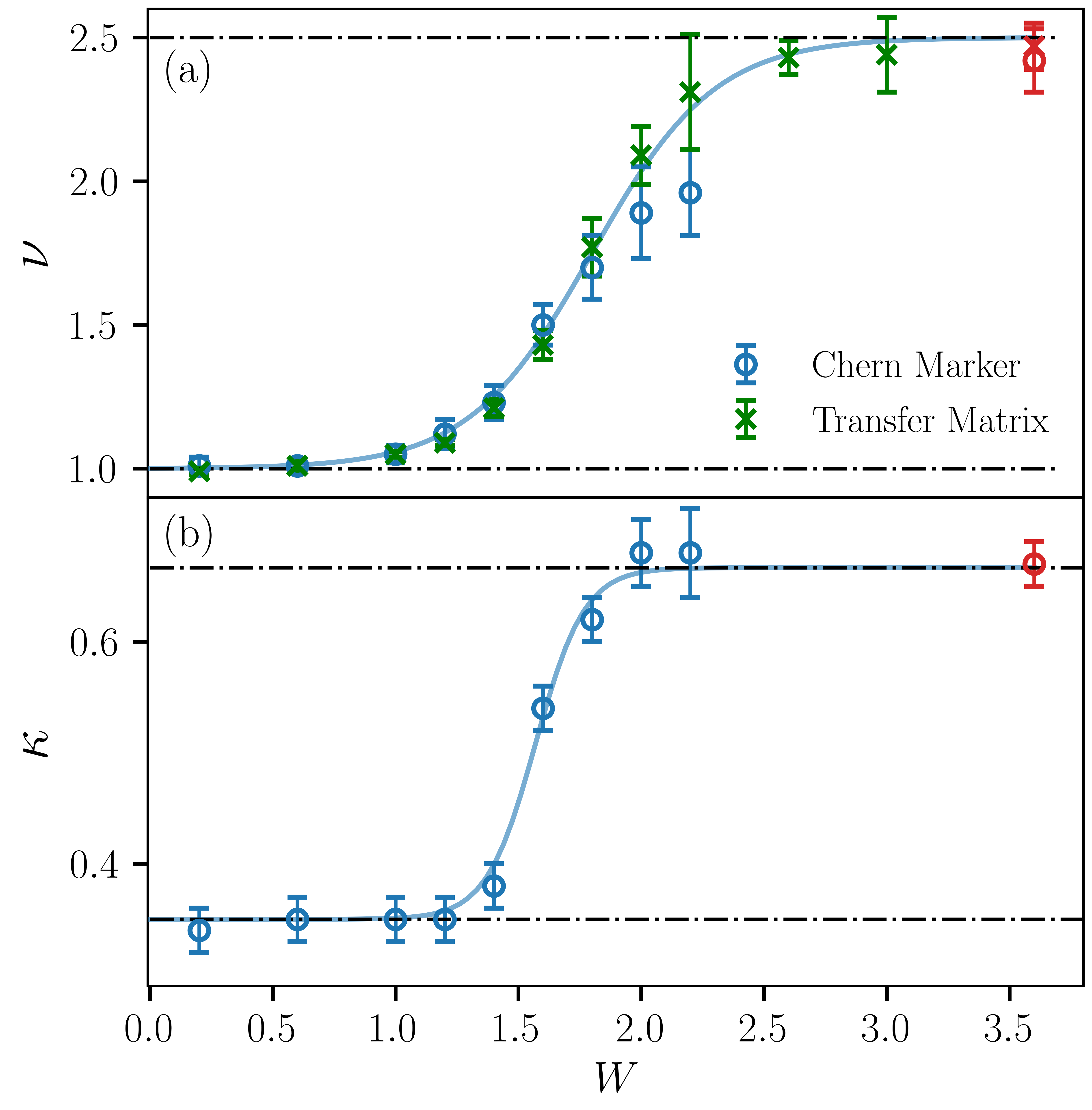}
   \caption{(a) Evolution of the correlation length exponent $\nu$ as a function of $W$ for the mass-driven transitions (vertical arrow) in Fig.~\ref{fig:MvsV_phi}. The results are extracted from the finite-size scaling of the Chern marker (blue circles) and transfer matrices (green crosses). The exponent interpolates between $\nu=1$ corresponding to free Dirac fermions and $\nu \sim 2.5$ (dashed lines), where $W\sim 3.6$ corresponds to the disorder-driven transition. The results for the latter are shown in red. The underlying continuous curve (light blue) is a guide to the eye only.
   (b) Evolution of the fluctuation exponent $\kappa$ for the same transitions in panel (a). The exponent interpolates between $\kappa \sim 0.35$ for $W=0.2$ and $\kappa \sim 0.65$ (red circle) for the disorder-driven transition, as indicated by the dashed lines.}\label{fig:ExponentVariation}
\end{figure}

Having discussed the scaling of the Chern marker and its fluctuations we now consider the variation of $\nu$ and $\kappa$ with $W$. In Fig.~\ref{fig:ExponentVariation}(a) we plot the variation of $\nu$ as we transit along the mass-driven phase boundary in Fig.~\ref{fig:MvsV_phi}. It can be seen that $\nu$ interpolates between that of free Dirac fermions with $\nu=1$ and the value $\nu \sim 5/2$ corresponding to the disorder-driven transitions. The results obtained via the Chern marker are in agreement with those obtained via the transfer matrix approach, although the error bars increase with $W$. The deviation between the results at strong disorder is attributed to finite-size effects in the Chern marker calculations, which are also performed in a different geometry; see Supplementary Material. The fluctuation exponent $\kappa$ shows a similar evolution as $\nu$, interpolating between $\kappa \sim 0.35$ and $\kappa \sim 0.65$; see Fig.~\ref{fig:ExponentVariation}(b). It is notable that Ref.~\cite{Sbierski_2021} also finds evidence for varying $\nu$ as a function of energy in a model of Dirac fermions. However, this is over a much smaller range of values, between $2.33$ and $2.53$. Here, we provide evidence for a very strong variation of the exponents over a wide range of parameters.

The results contained in Fig.~\ref{fig:ExponentVariation} raise a number of questions and scenarios. One possibility is that the disordered Haldane model exhibits a line of continuously varying exponents $\nu$ and $\kappa$, due to the presence of marginal perturbations \cite{Dresselhaus_2021, Zirnbauer_2021}. Another possibility is that the weak disorder regime shows very slow convergence to the thermodynamic limit and will ultimately flow to $\nu \sim 5/2$. Another scenario is the possibility of distinct fixed points at both weak and strong disorder to which the system will eventually flow. All of these scenarios may lead to the extraction of effective exponents, $\nu_\textrm{eff}$ and $\kappa_\textrm{eff}$, for the system sizes considered. It would be interesting to explore these possibilities in more detail. In closing, we note that a drift of the critical exponent $\nu$ was observed in early work on the site diluted Ising model \cite{Kuhn_1994}. This has been recently attributed to the effect of logarithmic corrections~\cite{Fytas_2010, Schrauth_2018}.

\section{Conclusions}
In this work we have explored the critical behavior of the disordered Haldane model using both the Chern marker and transfer matrix calculations. We provide evidence for disorder-driven transitions with $\nu \sim 5/2$. Our findings are also consistent with a line of fixed points with a continuously varying exponent which interpolates between $\nu=1$ and $\nu \sim 5/2$. It would be interesting to explore the latter in more detail both numerically and analytically. We have also introduced an exponent $\kappa$ associated with the power-law divergence of the Chern marker fluctuations in the vicinity of topological phase transitions. We provide numerical evidence for its variation along the mass-driven phase boundary. These results may provide a useful starting point to explore the drift in exponents found in other works on the IQHE \cite{Sbierski_2021, Dresselhaus_2021}.

\section{Acknowledgements}
We acknowledge helpful conversations with M. Foster, C. von Keyserlingk, R. K\"uhn, J. Pixley
and L. Privitera. The early stages of this work was supported by EPSRC
grants EP/J017639/1 and EP/K030094/1, by the Netherlands Organization
for Scientific Research (NWO/OCW), and by an ERC Synergy Grant. GM was
supported by the Royal Society under grant URF\textbackslash R\textbackslash 180004. NRC was
supported by EPSRC Grants EP/P034616/1 and EP/V062654/1 and by the
Simons Investigator Grant No. 511029. MJB was supported by the National
Renewable Energy Laboratory and thanks the EPSRC Centre for Doctoral
Training in Cross-Disciplinary Approaches to Non-Equilibrium Systems
(CANES) funded under grant EP/L015854/1. MJB, MDC and JM acknowledge the
Thomas Young Centre and the use of Create for numerical simulations. The data presented in this work is available upon request. For the purpose of open access, the authors have applied a creative commons attribution (CC BY) licence.

\clearpage

\beginsupplement

\onecolumngrid
\begin{center}
    \textbf{\large Supplementary Material\\ Topological Phase Transitions in the Disordered Haldane Model}
\end{center}
\vspace{0.3in}
\twocolumngrid

Here we provide further details of the Chern marker and transfer matrix calculations performed in the main text. We also provide additional results to support our findings.

\section{Real-space Chern Marker}
Topological order in non-interacting systems is often described by the
presence of a topologically non-trivial texture in the ground state
wavefunction. In the case of Chern insulators, this topological texture is usually
characterised by the global Chern index \cite{SM_Chern_1946}:
\begin{equation}
 C=-\frac{1}{\pi}{\rm Im}\sum_n^{\rm occ}\!\!\int_{BZ}\!\!\!\! d\v k\, 
\langle \partial_{k_x} u_{n \v k}| \partial_{k_y} u_{n \v k}\rangle,  
\label{eq:chern_number}
\end{equation}
where $u_{n \v k} (\v r) = e^{- i \v k \cdot \v r} \psi_{n \v k}(\v
r)$ is the periodic part of the occupied Bloch states for the $n$-th
band. Although $C$ is usually expressed in momentum space,
topological order can have consequences even when this is not
permitted, e.g. in finite-size systems with open boundaries or in the
presence of disorder. In view of this, a local topological
characteristic known as the Chern marker has been introduced by
Bianco and Resta \cite{SM_Resta2011}.
To see this, it is convenient to insert
a complete set of states into Eq.~(\ref{eq:chern_number}):
\begin{equation}
 C =-\frac{1}{\pi}{\rm Im}\sum_n^{\rm occ}\sum_m^{\rm 
unocc}\!\!\int_{BZ}\!\!\!\! d\v k\, \langle \partial_{k_x} u_{n \v 
k}|u_{m \v 
k}\rangle \langle u_{m \v k}| \partial_{k_y} u_{n \v k}\rangle, 
\end{equation}
where the missing terms are real. The derivatives in ${\bf k}$-space
can be recast in real space using
\begin{equation}
  \langle \psi_{m \v k}| \hat{\v r} | \psi_{n \v k} \rangle = i \langle 
u_{m 
\v k}|\partial_{\v k} u_{n \v k}\rangle
\end{equation}
for $m\neq n$; although the position operator is generically
ill-defined in the case of periodic boundary conditions, its
off-diagonal matrix elements are well defined. As such
\begin{align}
 C\!&=\!-\frac{A_c}{4\pi^3}{\rm Im}\sum_n^{\rm occ}\sum_m^{\rm unocc}\!\! 
\int_{BZ}\!\!\!\! d\v k\!\! \int_{BZ}\!\!\!\! d\v k\p \nonumber\\&\qquad\qquad\qquad\qquad\qquad \langle \psi_{n \v k}| 
\hat x | \psi_{m \v k\p} \rangle\langle \psi_{m \v k\p}| \hat y | \psi_{n \v k} 
\rangle \nonumber\\
 &=-\frac{4\pi}{A_c} {\rm Im}\,{\rm Tr}(\hat P\hat x \hat Q \hat y),
\label{eq:lcm}
\end{align}
where for $\v k \neq \v k\p$ the matrix elements vanish, and in the
second line $\hat P$ and $\hat Q=\hat I -\hat P$ are the projectors
onto the occupied and empty states, respectively. The pivotal point to
define the Chern marker is to recognize that the trace is independent
of the representation and can thus be taken in real
space. Eq.~(\ref{eq:lcm}) is finally obtained  using the cyclic
property of the trace and $\hat P^2=\hat P$ \cite{SM_Resta2011}. For
 free-electron systems, the ground state is uniquely determined by the
ground-state projector $P(\v r,\v r\p)=\braket{\v r|\hat P|\v
  r\p}$ which, for insulators, is exponentially decreasing with
$\abs{\v r-\v r\p}$. 
The Chern marker has proved to be efficient in
characterizing the topological phases of finite-size
systems in equilibrium \cite{SM_Resta2011} and out-of-equilibrium
\cite{SM_Privitera_2016}.  The Chern marker has also proven
successful in the presence of disorder~\cite{SM_Resta2011}.

\subsection*{Topological Phase Transition at Fixed Disorder}
\begin{figure}[h]
  \centering 
   \includegraphics[width=0.95\columnwidth]{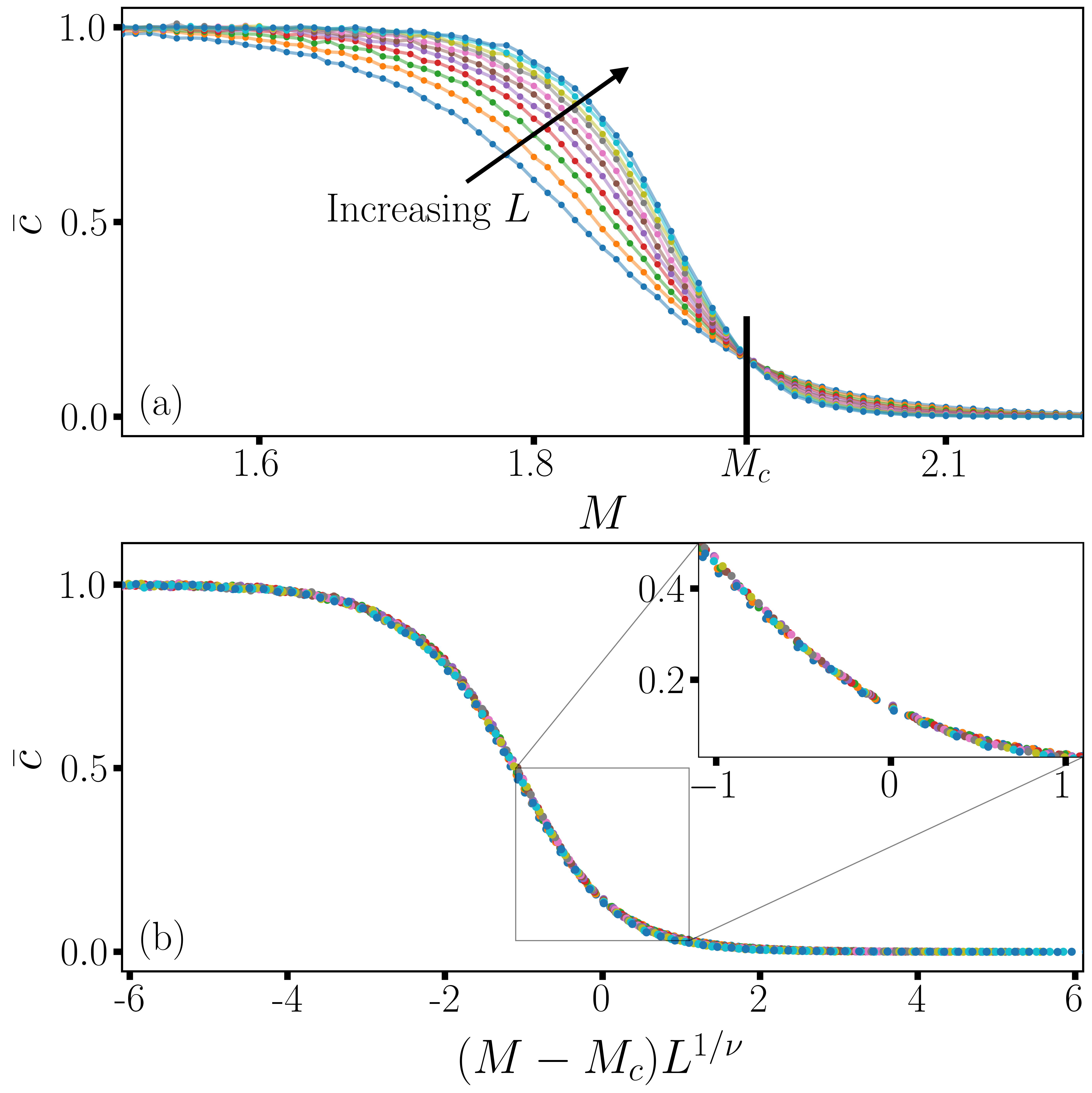}
   \caption{(a) Exact diagonalization results for the disorder
     averaged Chern marker $\bar c$ in the centre of a finite-size
     sample as a function of $M$, for $t_2=t_1/3$, $t_1=1$, $\varphi=\pi/2$
     and $W=1.4$. The results are obtained for $L=13, 15,...,
     35$, corresponding to $2L^2$ sites,
     averaged over $10^4$ disorder realizations. The data show a crossing point in the vicinity of
      $M_c\sim 1.95$ corresponding to the transition between the phases. (b) Scaling collapse of the data in (a) for the rescaling $(M-M_c)L^{1/\nu}$ of the horizontal
     axis with $\nu = 1.23$. The values of $M_c = 1.96 \pm 0.01$ and $\nu = 1.23 \pm 0.05$ are obtained by minimizing the mean-squared deviation in Eq.~(\ref{eq:MSE_CM}).
     Inset: magnified portion of the data in the vicinity of the critical point highlighting the quality of the collapse. The results are in good agreement with those obtained by the transfer matrix approach with $\nu = 1.21 \pm 0.03$; see the section below on intermediate disorder.}
     \label{fig:CM_W14}
\end{figure}

For the phase transitions at fixed disorder strength $W$, obtained by
changing $M$ in Fig.~3 of the main text, we find that the correlation
length exponent $\nu$ increases from unity with increasing disorder strength. Setting
$\varphi=\pi/2$ and taking $W=0.2,0.6,1,1.4,1.6, 1.8$ we find
$\nu=1.02(5),1.02(5),1.05(6), 1.23(0), 1.49(8)$ and $1.70(1)$ respectively, using the Chern marker. For $W \lesssim 1$ these values are
numerically close to the clean result with $\nu=1$~\cite{SM_Caio_2019}.
In Fig.~\ref{fig:CM_W14}(a) we show the disorder averaged topological marker $\bar{c}$ for the topological phase
transition with $W=1.4$ held fixed, and averaged over $10^4$ disorder
realizations. We consider diamond shaped samples with $L$ unit cells along each edge; see Fig.~1 of the main text. The data show a crossing point in the vicinity of the critical point at $M_c\sim 1.96$. The data exhibits scaling collapse when plotted as a function of $(M-M_c)L^{1/\nu}$ with $\nu = 1.23$ and $M_c = 1.96$; see Fig.~\ref{fig:CM_W14}(b). The parameters $\nu$ and $M_c$ can be obtained by minimising the mean-squared distance between the rescaled curves in Fig.~\ref{fig:CM_W14}(b). Explicitly, we minimize the square deviation 
\begin{equation}
    D=\frac{1}{(\# L)^2}\sum_{L,L'} \int_\Omega d\tilde{m} \; \Big(\bar{c}_L (\tilde{m} ) - \bar{c}_{L'} (\tilde{m}) \Big)^2,
\label{eq:MSE_CM}
\end{equation}
where $\bar{c}_L$ is the disorder averaged Chern marker and $\tilde{m} = (M-M_c)L^{1/\nu}$ for a system of size $L$. Here, $\# L$ is the number of system sizes used. The inset of Fig.~\ref{fig:CM_W14}(b) highlights the quality of the scaling collapse. 

\begin{figure}[h]
  \centering 
   \includegraphics[width=0.95\columnwidth]{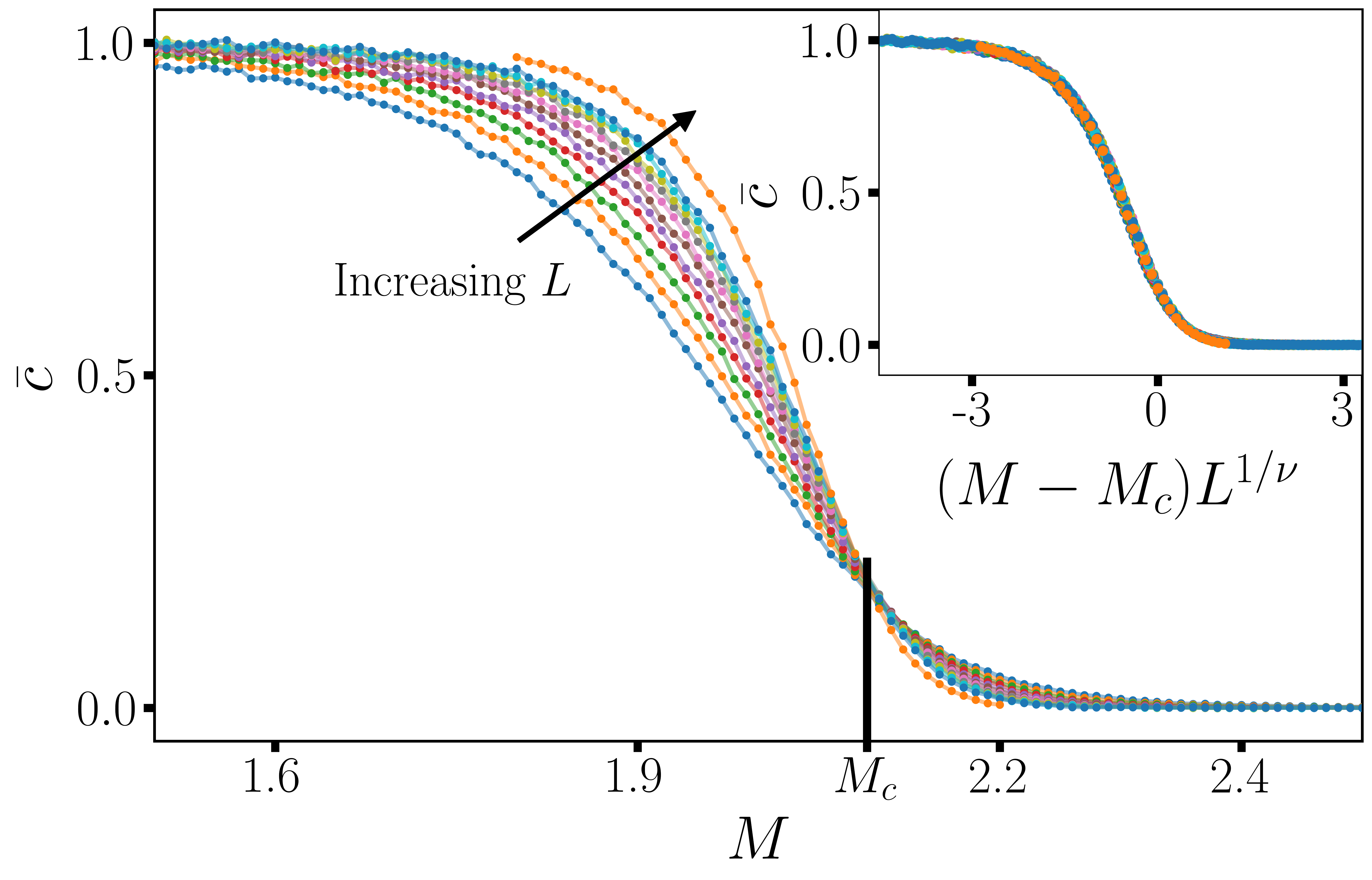}
   \caption{Exact diagonalization results for the disorder
     averaged Chern marker $\bar c$ in the centre of a finite-size
     sample as a function of $M$, for $t_2=t_1/3$, $t_1=1$, $\varphi=\pi/2$
     and $W=1.8$. The results are obtained for $L=13, 15,...,
     35$ and $L=49$, corresponding to $2L^2$ sites,
     averaged over $10^4$ disorder realizations. Inspection of the plot
     yields $M_c\sim 2.1$, as the critical value for the topological
     transition associated with the crossing point. Inset: Collapse of the data in
     the main figure for the rescaling $(M-M_c)L^{1/\nu}$ of the horizontal
     axis with $\nu = 1.70$. The values of $M_c = 2.09 \pm 0.01$ and $\nu = 1.70 \pm 0.09$ are obtained by minimizing the squared deviation between the rescaled curves. The result is in agreement with $\nu=1.77 \pm 0.10$ obtained via the transfer matrix approach.}
     \label{fig:CM_W18}
\end{figure}

In Fig.~\ref{fig:CM_W18} we show the disorder averaged Chern marker on transiting from the topological to the non-topological phase with $W=1.8$. The data show a crossing point in the vicinity of $M_c \sim 2.1$. The data exhibit scaling collapse when plotted as a function of $(M-M_c) L^{1/\nu}$ with $\nu = 1.70$ and $M_c=2.09$; see the inset. These values are obtained by minimizing the mean-squared displacement in Eq.~(\ref{eq:MSE_CM}). The data collapse onto a single curve verifying the non-trivial results.

\section{Transfer matrix method}
\begin{figure}[ht]
\includegraphics[width=.4\textwidth]{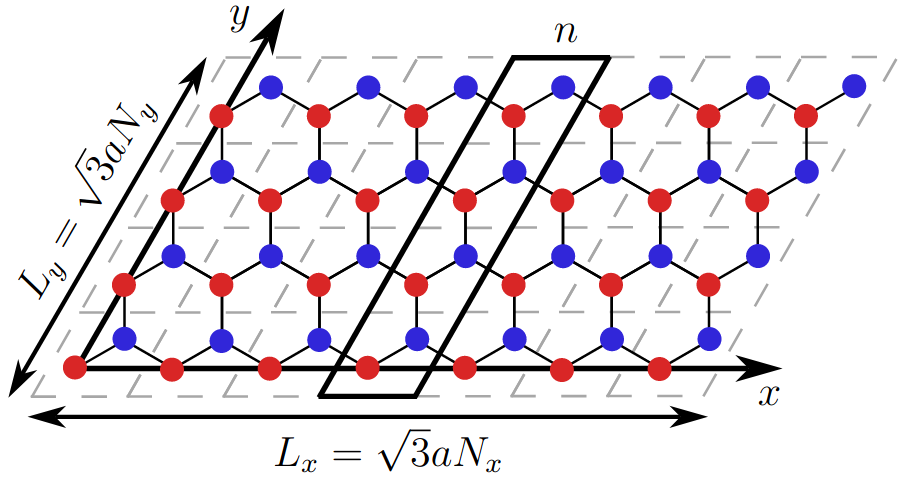}
\caption{Honeycomb lattice structure of the Haldane model where the red and blue dots indicate the two sublattices. The unit cells are shown by dashed lines. The lattice is divided into one-dimensional slices indexed by an integer $n$, as indicated by the solid black line. The transfer matrix $\mathbb{T}$ connects the wavefunctions in adjacent slices.}
\label{fig:unit_cell}
\end{figure}

In the main text we extract the correlation length exponent $\nu$ for the Haldane model via the real-space Chern marker and via the transfer matrix approach. Here, we provide further details on the latter. The derivation of the transfer matrix starts from a real-space discretization of the Schr\"odinger equation, $\hat{H} \ket{\psi} = E\ket{\psi}$ \cite{SM_MacKinnon_1981,SM_Pendry_1992}. In the case of a two-dimensional lattice, the system is split into one-dimensional slices, indexed by the position $n$ along the side of length $L_x$. In the case of the Haldane model $L_x = \sqrt{3} a N_x$, where $N_x$ is the number of slices in the $x$-direction and $a$ is the nearest neighbor lattice spacing; see Fig.~\ref{fig:unit_cell}. For Hamiltonians with short-range hopping the Schr\"odinger equation reduces to
\begin{equation}
    \mathbb{J}\psi_{n+1} + \mathbb{M}\psi_n +\mathbb{J}^\dagger \psi_{n-1} = E \psi_n,
    \label{eq:SchEqSlice}
\end{equation}
where $\psi_n = \braket{n\rvert\psi}$ is the real-space wavefunction for slice~$n$. Here, $\mathbb{J}= \braket{n \lvert \hat{H} \rvert n+1}$ is the hopping matrix for nearest neighbour slices and $\mathbb{M}=\braket{n \lvert \hat{H} \rvert n}$ is a local contribution within a slice \cite{SM_Chua_2016}. The matrices $\mathbb{J}$ and $\mathbb{M}$ are evaluated for a fixed $n$, and have dimension $N_y n_c \times N_y n_c$, where $n_c$ is the number of sites in the unit cell and $N_y$ is the number of unit cells per slice. For the Haldane model $n_c=2$, as shown in Fig.~\ref{fig:unit_cell}.

\noindent
The Schr\"odinger equation (\ref{eq:SchEqSlice}) can be recast as
\begin{equation}
    \begin{pmatrix}
        \psi_{n+1} \\
        \psi_n
    \end{pmatrix} =
        \mathbb{T}_n\begin{pmatrix}
        \psi_{n} \\
        \psi_{n-1}
    \end{pmatrix},
\end{equation}
where the transfer matrix $\mathbb{T}_n$ is given by
\begin{equation}
    \mathbb{T}_n=\begin{pmatrix}
        \mathbb{J}^{-1}(E\Id -\mathbb{M}) & -\mathbb{J}^{-1}\mathbb{J}^\dagger\\
        \Id & 0
    \end{pmatrix}.
    \label{eq:TrMat}
\end{equation}
 This is a square matrix of dimension $2N_yn_c \times 2N_yn_c$. For a fixed value of $E$, the diagonalisation of $\mathbb{T}_n$ is faster than that of $\hat{H}$ due to the reduction in size. The transfer matrix $\mathbb{T}$ across the whole system is obtained by multiplying the transfer matrices for each slice:
\begin{equation}
    \mathbb{T} = \prod_{n=1}^{N_x} \mathbb{T}_n,
    \label{eq:TM_def}
\end{equation}
where we set $\psi_{N_{x}+1}=\mathbf{0}$ and $\psi_0 = \mathbf{0}$. In the absence of disorder the transfer matrices $\mathbb{T}_n$ coincide and $\mathbb{T}=(\mathbb{T}_1)^{N_x}$. In general, $\mathbb{T}$ is non-Hermitian, as follows from Eq.~(\ref{eq:TrMat}). It is therefore convenient to define the Hermitian matrix \cite{SM_Mackinnon_1983_ST}
\begin{equation}
    \Omega = \ln{\bigl(\mathbb{T}^\dagger \mathbb{T}\bigr)}.
\end{equation}
The eigenvalues of $\Omega$ come in pairs with opposite signs, $\pm\lambda_j$, where $j=1,...,N_y n_c$. This reflects the conservation of probability flux.
The inverse correlation length is obtained from the smallest positive eigenvalue:
\begin{equation}
    \xi ^{-1} = \lim_{N_x \rightarrow \infty} \frac{\min_j \lvert \lambda_j \rvert }{2L_x}.
    \label{eq:corr_length}
\end{equation}
In practice, this is obtained for a large but finite $N_x$.

\section{Boundary Conditions}
\begin{figure}[ht]
    \includegraphics[width=0.35\textwidth]{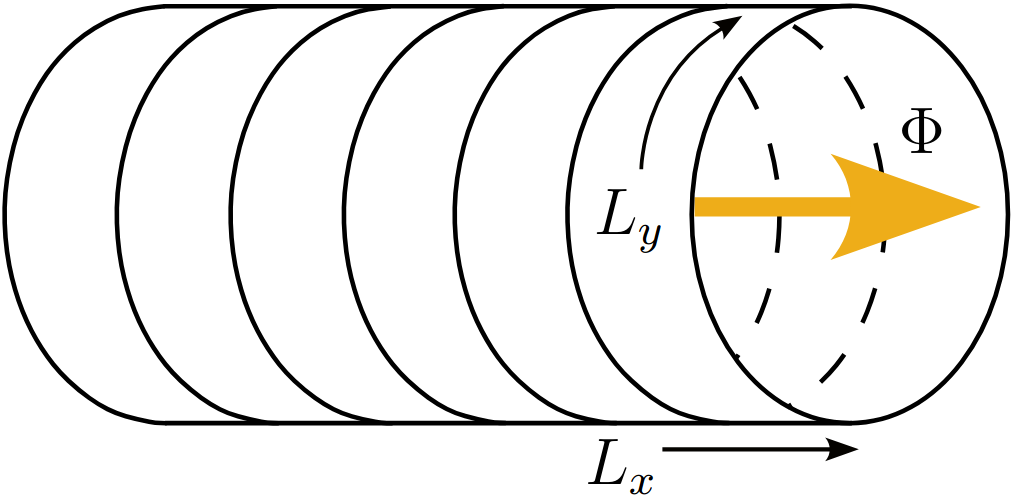}
    \caption{Cylinder geometry used for the transfer matrix calculations, with $L_x \gg L_y$. In the case of the Haldane model, it is convenient to choose twisted boundary conditions. This can be achieved by threading a magnetic flux $\Phi$ through the system with periodic boundary conditions in the $y$-direction.}
    \label{fig:Cylinder}
\end{figure}

\begin{figure}[ht]
    \includegraphics[width=0.3\textwidth]{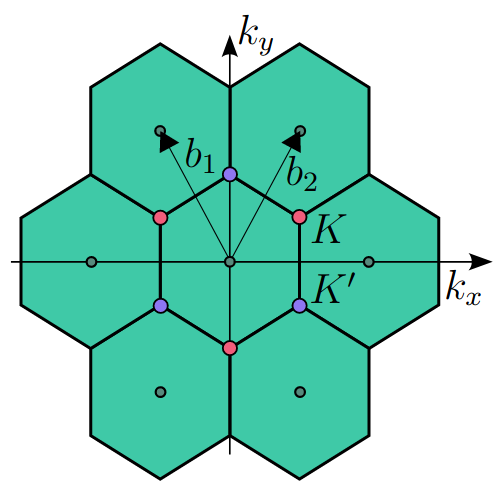}
    \caption{Reciprocal lattice of the Haldane model showing the Dirac points at $K=\frac{2\pi}{3a} (1,1/\sqrt{3})$ and $K'=\frac{2\pi}{3a}(1,-1/\sqrt{3})$. For system sizes $N_y$ that are multiples of $3$ the allowed values of $k_y$ includes the $y$-component corresponding to $K'$ as shown in Fig.~\ref{fig:TMIssue} . This leads to the distinct linear behavior shown in Fig.~\ref{fig:TMIssue2}.}
    \label{fig:BrillZone}
\end{figure}

\begin{figure}[h]
\includegraphics[width=.43\textwidth]{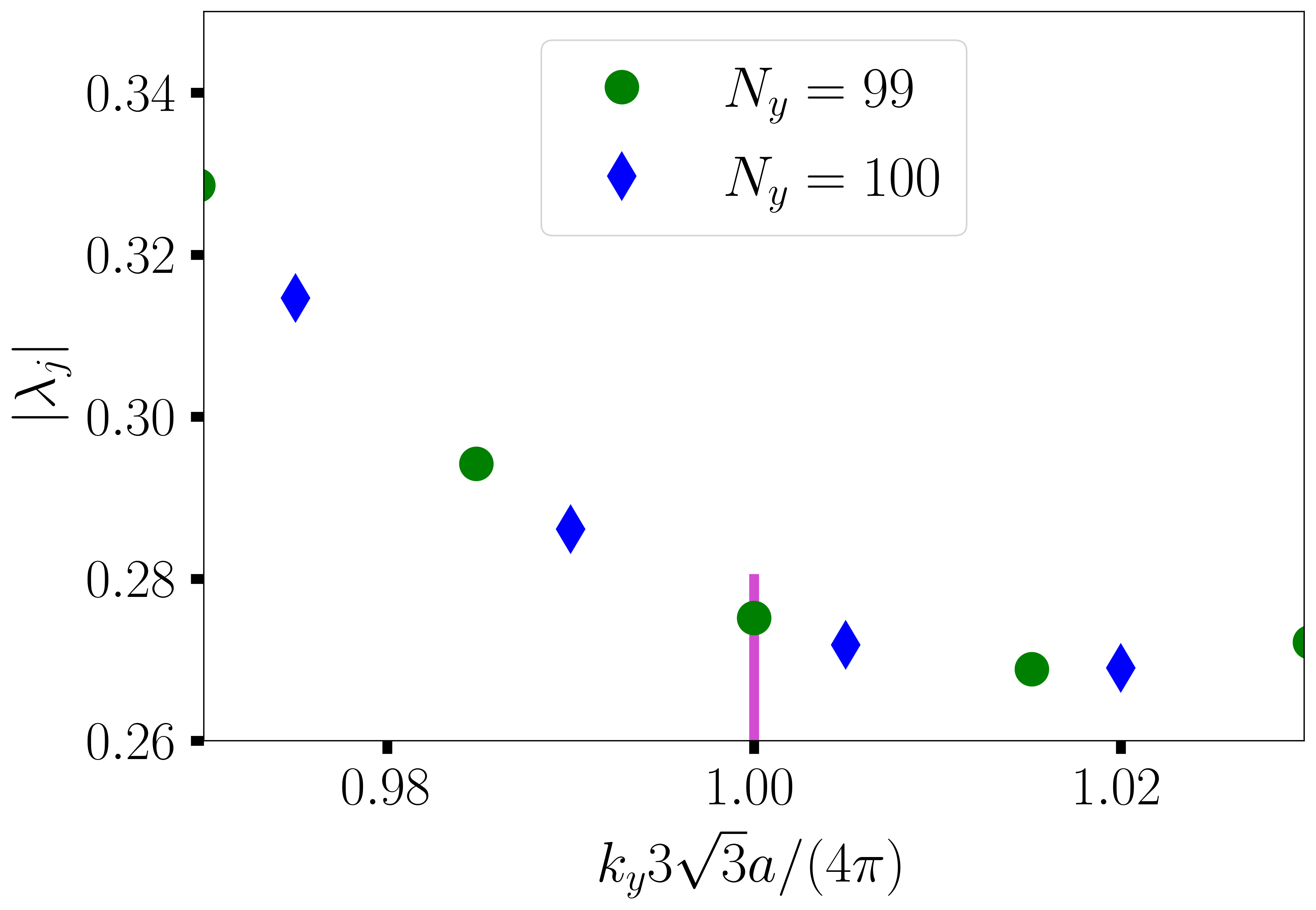}
\caption{Eigenvalues of $\Omega$ for the clean Haldane model with PBCs in the $y$-direction. We set $t_1=3t_2=1$, $\varphi=\pi/2$, corresponding to a point in the topological phase with $C=1$. For system sizes $N_y$ that are multiples of $3$ the spectrum of $\Omega$ includes the $y$-component of the Dirac momentum (vertical line). On transiting from the topological to the non-topological phase the lowest eigenvalue corresponding to the Dirac momentum goes to zero at the critical point as illustrated in Fig.~\ref{fig:TMIssue2}. In order to treat all system sizes on an equal footing we impose TBCs to shift the momenta away from the Dirac point.}
\label{fig:TMIssue}
\end{figure}

\begin{figure}[h]
\includegraphics[width=.43\textwidth]{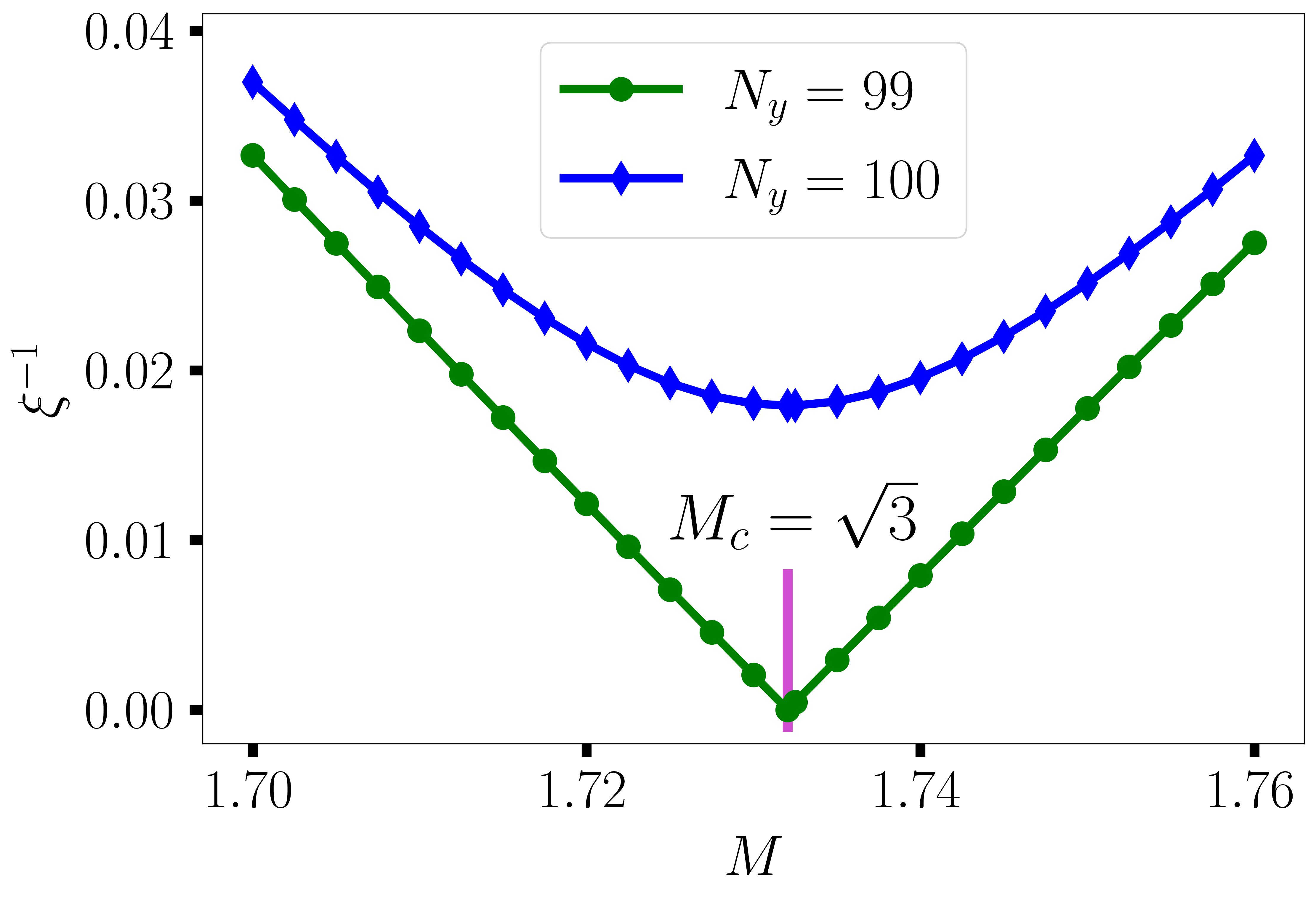}
\caption{Evolution of the inverse correlation length $\xi^{-1}$ of the clean Haldane model on transiting from the topological to the non-topological phase. We set $t_1=3t_2=1$, $\varphi=\pi/2$ and impose PBCs in the $y$-direction. For system sizes $N_y$ that are multiples of $3$ the gap closes linearly in the vicinity of the critical point at $M_c=\sqrt{3}$ (vertical line) due to the inclusion of the Dirac point. For system sizes $N_y$ that are not multiples of $3$, the gap closing is rounded as the Dirac momentum is not included in the $k$-space grid. This may be regarded as a finite-size effect. The evolution of $\xi^{-1}$ with increasing system size is illustrated in Fig.~\ref{fig:TMIssue3}.}
\label{fig:TMIssue2}
\end{figure}

\begin{figure}[ht]
\includegraphics[width=.43\textwidth]{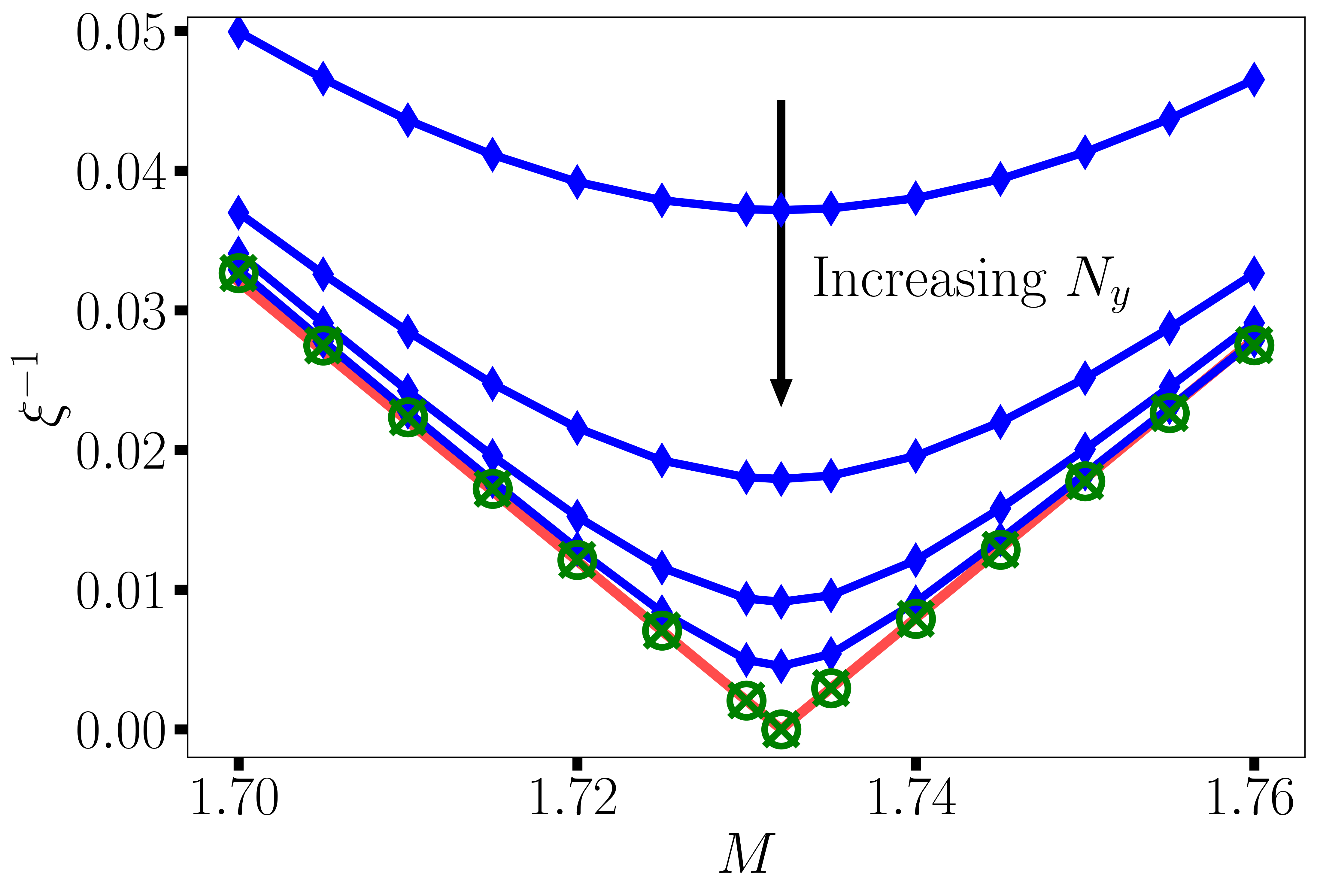}
\caption{Evolution of $\xi^{-1}$ with increasing system size for the topological phase transition illustrated in Fig.~\ref{fig:TMIssue2}. The results for $N_y=50,100,200,400$ (blue) that are not multiples of $3$ converge to the linear gap-closing of the low-energy Dirac theory (red line). The results for relatively small system sizes $N_y = 12$ (open circles) and $N_y=24$ (crosses) that are multiples of $3$ are in very good agreement with the linear prediction of the Dirac theory. This is due to the inclusion of the Dirac momentum as illustrated in Fig.~\ref{fig:TMIssue}. We have checked that similar behaviour is observed for other values of $\varphi$ including $\varphi=\pi/6, \pi/4$.}
\label{fig:TMIssue3}
\end{figure}

To compute the correlation length from Eq.~(\ref{eq:corr_length}) one typically chooses a strip geometry where $L_x \gg L_y$. It is often convenient to impose periodic boundary conditions (PBCs) in the $y$-direction, as shown in Fig.~\ref{fig:Cylinder}. As we will discuss in more detail below, for the Haldane model, it turns out to be more convenient to use twisted boundary conditions (TBCs). In addition, the matrix $\mathbb{J}$ has two vanishing eigenvalues and is not invertible. One approach is to separate the singular and non-singular contributions following the general treatment of Ref.~\cite{SM_Chua_2016}. For the model with PBCs in the $y$-direction, this leads to distinct physical behavior when $N_y$ is a multiple of $3$. In this case, the allowed momenta $k_y=2\pi j/L_y$, with $j=0,...,N_y-1$, include the $y$-momentum of the Dirac point at $(0,4\pi / (3\sqrt{3}a))$; see Fig.~\ref{fig:BrillZone}. This is further illustrated in Fig.~\ref{fig:TMIssue} which shows the eigenvalues of $\Omega$ plotted as a function of the rescaled momentum in the $y$-direction. For the particular choice of $N_y=99$ it can be seen that one of the eigenvalues corresponds to the Dirac momentum $k_y = 4\pi / (3\sqrt{3}a)$. The impact of this can be seen in Fig.~\ref{fig:TMIssue2} which shows the variation of the inverse correlation length on passing from the topological to the non-topological phase. The linear gap closing for $N_y=99$ is consistent with the inclusion of the Dirac point. We further consider the finite-size scaling of systems with $N_y \textrm{ mod } 3 \neq 0$ and show that the Dirac point is approached with increasing $N_y$. This is demonstrated in Fig.~\ref{fig:TMIssue3} which shows the evolution of the inverse correlation length $\xi^{-1}$ with increasing system size. It can be seen that the results converge towards the linear gap-closing expected on the basis of the low-energy Dirac Hamiltonian:
\begin{equation}
    \hat{H}(p,\gamma) = \sum_{\alpha} \hat{H}_\alpha; \quad \hat{H}_\alpha=
    \begin{pmatrix}
        m_{\alpha} c^2 & -cpe^{\iu \alpha \gamma}\\
        -cpe^{-\iu \alpha \gamma} & -m_{\alpha} c^2
    \end{pmatrix},
    \label{eq:DiracHam}
\end{equation}
where $\alpha=\pm 1$ labels the Dirac point. Here, $c=3t_1a/(2\hbar)$ is the effective speed of light, $pe^{\iu \alpha \gamma} = p_x + \iu \alpha p_y$ is the $2$D momentum $(p_x,p_y)$ mapped onto the complex plane, and $m_{\alpha} = (M-3\sqrt{3}\alpha t_2 \sin{\varphi})/c^2$ is the effective mass.
The energy bands in the vicinity of the critical point at $M_c=3\sqrt{3}\alpha t_2 \sin{\varphi}$ are given by $E_\pm (M) = \pm \epsilon |M-M_c|^{\nu}$ (where $\pm$ refers to the upper and lower bands respectively) with $\nu=1$ and $\epsilon=1$. This yields the inverse correlation length $\xi^{-1} = |M-M_c|$, in agreement with the transfer matrix approach; see Fig.~\ref{fig:TMIssue3}.

In order to treat systems with different $L_y$ on an equal footing, it is convenient to impose Twisted Boundary Conditions (TBCs) in the $y$-direction such that
\begin{equation}
    \psi(x,y) = e^{i\theta}\psi(x,y+L_y).
\end{equation}
The twist can be generated by threading a magnetic flux $\Phi= \frac{\hbar}{e}\theta$ through a system with PBCs \cite{SM_Zawadzki_2017}, as shown in Fig.~\ref{fig:Cylinder}. This shifts the wavevector such that $k_y\rightarrow k_y-\theta/L_y = (2\pi n -\theta)/L_y$, for $n=0, ..., N_y-1$, where the twist-angle $\theta$ is defined modulo $2\pi$. This renders $\mathbb{J}$ invertible and the transfer matrix (\ref{eq:TrMat}) can be used directly. It is convenient to choose $\theta$ so that the eigenvalues of $\Omega$ are symmetrically distributed around the Dirac point at $k_y=4\pi/(3\sqrt{3}a)$ for $\varphi = \pi/2$. For $N_y = 0,1,2 \mod 3$ we set $\theta =\pi, 2\pi/3$ and $\pi/3$ respectively.

\section{Finite-size scaling}
\begin{figure}[ht]
\includegraphics[width=.43\textwidth]{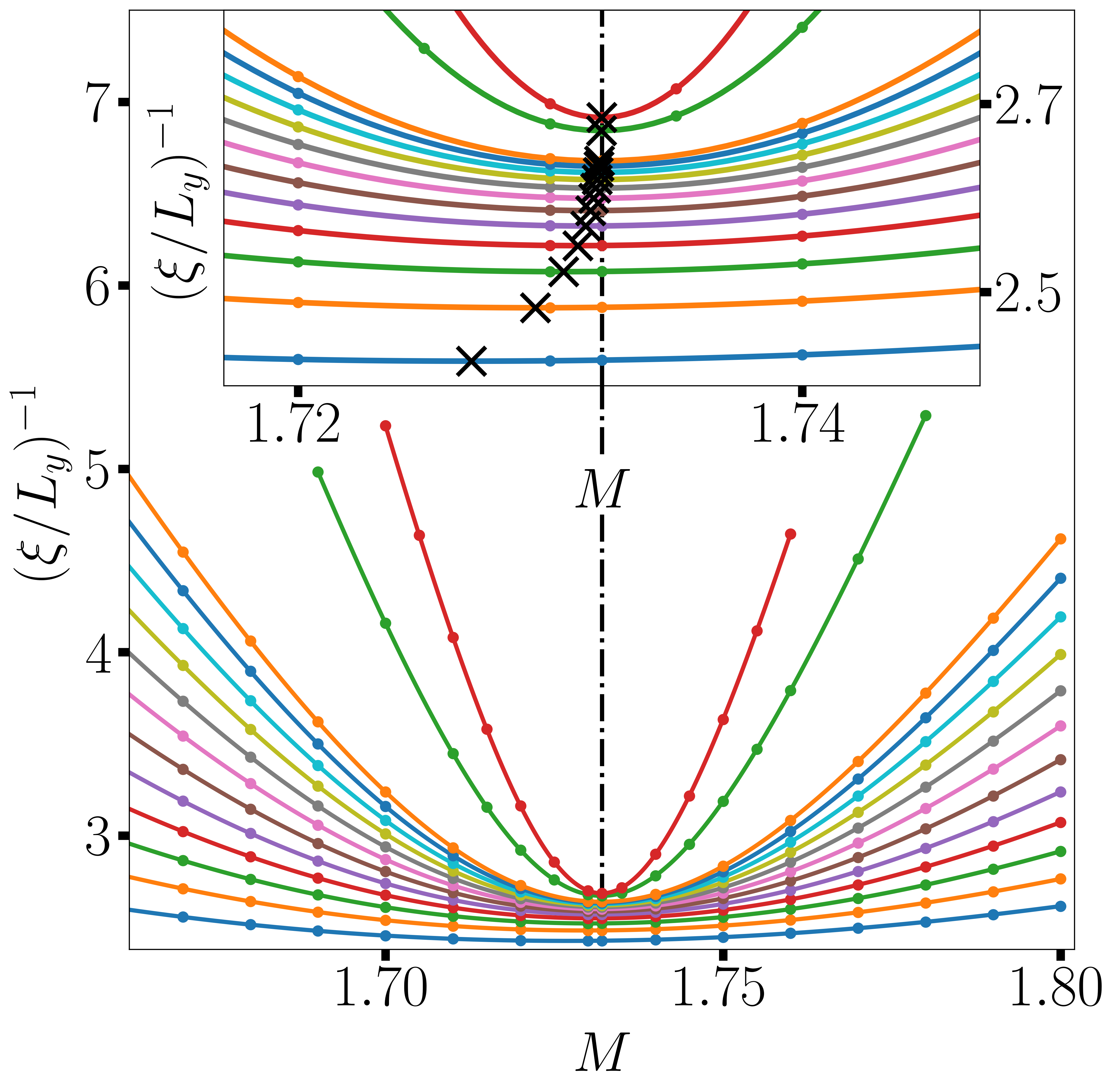}
\caption{Evolution of the dimensionless gap $\Lambda =(\xi/L_y)^{-1}$ across the topological ($C=1$) to non-topological ($C=0$) phase transition for system sizes $N_y=15,19,...,59, 99,139$. We set $t_1=1, t_2=1/3,\varphi=\pi/2$ and impose TBCs. The critical region shrinks as the system size increases. Inset: A zoomed-in plot of $\Lambda$ around the critical point $M_c=\sqrt{3}$ for the same system sizes. The finite-size critical point $M_c(N_y)$ associated with the minimum of $\Lambda$ (black crosses) drifts towards the critical point $M_c$ with increasing system size. This drift is further investigated in Fig.~\ref{fig:TrendClean}.}
\label{fig:raw_data}
\end{figure}

In order to extract the correlation length exponent $\nu$ for the topological phase transition illustrated in Fig.~\ref{fig:TMIssue2}, we perform a finite-size scaling analysis for the correlation length. In the first instance we consider the simple scaling relation 
\begin{equation}
    \xi^{-1} \sim L_y^{-1} f(m N_y^{1/\nu}),
    \label{eq:naive_ansatz}
\end{equation}
where $m = (M-M_c)/M_c$ is the dimensionless distance from the critical point. As we will discuss below it is also necessary to include corrections to Eq.~(\ref{eq:naive_ansatz}) due to irrelevant operators. On the basis of the na\"ive ansatz in Eq.~(\ref{eq:naive_ansatz}) it is natural to investigate the evolution of the dimensionless ratio $\Lambda=(\xi/L_y)^{-1}$ as shown in Fig.~\ref{fig:raw_data}. 

\begin{figure}[ht]%
\includegraphics[width=.48\textwidth]{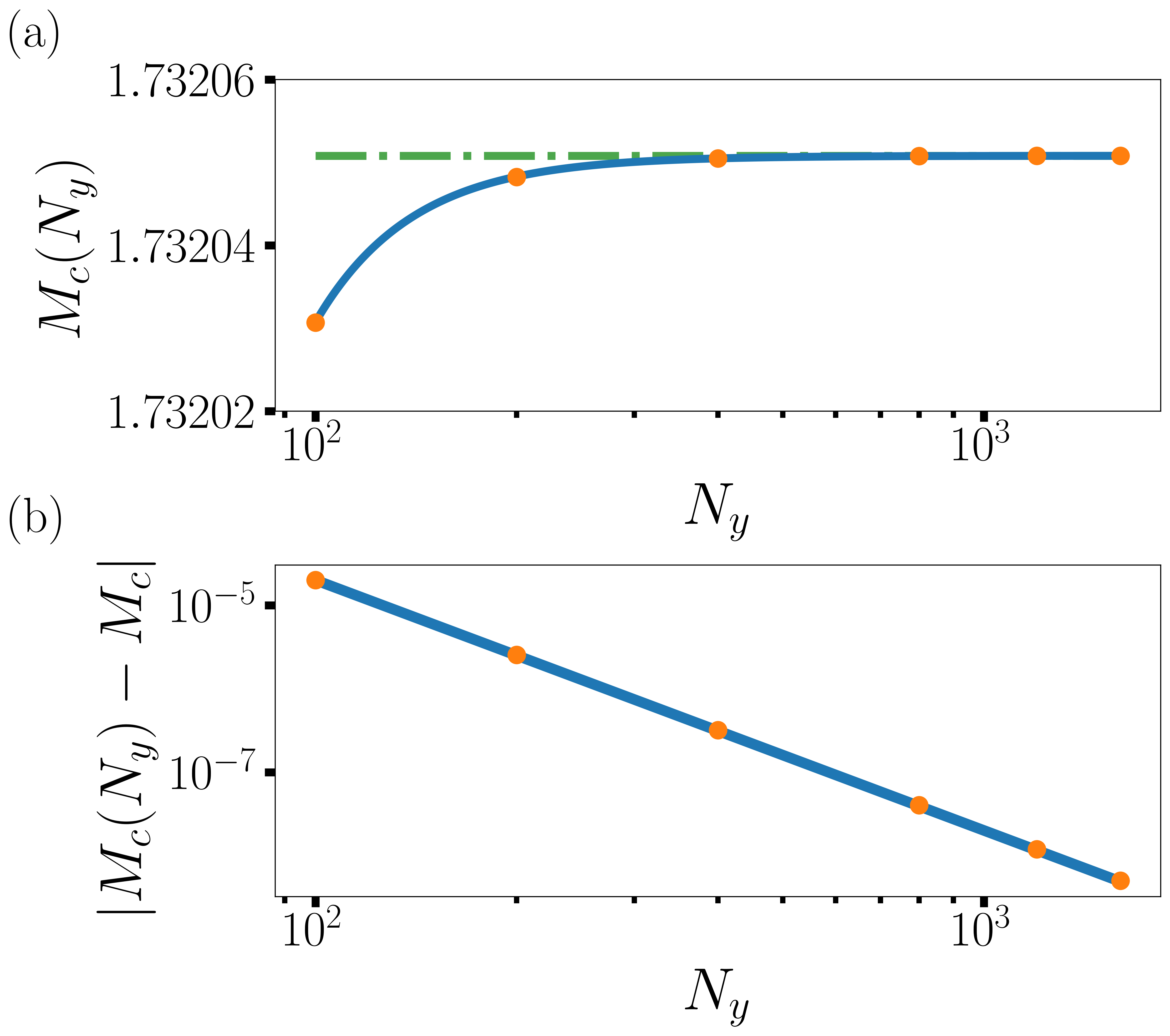}
\caption{(a) Evolution of the finite-size critical point $M_{c}(N_y)$ with increasing system size $N_y$ for the clean Haldane model with $t_1 = 3t_2 = 1$ and $\varphi=\pi/2$. The blue line corresponds to Eq.~(\ref{eq:Xdrift}) with $M_c=\sqrt{3}$, $\lambda = 3$ and $\mathcal{A} = e^3$. The departure of $\lambda$ from $1/\nu = 1$ indicates the presence of irrelevant corrections to Eq.~(\ref{eq:naive_ansatz}). (b) Deviation of the finite-size critical point $M_c(N_y)$ from that in the thermodynamic limit $M_c=M_c(\infty)$. The results are in excellent agreement with the power-law scaling in Eq.~(\ref{eq:Xdrift}) with $\lambda = 3$ (blue line)).}
\label{fig:TrendClean}
\end{figure}

\begin{figure}[ht]
\includegraphics[width=.43\textwidth]{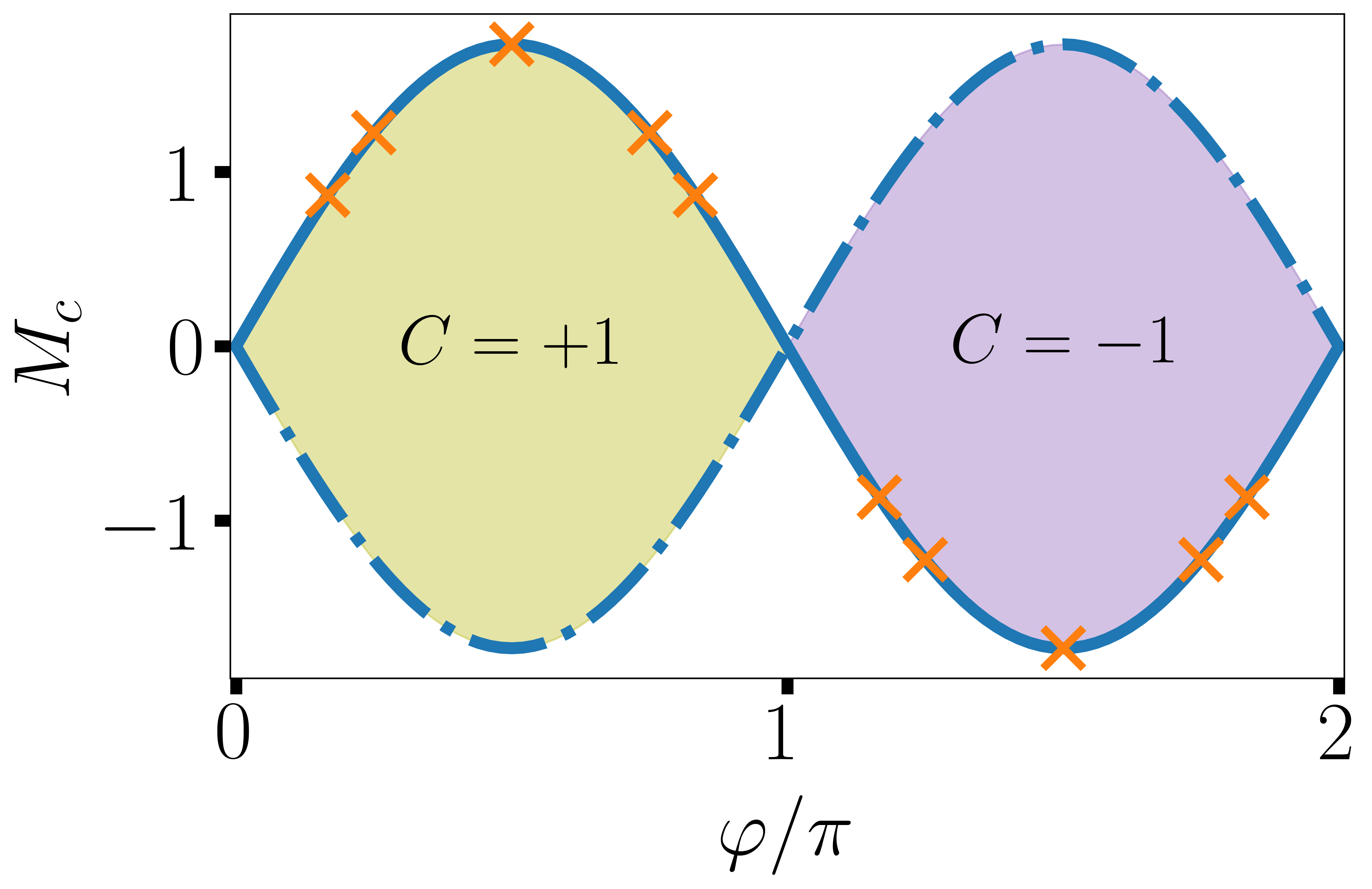}
\caption{Evolution of the infinite-size critical point $M_c$ as a function of $\varphi$ obtained from the finite-size scaling relation in Eq.~(\ref{eq:Xdrift}). The results are in excellent agreement with the field-theory result $M_c(\varphi)=\sqrt{3} \sin{\varphi}$ corresponding to the gap-closing of a single Dirac point (solid blue line). The closing at the other Dirac point is indicated by a dashed line.}
\label{fig:phase_diagram}
\end{figure}

\begin{figure}[ht]
\includegraphics[width=.45\textwidth]{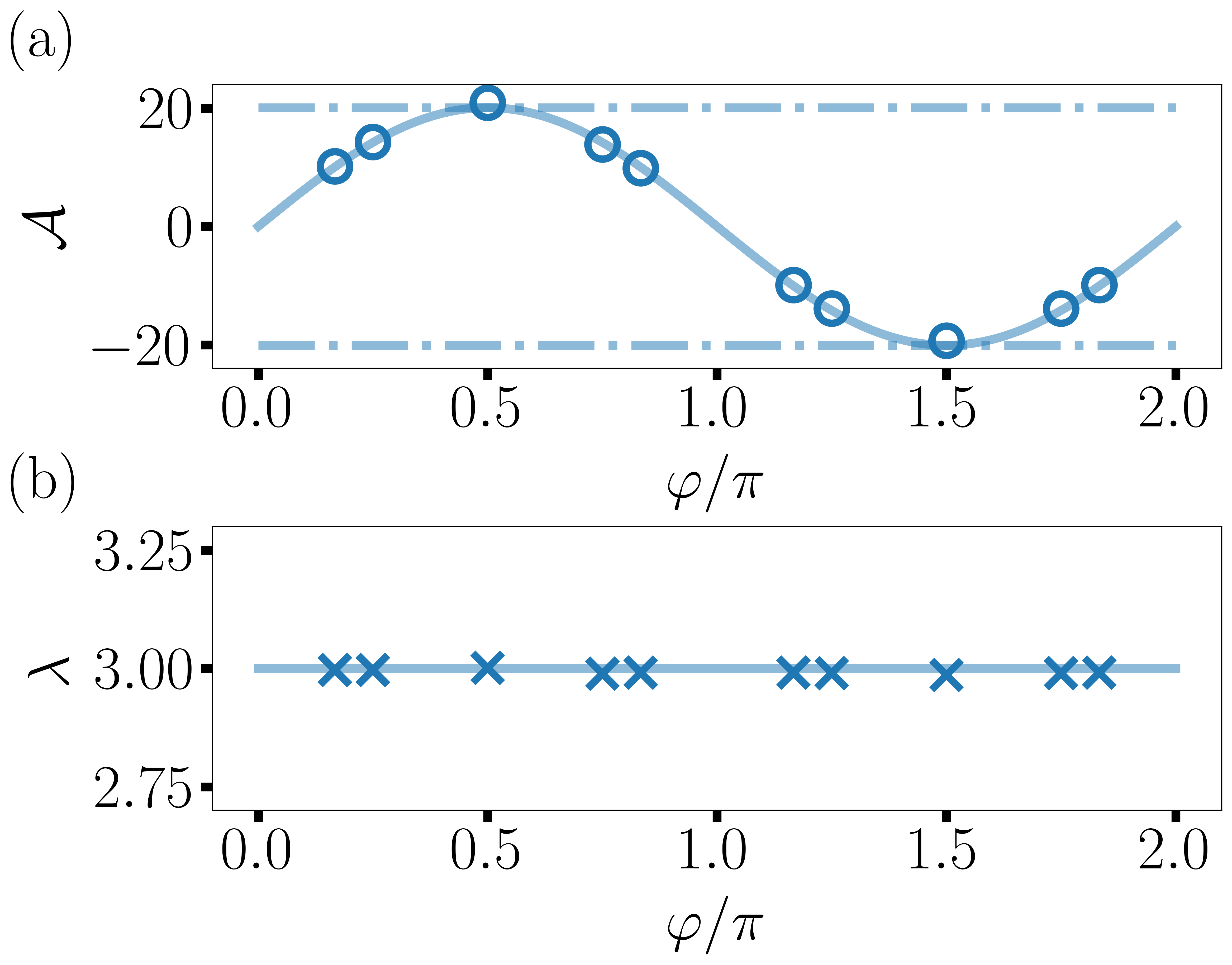}
\caption{(a) Evolution of the prefactor $\mathcal{A}$ (circles) from Eq.~(\ref{eq:Xdrift}) as a function of $\varphi$ for $t_1=3t_2=1$. The value of $\mathcal{A}$ varies sinusoidally (solid line) between values $\pm e^3$ (dashed lines). The variation is proportional to the critical value $M_c(\varphi)$, as shown in Fig.~\ref{fig:phase_diagram}). (b) Evolution of the shift exponent $\lambda$ (crosses) as a function of $\varphi$. The exponent $\lambda$ stays constant suggesting a universal value of $\lambda=3$ (solid line) for the clean Haldane model.}
\label{fig:A_lam_vs_phi}
\end{figure}

The data show a clear minimum in the vicinity of the critical point at $M_c =\sqrt{3}$ which follows from the Dirac theory for $t_1=3 \, t_2 = 1$ and $\varphi = \pi/2$; see Fig.~\ref{fig:raw_data}. The inset shows a zoomed-in portion which shows the finite-size approach to the thermodynamic result, $M_c=\sqrt{3}$. This is further illustrated in Fig.~\ref{fig:TrendClean} which shows the evolution of the finite-size critical point $M_c(N_y)$ (corresponding to the minima in Fig.~\ref{fig:raw_data}) with increasing system size. The results asymptote towards the field theory prediction $M_c=\sqrt{3}$. The evolution is well described by a power law 
\begin{equation}
    M_{c}(N_y)=M_c - \mathcal{A} N_y^{-\lambda},
    \label{eq:Xdrift}
\end{equation}
where $\lambda$ is the shift exponent \cite{SM_Fisher_1972} and $M_c=\sqrt{3}$. As can be seen in Fig.~\ref{fig:TrendClean}, the results are compatible with $\lambda=3$ and $\mathcal{A} = 20.1 \pm 0.1$; the latter is empirically close to $\mathcal{A} \simeq e^3$ as inferred from the logarithmic plot. In Fig.~\ref{fig:phase_diagram} we plot the evolution of $M_c$ as a function of $\varphi$. It can be seen that the results are in an excellent agreement with the field-theory prediction $M_c = \sqrt{3}\sin{\varphi}$ for $t_1 = 1$ and $t_2 = 1/3$. In a similar way we can track the evolution of $\lambda$ and $\mathcal{A}$ for different values of $\varphi$. The prefactor $\mathcal{A}$ also varies sinusoidally as shown in Fig.~\ref{fig:A_lam_vs_phi}~(a). In contrast, the exponent $\lambda$ stays constant as illustrated in Fig.~\ref{fig:A_lam_vs_phi}~(b). Combining these results, Eq.~(\ref{eq:Xdrift}) can be recast as 
\begin{equation}
    M_{c}(N_y)=M_c(1 - \tilde{\mathcal{A}} N_y^{-3}),
    \label{eq:Xdrift_simple}
\end{equation} 
where $M_c = \sqrt{3} \sin{\varphi}$ and $\tilde{\mathcal{A}}=e^3/ \sqrt{3}$ for fixed $t_1 = 1$ and $t_2 = 1/3$.

\subsection{Irrelevant contribution}
On the basis of the na\"ive ansatz given by Eq.~(\ref{eq:naive_ansatz}) one would expect the relation $\lambda=1/\nu$ to hold \cite{SM_Cardy_1996}. This can be seen by minimizing the na\"ive scaling relation $\Lambda = f(m N_y^{1/\nu})$ and applying $f^{-1}$ which renders $M_c(N_y)=M_c(1 +f^{-1}(\Lambda_{min}) N_y^{-1/\nu})$. In the case of the low-energy Dirac theory with $\nu=1$ this would yield $\lambda=1$; this differs from the extracted value of $\lambda = 3$, as shown in Fig.~\ref{fig:TrendClean}. As we will see below, this discrepancy is due to the absence of irrelevant scaling variables in Eq.~(\ref{eq:naive_ansatz}). The need for irrelevant corrections has been found in other disordered systems \cite{SM_Slevin_1999,SM_Slevin_2009, SM_Beck_2021,SM_Sbierski_2021}. These corrections are also required to accommodate the vertical drift of $\Lambda$ at the critical point at $M_c=\sqrt{3}$ which is present in Fig.~\ref{fig:raw_data}.
In view of this, we generalise the ansatz in Eq.~(\ref{eq:naive_ansatz}) by taking into account a single irrelevant contribution:
\begin{equation}
    \Lambda = F(m N_y^{1/\nu}, \psi N_y^{-y}),
    \label{eq:FRG}
\end{equation}
where $y>0$ is an exponent associated with the irrelevant field $\psi(m)$. In the case of a finite-size system, the scaling function $F$ is analytic and can be Taylor expanded around the vanishing irrelevant contribution $\psi N_y^{-y}=0$:
\begin{equation}
    \Lambda = F_0(m N_y^{1/\nu}) + \psi N_y^{-y} F_1( m N_y^{1/\nu}) + \mathcal{O}( (\psi N_y^{-y})^2).
    \label{eq:Taylor_irr}
\end{equation}
This allows the drift shown in Figs.~\ref{fig:TrendClean}(a) and \ref{fig:TrendClean}(b) to converge to the critical point $M_c$ with an exponent $\lambda \neq 1$. We note that $m\sim N_y^{-\lambda}$ at the minimum of $\Lambda$. In order to keep the argument $m N_y^{1/\nu}$ from Eq.~(\ref{eq:Taylor_irr}) finite at the minimum, we demand $N_y^{-\lambda + 1/\nu} < \infty$, yielding $\lambda \geq 1/\nu$.

\subsection{Extraction of the correlation length exponent $\nu$}
\begin{figure}[ht]
    \includegraphics[width=.43\textwidth]{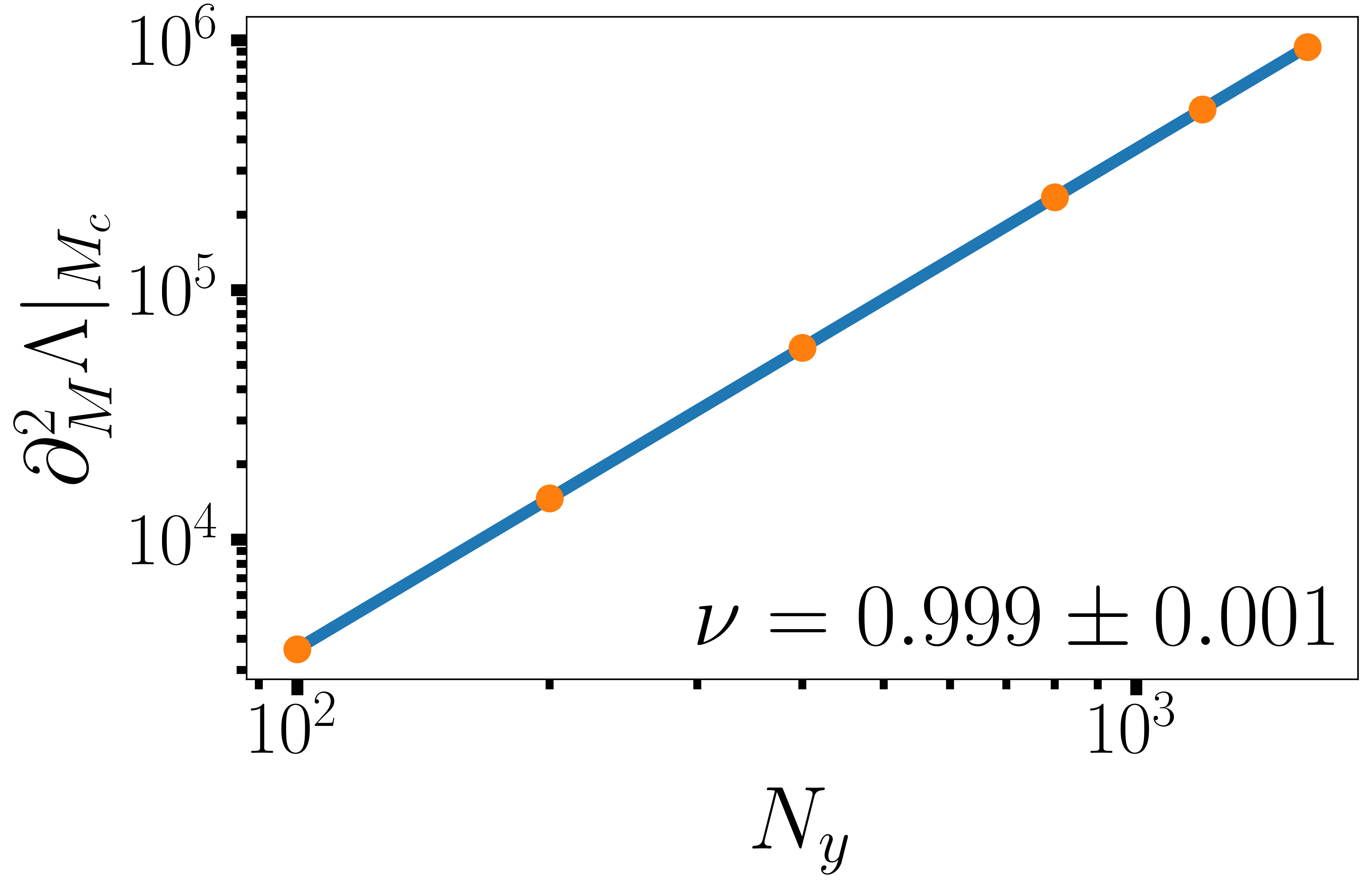}
    \caption{Finite-size scaling of the second derivative of $\Lambda$ at the (infinite-size) critical point at $M_c = \sqrt{3}$ for the clean Haldane model with $t_1 = 3 t_2 = 1$ and $\varphi=\pi /2$. The results are in agreement with the leading-order prediction $\partial^2_m\Lambda\vert_{M_c} \sim N_y^{2/\nu}$ obtained from Eq.~(\ref{eq:sec_der}). The value of $\nu = 0.999 \pm 0.001$ agrees with the field-theory prediction of $\nu = 1$.}
    \label{fig:2nd_der_clean}
\end{figure}

Due to the quadratic variation of $\Lambda$ in the vicinity of the critical point, as illustrated in Fig.~\ref{fig:raw_data}, it is numerically preferable to focus on the scaling of the second derivative $\partial^2_m\Lambda$ with system size. This can be obtained by explicit differentiation of Eq.~(\ref{eq:Taylor_irr}):
\begin{equation}
    \partial^2_m\Lambda\vert_{M_c} = N_y^{2/\nu}\partial^2_m F_0(0) +N_y^{-y}\partial_m^2 \left(\psi(m) F_1(mL^{1/\nu})\right) \Big\vert_{m=0}.
    \label{eq:sec_der}
\end{equation}
The first term in Eq.~(\ref{eq:sec_der}) scales as $\partial^2_m\Lambda\vert_{M_c} \sim N_y^{2/\nu}$ which is quadratic for $\nu=1$. The second term scales as $N_y^{2/\nu - y}$ with $y>0$, which is subleading in the thermodynamic limit. For the large system sizes explored in Fig.~\ref{fig:2nd_der_clean}, the subleading corrections are indeed found to be insignificant. The first term yields $\nu =0.999 \pm 0.001 \simeq 1$, as illustrated in Fig.~\ref{fig:2nd_der_clean}. This is in agreement with the correlation length exponent $\nu=1$ of the Dirac theory.

\subsection{Extraction of the irrelevant scaling exponent y}
\begin{figure}[ht]
    \includegraphics[width=.43\textwidth]{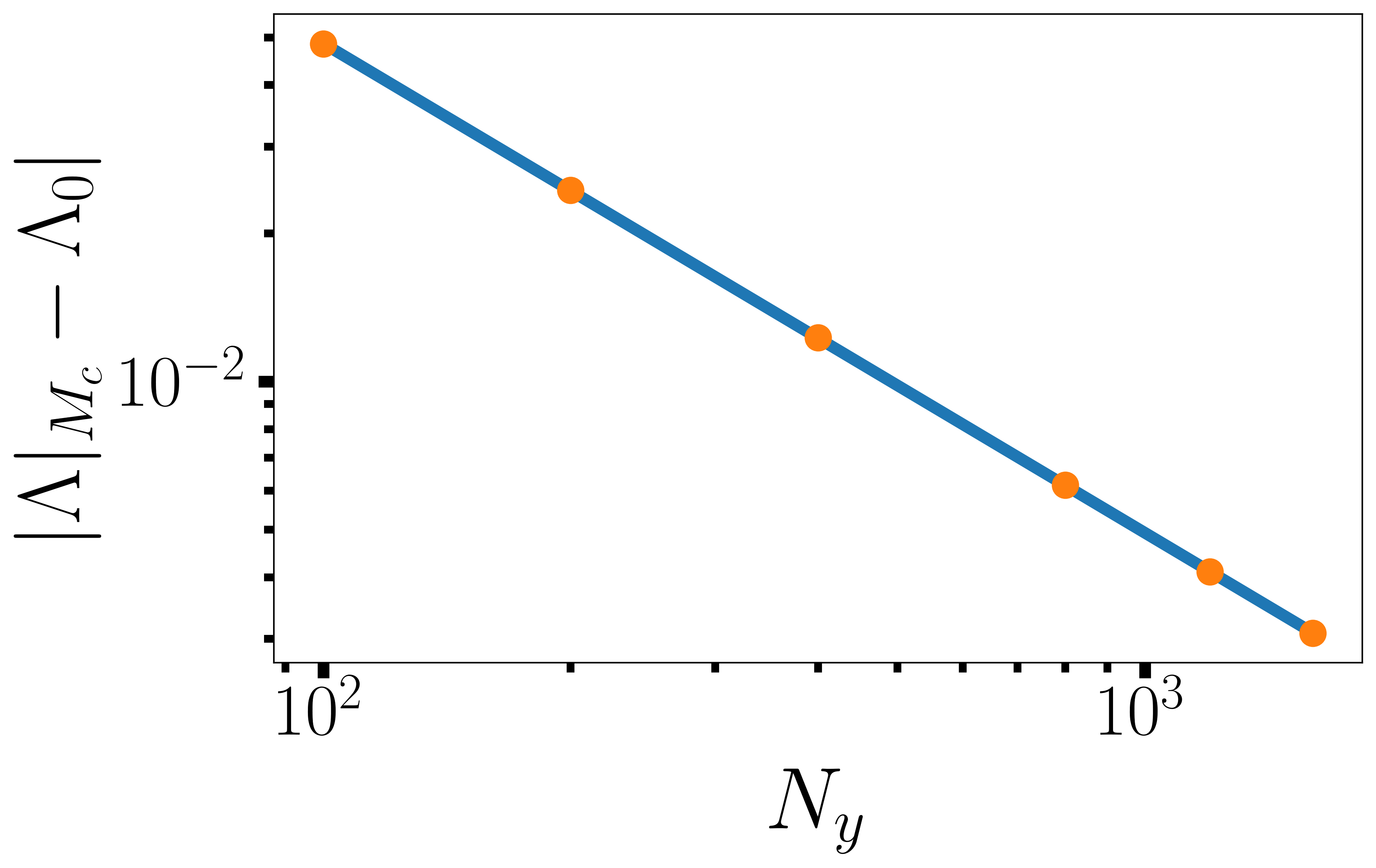}
    \caption{Finite-size scaling of the deviation between the dimensionless gap $\Lambda =(\xi/L_y)^{-1}$ at the critical point and the field-theory prediction $\Lambda_0=3\pi/(2\sqrt{3})$ for $t_1=3 \, t_2 =1 $ and $\varphi = \pi/2$. The power-law scaling of $\vert \Lambda\vert_{M_c} - \Lambda_0\vert \sim N_y^{-y}$ yields $y=1$ (blue line) and is consistent with the prediction of $\Lambda_0$ in the thermodynamic limit.}
    \label{fig:upward_trend_clean}
\end{figure}

In order to extract the scaling exponent $y$ of the irrelevant field $\psi$ we focus on the vertical shift of the dimensionless gap at the critical point, $\Lambda\vert_{M_c}$. This describes the vertical shift in Fig.~\ref{fig:raw_data}. The scaling with system size can be inferred from Eq.~(\ref{eq:Taylor_irr}) by setting $m = 0$:
\begin{equation}
    \Lambda\vert_{M_c} = F_0(0) - N_y^{-y}\psi(0) F_1(0).
    \label{eq:upward_trend}
\end{equation}
This is of a similar form to Eq.~(\ref{eq:Xdrift}). In the thermodynamic limit $N_y \rightarrow \infty$, Eq.~(\ref{eq:upward_trend}) converges to $\Lambda\vert_{M_c} =F_0(0) \equiv \Lambda_0$. In the case of the clean Haldane model, the limiting value of $\Lambda_0$ can be inferred from the Dirac theory. To see this, we note that the dispersion relation in the vicinity of the Dirac point is given by $E_+(\delta k_y)= \vert c \hbar \delta k_y \vert$ with $c=3t_1 a/ (2\hbar)$ and $\delta k_y$ is the distance from the Dirac momentum. In the presence of TBCs, the minimum of the dispersion relation occurs at a value of $\delta k_y$ that is half of the momentum spacing. The minimum value of $E_+$ is given by $\min E_+ = 3\pi / (2\sqrt{3} N_y)$ for $t_1=1$, yielding $\Lambda_0 = 3\pi/(2\sqrt{3})\simeq 2.72$. 
On the basis of Eq.~(\ref{eq:upward_trend}) it can be seen that the finite-size deviation from $\Lambda_0$ scales as $\bigl \vert \Lambda\vert_{M_c} - \Lambda_0 \bigr\vert \sim N_y^{-y}$. The data shown in Fig.~\ref{fig:upward_trend_clean} is consistent with $\Lambda_0 = 3\pi/(2\sqrt{3})$ and the exponent $y=1$.

\section{Scaling relation}
\begin{figure}[ht]
\includegraphics[width=.43\textwidth]{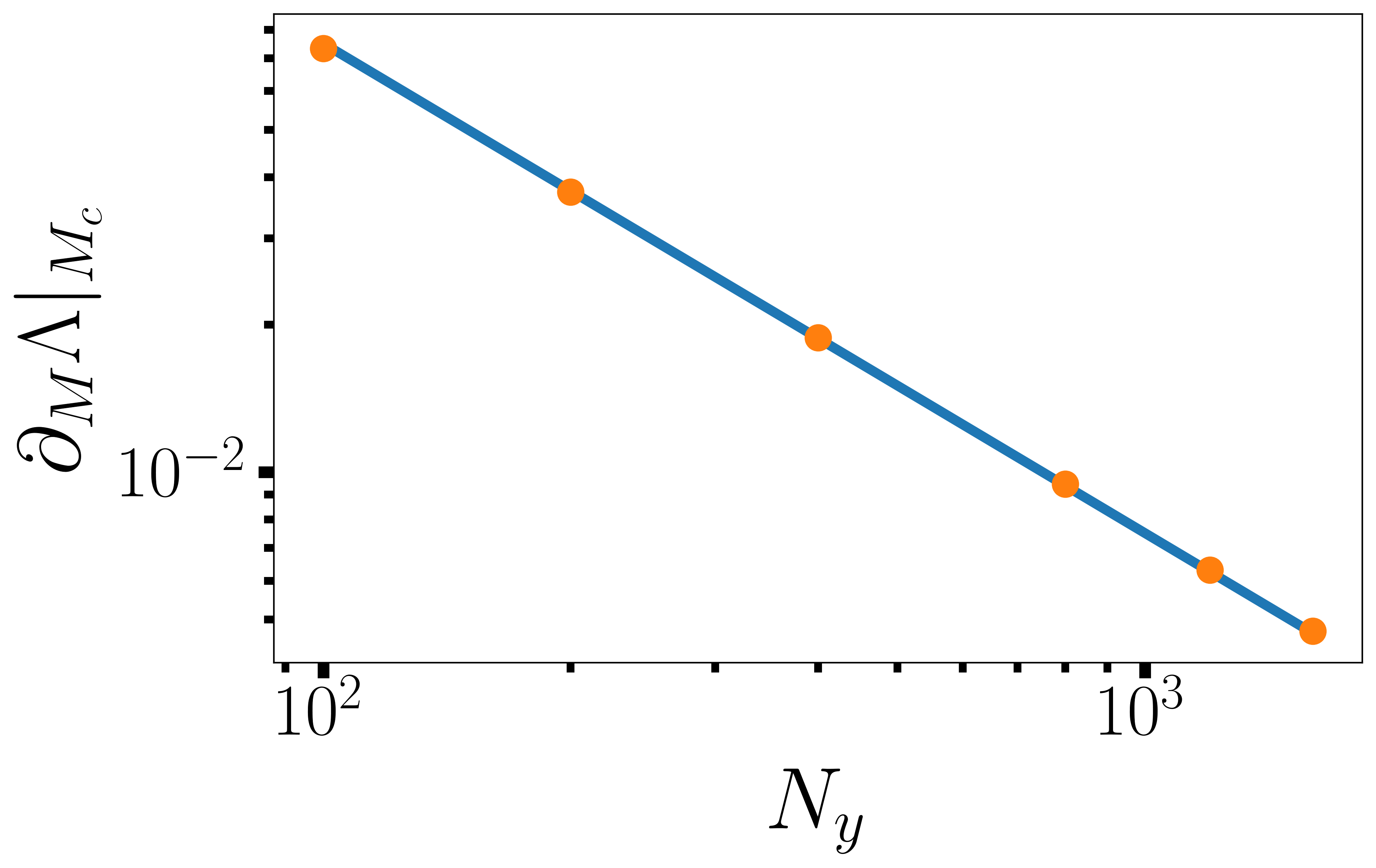}
\caption{Finite-size scaling of the first derivative of $\Lambda$ at the (infinite-size) critical point at $M_c = \sqrt{3}$ for the clean Haldane model with $t_1 = 3 t_2 = 1$ and $\varphi=\pi /2$. The derivative decays as a power-law $\partial_m \Lambda \vert_{m=0} \sim N_y^{-1}$, vanishing in the thermodynamic limit. The value of the exponent matches the value of the irrelevant exponent $y=1$.}
\label{fig:1st_der_clean}
\end{figure}

A notable feature of the results obtained above is that the values $\lambda=3$, $y=1$ and $\nu =1$ satisfy the relation $\lambda = 2/\nu + y$. The latter can be seen from the condition for the finite-size critical point, $\partial_m \Lambda \vert_{m_c(N_y)}=0$. Taylor expansion of $\partial_m\Lambda$ in powers of $m$ around the critical point $m=0$ yields
\begin{equation}
    0=\partial_m\Lambda \Big\vert_{m_c(N_y)} =  \partial_m \Lambda\Big\vert_{m=0} + m_c(N_y) \partial^2_m \Lambda \Big\vert_{m=0},
    \label{eq:minimum_Taylor}
\end{equation}
with $m_c(N_y) \sim N_y^{-\lambda}$. The scaling of $\partial_m \Lambda$ and $\partial^2_m \Lambda$ at the critical point $m=0$ can be obtained from Eq.~(\ref{eq:Taylor_irr}) by explicit differentiation. We find that the first derivative at $m=0$ vanishes in the thermodynamic limit as $\partial_m \Lambda\vert_{m=0} \sim N_y^{-y}$, as illustrated in Fig.~\ref{fig:1st_der_clean}. Substituting Eq.~(\ref{eq:sec_der}) into Eq.~(\ref{eq:minimum_Taylor}) yields the scaling relation $\lambda = 2/\nu + y$, in agreement with the calculated values of the exponents $\lambda=3$ and $\nu = y = 1$.

\subsection{Scaling collapse}
\begin{figure}[ht]
\includegraphics[width=.43\textwidth]{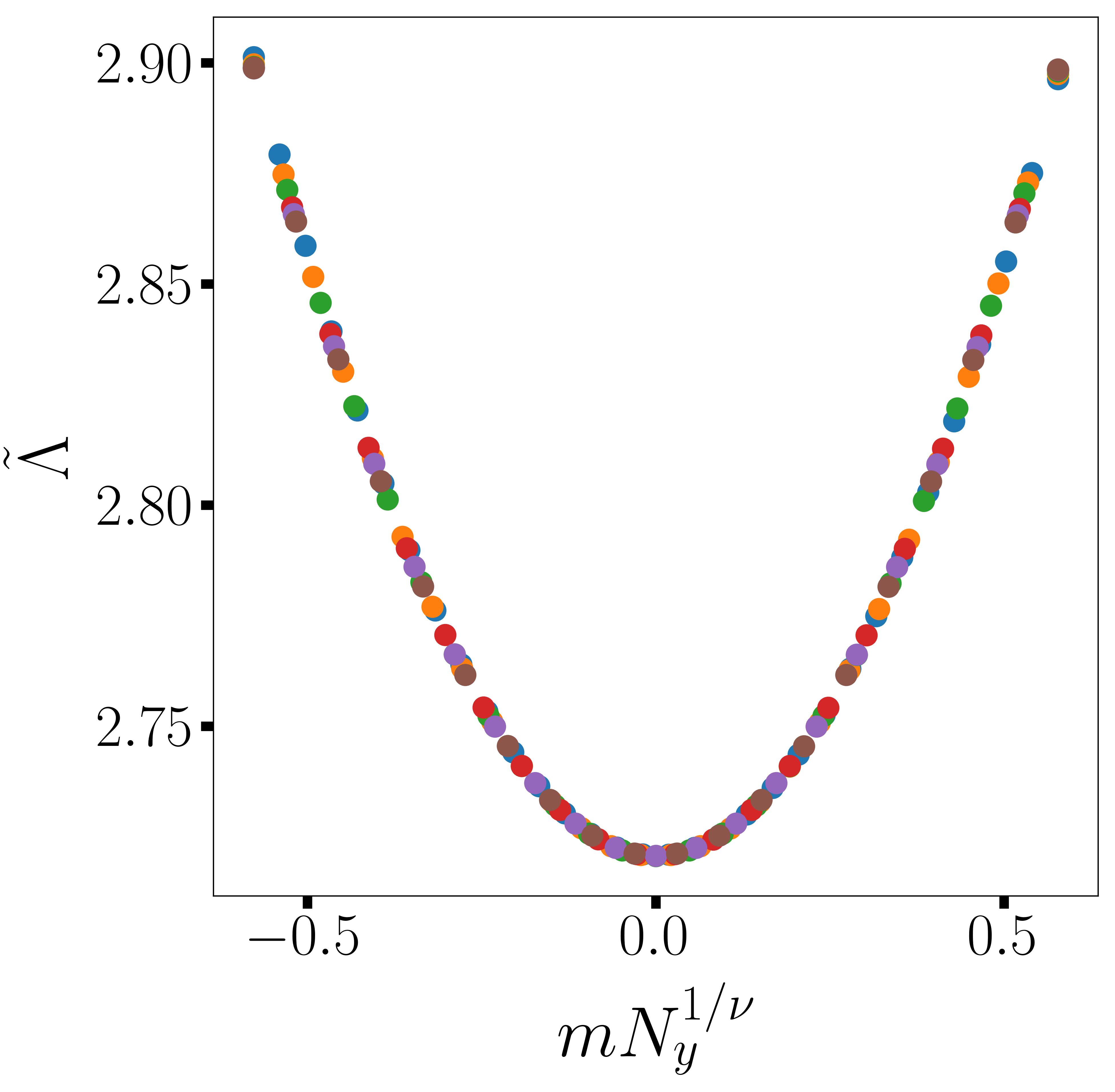}
\caption{Scaling collapse of the shifted variable $\tilde{\Lambda}$ which removes the irrelevant contribution via Eq.~(\ref{eq:lambda_corr}). The corrections have been removed on the basis of a Taylor expansion of $F_1$, using the previously extracted exponents $\nu=y=1$ and $\lambda=3$.}
\label{fig:Collape_clean}
\end{figure}

To validate our results, we collapse the data shown in Fig.~\ref{fig:raw_data} onto a single curve; see Fig.~\ref{fig:Collape_clean}. Explicitly, we consider the shifted variable $\tilde{\Lambda}$:
\begin{equation}
    \tilde{\Lambda} = \Lambda -  \psi N_y^{-y} F_1( m N_y^{1/\nu})
    \label{eq:lambda_corr}
\end{equation}
to remove the vertical shift in Fig.~\ref{fig:raw_data}. It follows that $\tilde{\Lambda}$ is a function of $m N_y^{1/\nu}$ as shown in Fig.~\ref{fig:Collape_clean}. In practice, the irrelevant contribution $F_1$ described by Eq.~(\ref{eq:lambda_corr}) can be Taylor expanded up to the second order in $m$ and removed appropriately.

\section{Disordered Haldane model}
Having investigated the critical properties of the clean Haldane model, we now consider the model in presence of on-site disorder. In the case of an infinite strip $N_x \rightarrow \infty$, the eigenvalues of the transfer matrix for a fixed realisation of disorder converge to their disorder averages; this is due to the consecutive multiplication of single-slice transfer matrices \cite{SM_Kramer_2010}. In practice, we use a long but finite strip with $N_x=10^6$ or $N_x=10^7$.

\subsection{Weak Disorder}
\begin{figure}[ht]
\includegraphics[width=.43\textwidth]{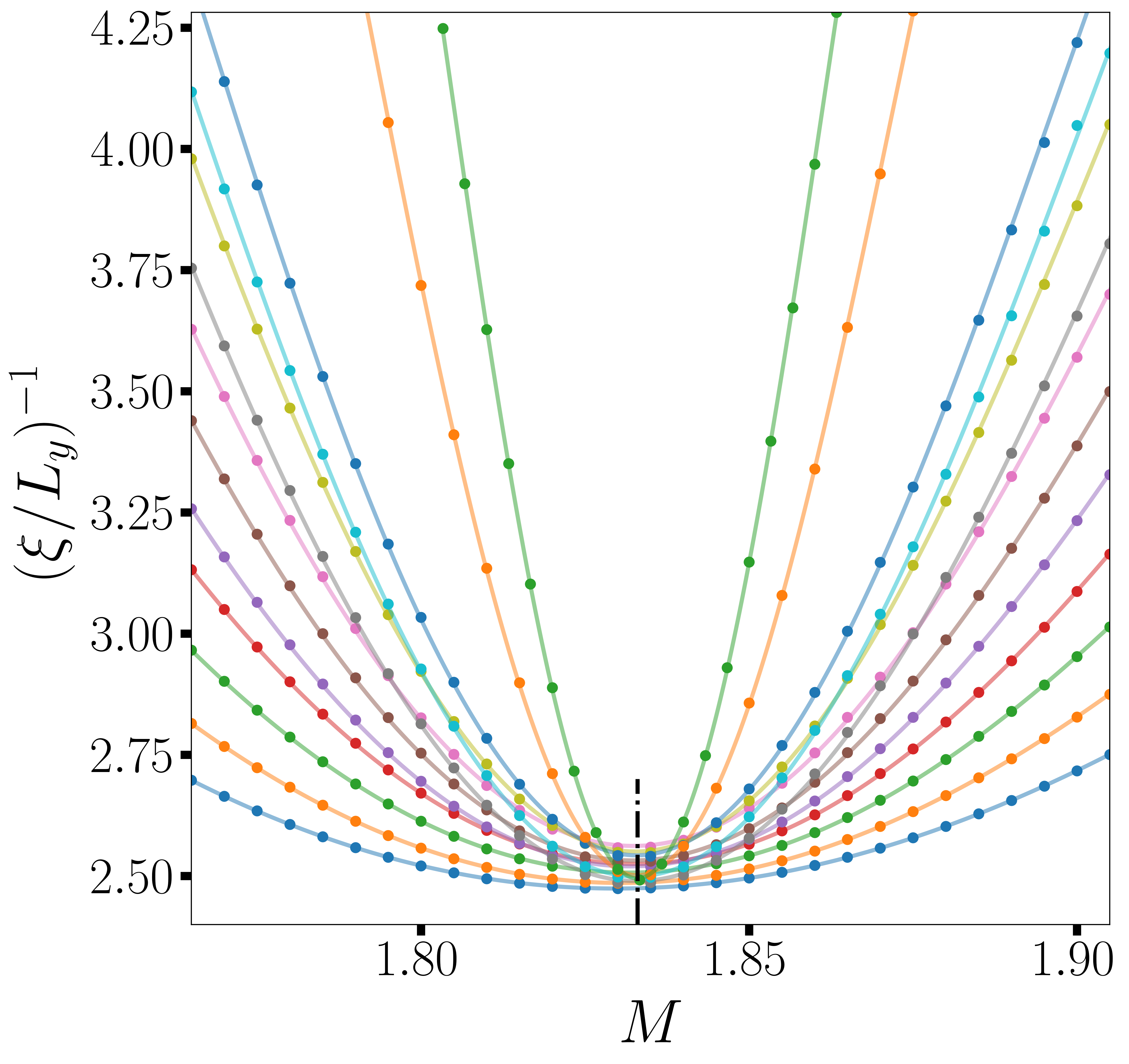}
\caption{Evolution of the dimensionless inverse gap $\Lambda$ for disorder strength $W=1$ on transiting from the topological to the non-topological phase for system sizes $N_y= 19,23,27,..., 59$ and $N_y=99, 139$. We set $t_1=3t_2=1$ and $\varphi=\pi/2$ and impose TBCs. The results are obtained from $N_x=10^6$ transfer matrix multiplications using Eq.~(\ref{eq:TM_def}). The finite-size critical point $M_c(N_y)$ associated with the minimum of $\Lambda$ drifts towards $M_c(\infty) = 1.833$ with increasing system size; see Fig.~\ref{fig:w10_Xdrift}.}
\label{fig:w10_raw_data}
\end{figure}

\begin{figure}[ht]
\includegraphics[width=.43\textwidth]{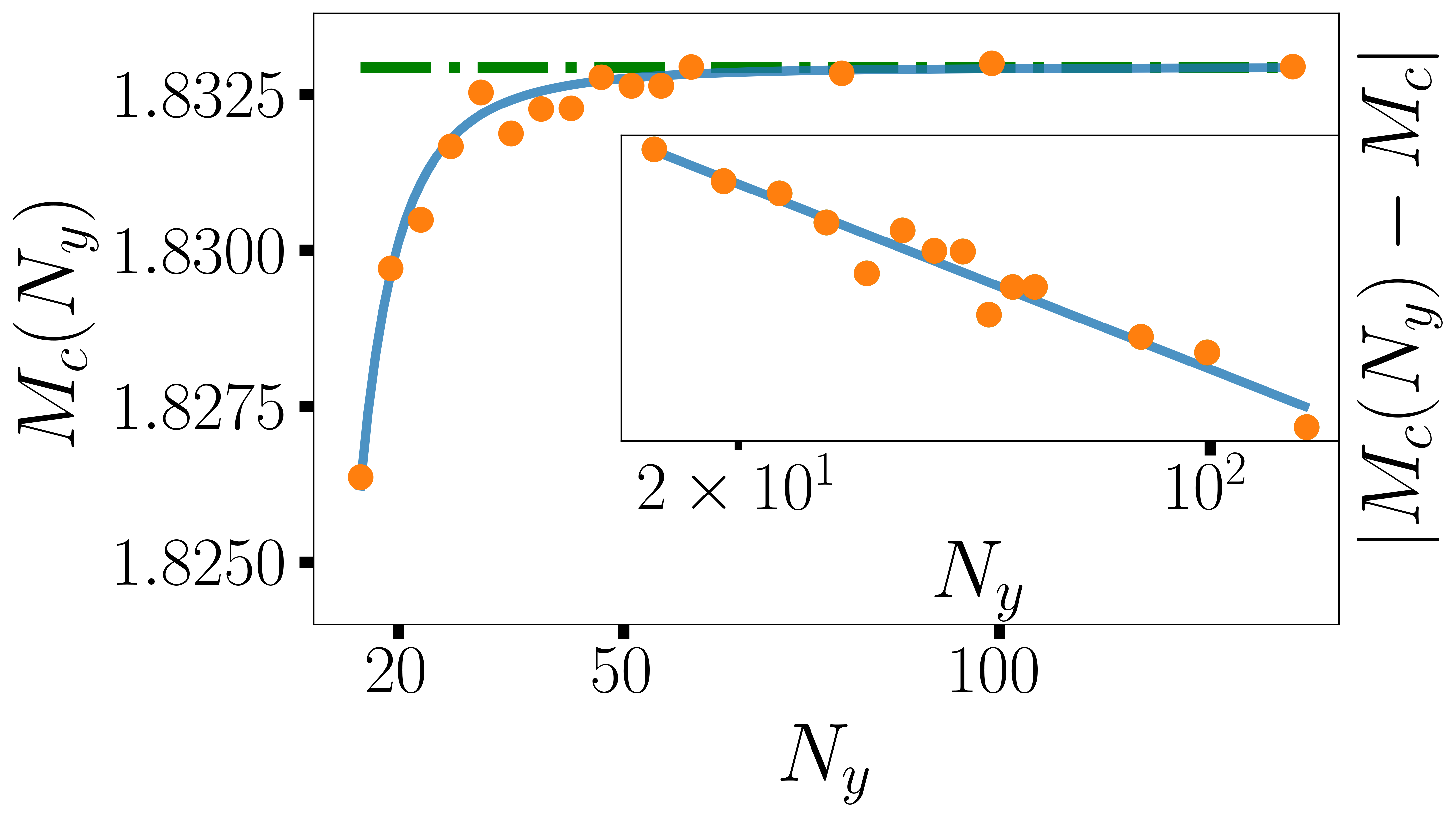}
\caption{Drift of the minimum of the dimensionless gap $\Lambda$ as a function of the system size $N_y$ for the disordered Haldane model with disorder strength $W=1$. We set $t_1=3t_2=1$ and $\varphi=\pi/2$ and impose TBCs. The data asymptotes towards $M_c(\infty)= 1.833 \pm 0.002$ on the basis of Eq.~(\ref{eq:Xdrift}) (solid line). The location of the critical point differs from the clean case where $M_c=\sqrt{3} \simeq 1.73$. Inset: The approach to the critical point is consistent with the power-law (\ref{eq:Xdrift}) with the shift exponent $\lambda = 2.7\pm 0.3$.}
\label{fig:w10_Xdrift}
\end{figure}

\begin{figure}[ht]
\includegraphics[width=.43\textwidth]{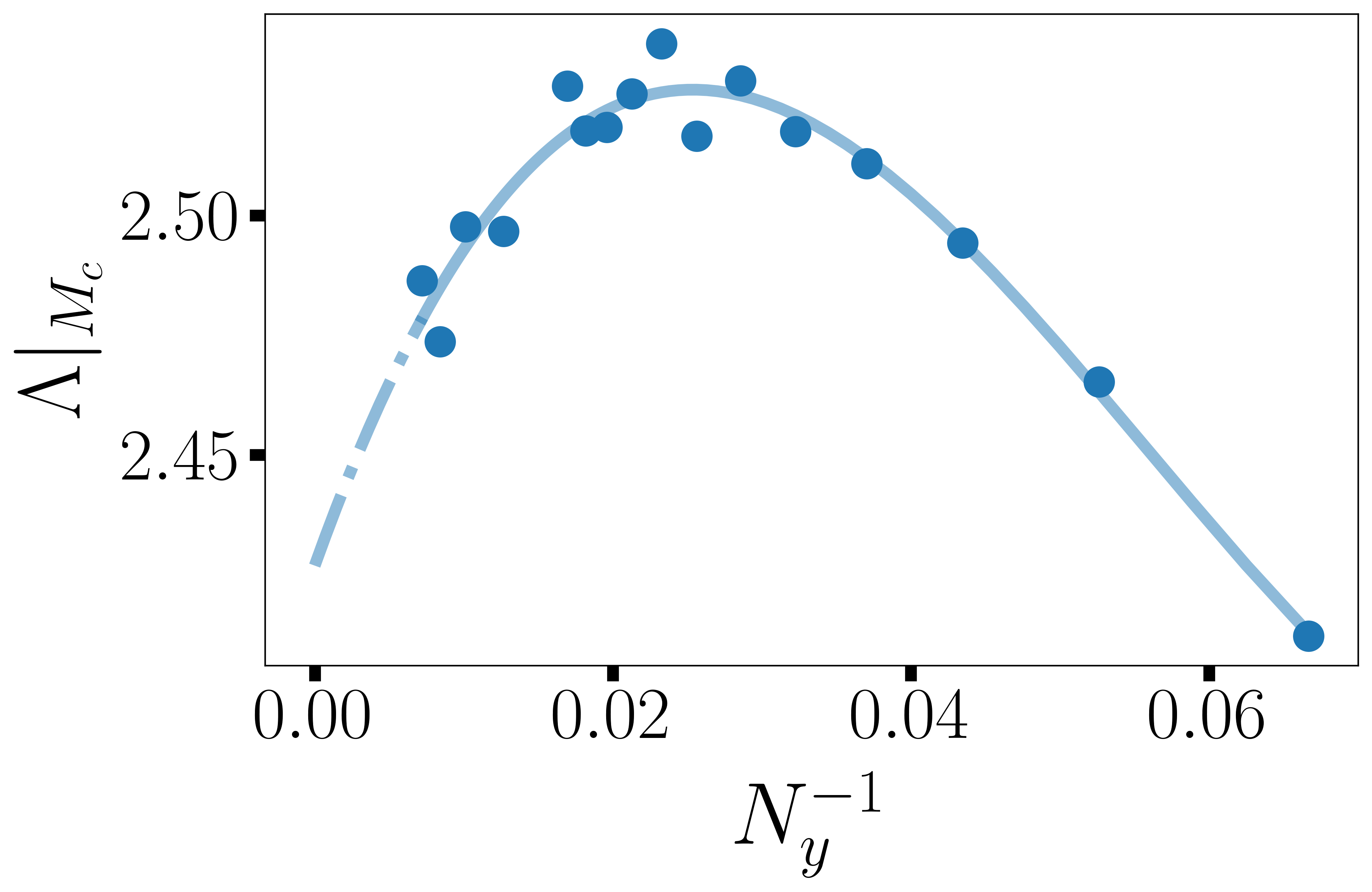}
\caption{Evolution of the dimensionless gap $\Lambda\vert_{M_c}$ as a function of the inverse system size $N_y^{-1}$ in the disordered Haldane model with disorder strength $W=1$. We set $t_1=3t_2=1$ and $\varphi=\pi/2$ and impose TBCs. The evolution is non-monotonic (dots) which complicates the extraction of the exponent $y$. The non-monotonicity is potentially due to the higher-order corrections in Eq.~(\ref{eq:Taylor_irr}). On the basis of a na\"ive third order polynomial fit in $N_y^{-1}$ (solid line) we extract $\Lambda_0 = 2.43 \pm 0.02$. A crude estimate of $y$ can be obtained by a power-law fit to the largest system size data.}
\label{fig:w10_Ydrift_rescaled}
\end{figure}

\begin{figure}[ht]
\includegraphics[width=.43\textwidth]{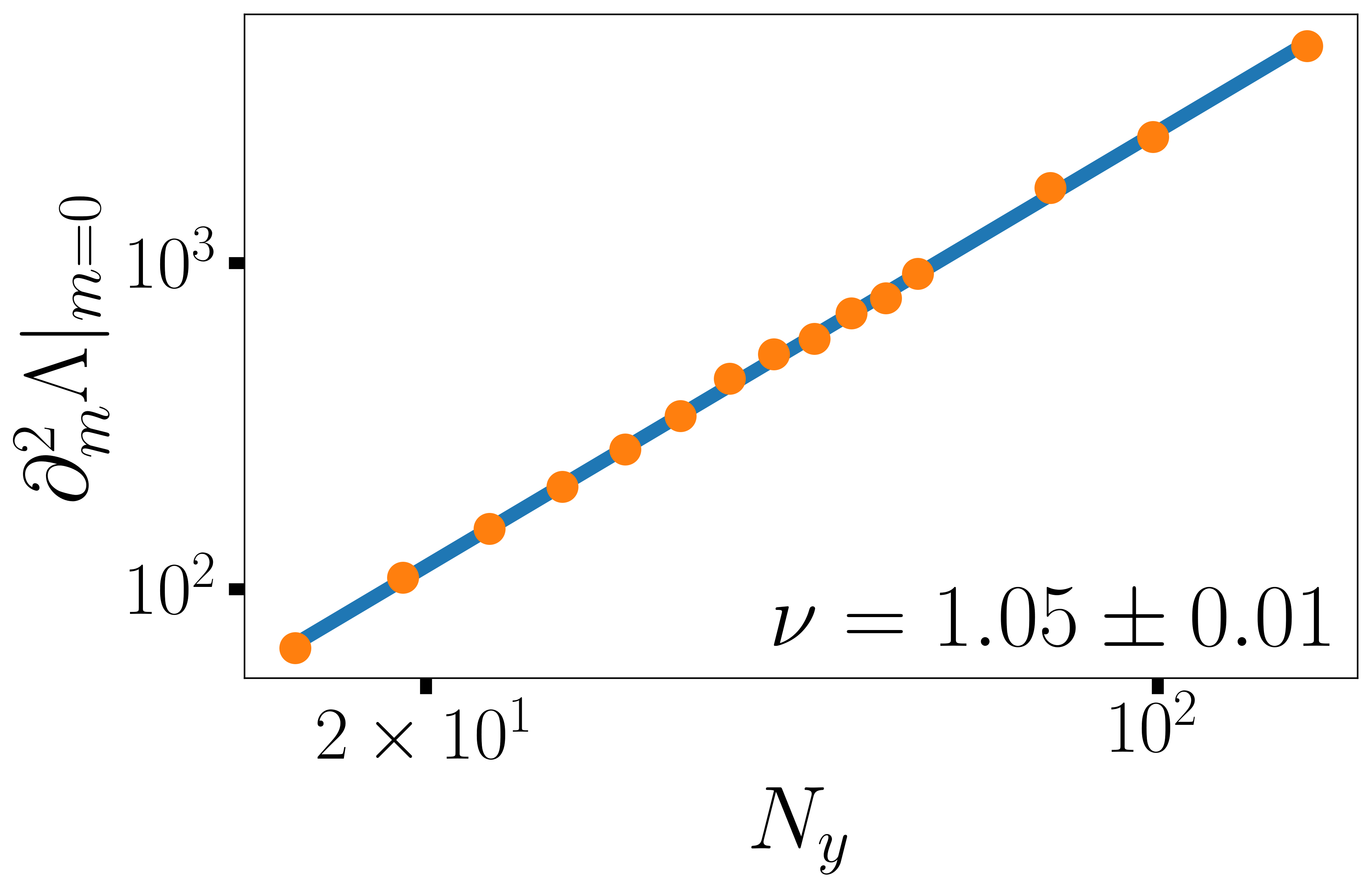}
\caption{Finite-size scaling of the second derivative of the inverse gap $\partial_m^2 \Lambda$ at the critical point. We consider the disordered Haldane model with $W=1$, $t_1=3t_2=1$, $\varphi =\pi/2$ and imposed TBCs. The results are consistent with the power-law (\ref{eq:sec_der}) with $\nu=1.05 \pm 0.01$ without the need for subleading corrections. The value of $\nu$ differs from that of the clean Haldane model where $\nu= 1$. This is in agreement with the result $\nu =1.05 \pm 0.03$ obtained via the Chern Marker.}
\label{fig:w10_2nd_der}
\end{figure}

In Fig.~\ref{fig:w10_raw_data} we plot the evolution of the dimensionless inverse gap $\Lambda$ for disorder strength $W=1$, on transiting from the topological to the non-topological phase. As found in the clean case, the data show a clear minimum which drifts with the system size $N_y$. On the basis of Eq.~(\ref{eq:Xdrift}) we find $M_c(\infty) = 1.833 \pm 0.002$, and the shift exponent $\lambda=2.7 \pm 0.3$, as illustrated in Fig.\ref{fig:w10_Xdrift}. The location of the critical point differs from the critical point of the clean system at $M_c = \sqrt{3} \simeq 1.73$.
In contrast to the clean case, the vertical drift of $\Lambda$ at the critical point $M_c = 1.833$ is non-monotonic, as shown in Fig.~\ref{fig:w10_Ydrift_rescaled}. This complicates the extraction of the irrelevant exponent $y$ via Eq.~(\ref{eq:upward_trend}). Nonetheless we can extract the correlation length exponent $\nu$ via the scaling of the second derivative $\partial_m^2 \Lambda$ at the critical point $M_c=1.833$ using Eq.~(\ref{eq:sec_der}). As in the clean case, we find that $\nu$ can be obtained from the leading term. This yields $\nu = 1.05 \pm 0.01$ as illustrated in Fig.~\ref{fig:w10_2nd_der}. The value of the exponent differs from that of the clean Haldane model, where $\nu=1$. 

\begin{figure}[ht]
\includegraphics[width=.41\textwidth]{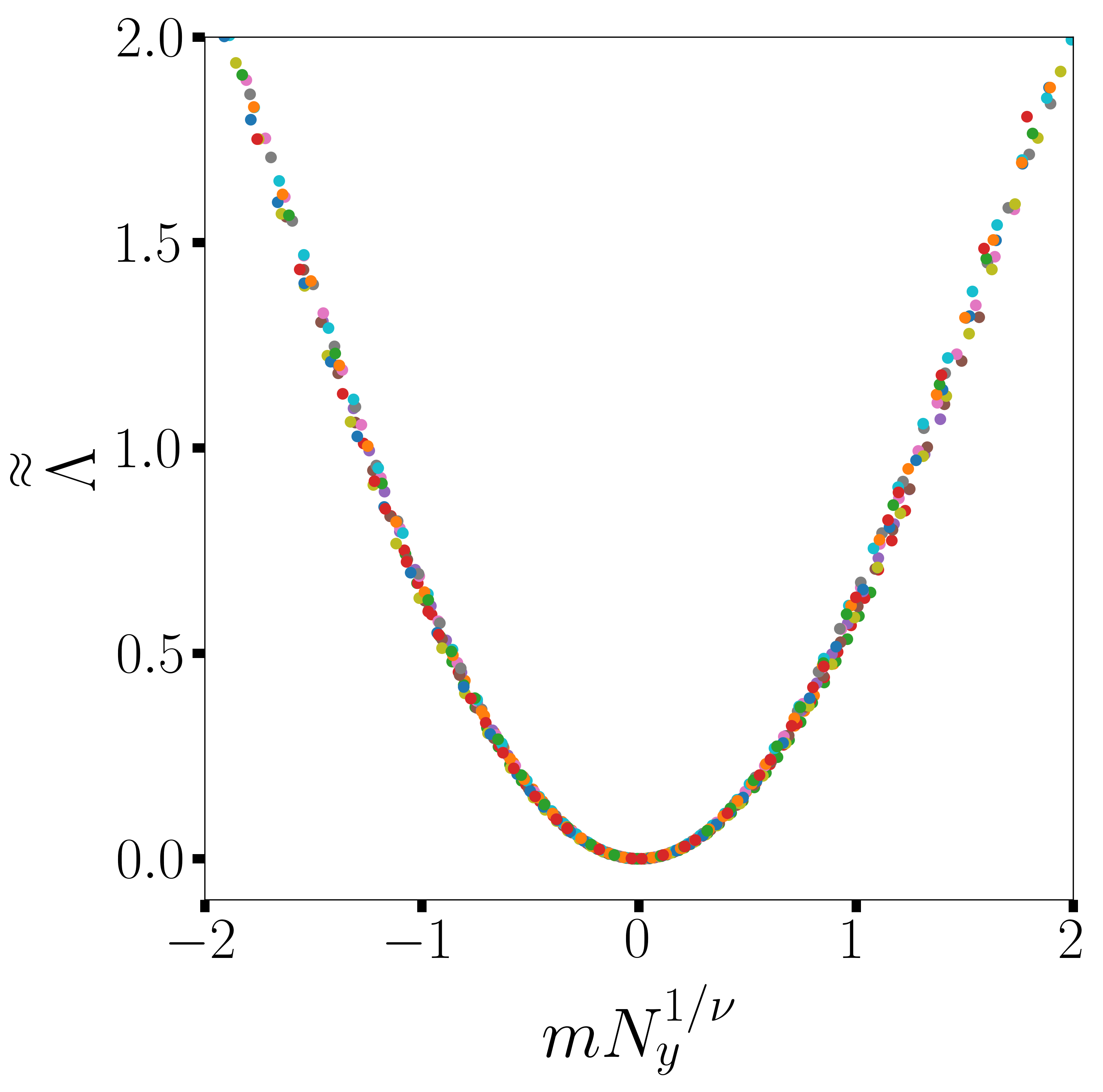}
\caption{Scaling collapse of $\dtilde{\Lambda}$ defined in Eq.~(\ref{eq:lam_cent}) for the disordered Haldane model with disorder strength $W=1$. We set $t_1=3t_2=1$, $\varphi =\pi/2$ and imposed TBCs. The data collapses onto a single curve when plotted as a function of $m N_y^{1/\nu}$ with $\nu=1.05$. The exponent differs from that in the clean Haldane model where $\nu=1$.}
\label{fig:w10_collapse}
\end{figure}

In order to verify the extracted value of $\nu\neq 1$, we consider scaling collapse of the data. Due to the non-monotonic vertical shift shown in Fig.~\ref{fig:w10_raw_data}, we focus on the scaling in the horizontal direction. This can be done by shifting the minimum of each of the curves in Fig.~\ref{fig:w10_raw_data} to a common origin. Explicitly we define:
\begin{equation}
    \dtilde{\Lambda} = \Lambda(M-M_c(N_y)) -  \Lambda_\textrm{min},
    \label{eq:lam_cent}
\end{equation}
where $\Lambda_{min}$ is the value of the inverse gap at the minimum. Replotting $\dtilde{\Lambda}$ as a function of $m N_y^{1/\nu}$ with $\nu=1.05$, it can be seen that the data collapse onto a single curve as shown in Fig.~\ref{fig:w10_collapse}. This is consistent with a correlation length exponent that differs from the clean Haldane model where $\nu=1$.

\subsection{Strong Disorder}
\begin{figure}[ht]
\includegraphics[width=.43\textwidth]{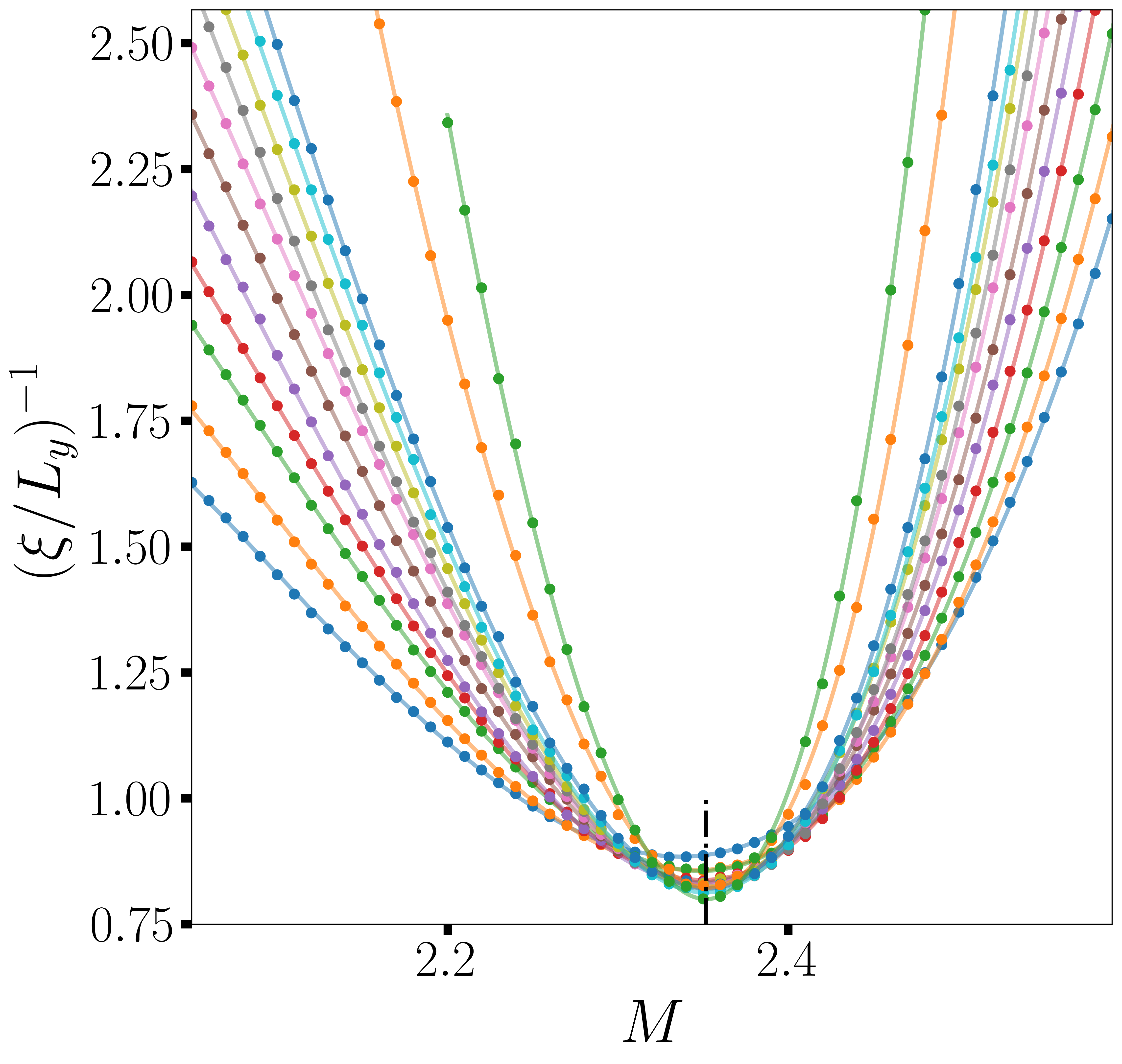}
\caption{Evolution of the dimensionless inverse gap $\Lambda$ for the disordered Haldane model with $W=2.6$ on transiting from the topological to the non-topological phase. We set $t_1=3t_2=1$ and $\varphi=\pi/2$ and impose TBCs. The data are obtained are obtained from $N_x=10^6$ transfer matrix multiplications in Eq.~(\ref{eq:TM_def}). We consider system sizes $N_y= 19,23,27,..., 59$ and $N_y=99, 139$. The minimum of $\Lambda$, associated with the finite-size critical point, drifts towards the infinite-size critical point at $M_c = 2.352 \pm 0.002$, as shown in Fig.~\ref{fig:w26_Xdrift}. }
\label{fig:w26_raw_data}
\end{figure}

\begin{figure}[ht]
\includegraphics[width=.43\textwidth]{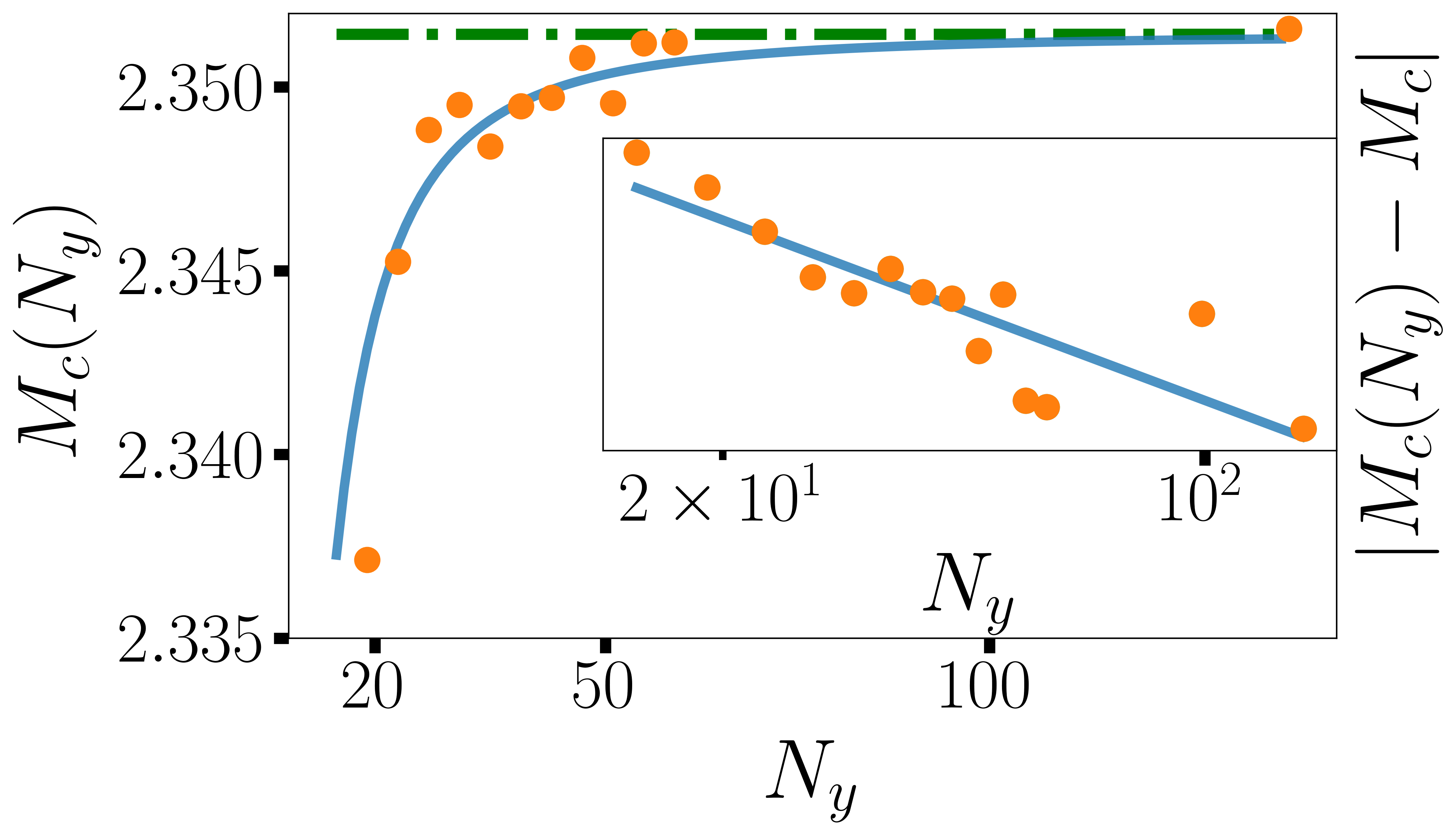}
\caption{Drift of the finite-size critical point $M_c(N_y)$ associated with the minimum of the dimensionless gap $\Lambda$ with the system size $N_y$ for the disordered Haldane model with disorder strength $W=2.6$. We set $t_1=3t_2=1$ and $\varphi=\pi/2$ and impose TBCs. The infinite-size critical point is estimated at $M_c = 2.352 \pm 0.002$ on the basis of Eq.~(\ref{eq:Xdrift}). The uncertainty is obtained from the covariance matrix of the least squared error estimation. Inset: Departure of the finite-size critical point $M_c(N_y)$ from the infinite-size critical point at $M_c(\infty)=2.352$. The departure vanishes as a power-law (\ref{eq:Xdrift}) with the shift exponent $\lambda = 2.1\pm 0.4$.}
\label{fig:w26_Xdrift}
\end{figure}

Having found a non-trivial value of the correlation length exponent $\nu \neq 1$ in the disordered Haldane model with $W=1$, we now consider a stronger disorder strength, $W=2.6$. The inverse correlation length $\xi^{-1}$ is extracted from $N_x = 10^6$ transfer matrix multiplications in Eq.~(\ref{eq:TM_def}). In Fig.~\ref{fig:w26_raw_data}, we plot the dimensionless gap $\Lambda = (\xi/L_y)^{-1}$ on transiting from the topological to the non-topological phase. The minimum of $\Lambda$ drifts with increasing system size $N_y$ towards the critical point at $M_c = 2.352 \pm 0.002$ as illustrated in Fig.~\ref{fig:w26_Xdrift}. The deviation of the finite-size critical point from the infinite-size critical point is described by the power law (\ref{eq:Xdrift}) with the shift-exponent $\lambda=2.1 \pm 0.4$; see inset of Fig.~\ref{fig:w26_Xdrift}.

\begin{figure}[ht]
\includegraphics[width=.45\textwidth]{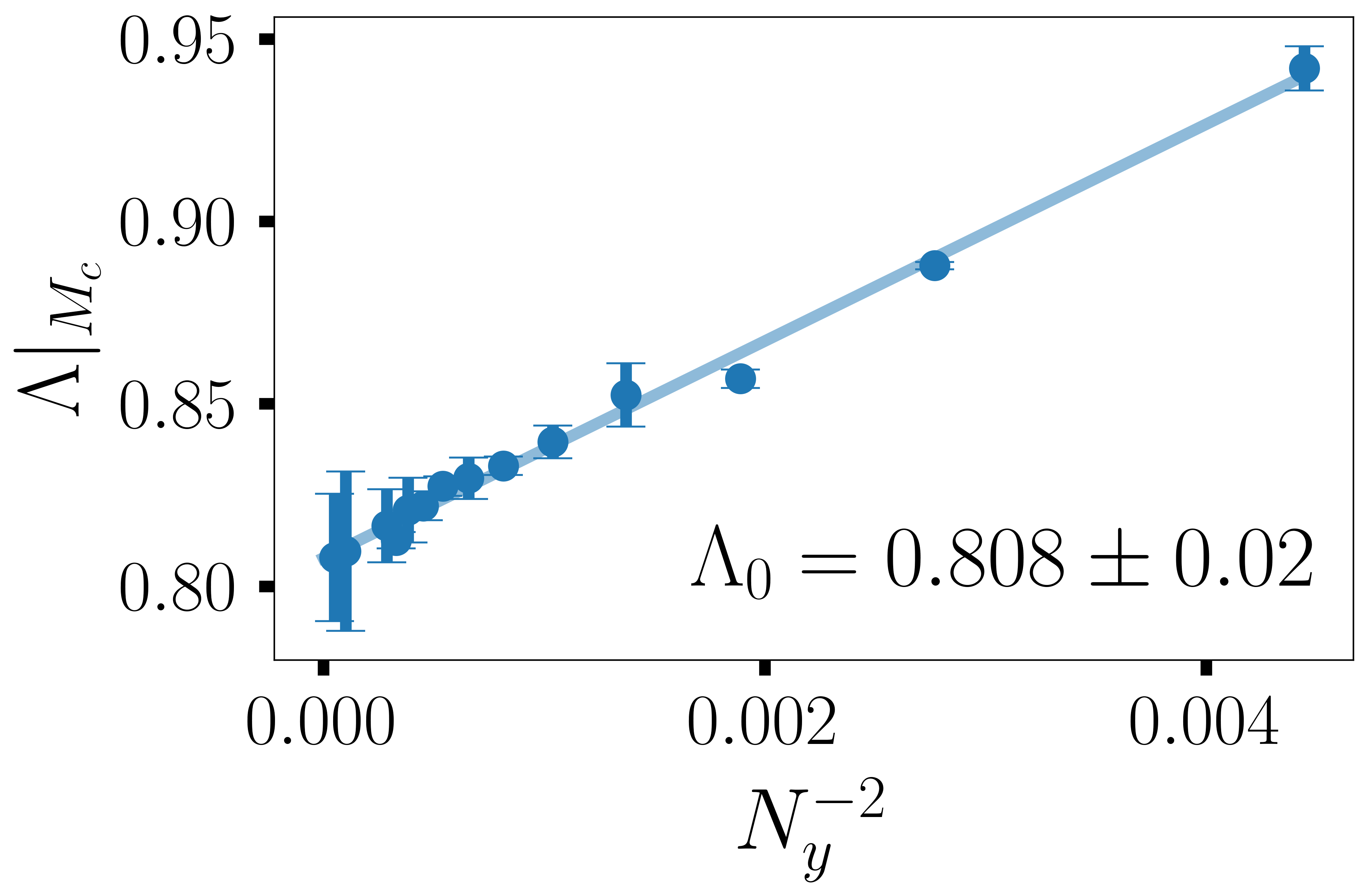}
\caption{Evolution of the dimensionless gap ratio at the critical point $\Lambda \vert_{M_c}$ with system size in the disordered Haldane model with $W=2.6$. We set $t_1=3t_2=1$ and $\varphi=\pi/2$ and impose TBCs. The drift is well described by Eq.~(\ref{eq:upward_trend}) with the exponent $y=2$. We estimate the limiting value $\Lambda_0 = 0.81\pm 0.02$ from the intercept of the linear fit (blue line). This is consistent with other results for the plateau transitions in the Integer Quantum Hall effect \cite{SM_Slevin_2009,SM_Gruzberg_2017,SM_Sbierski_2021}.}
\label{fig:w26_upward_drift}
\end{figure}

The vertical drift of $\Lambda$ at the critical point $M_c = 2.352$ with system size is monotonic and downwards for $W=2.6$. The drift is well described by Eq.~(\ref{eq:upward_trend}) with $y=2$, as illustrated in Fig.~\ref{fig:w26_upward_drift}. We estimate the limiting value of the infinite-size dimensionless gap ratio as $\Lambda_0 = 0.81 \pm 0.02$. This is close to values previously reported for the plateau transitions for the Quantum Hall Effect \cite{SM_Slevin_2009,SM_Gruzberg_2017,SM_Sbierski_2021}.

\begin{figure}[ht]
\includegraphics[width=.43\textwidth]{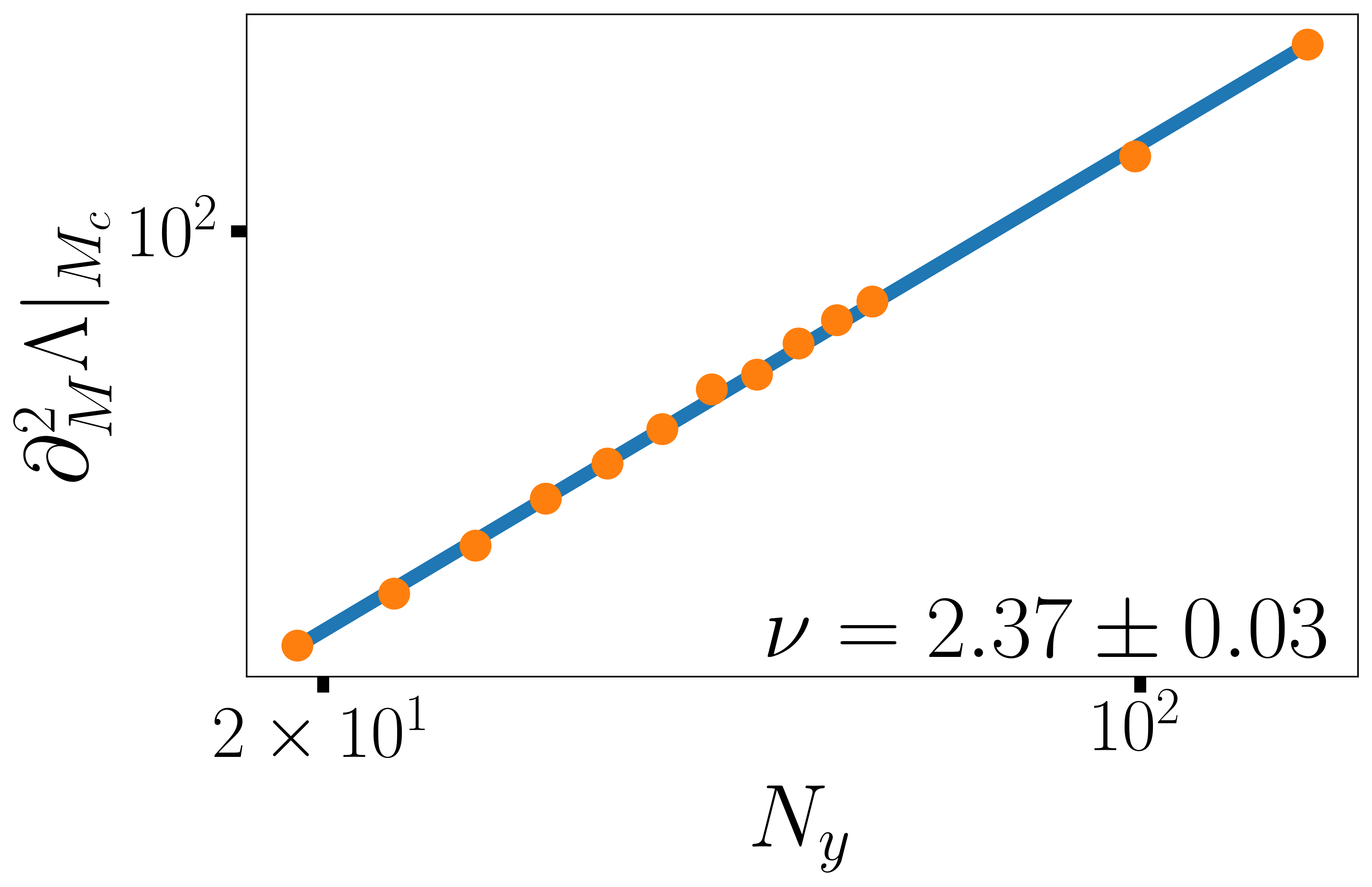}
\caption{Finite-size scaling of the second derivative $\partial_M^2 \Lambda \vert_{M_c}$ at the critical point at $M_c = 2.352$ for the disordered Haldane model with disorder strength $W=2.6$. The divergence of the second derivative is well describes by a power-law $\partial_M^2 \Lambda \vert_{M_c} \sim N_y^{2/\nu}$ with $\nu=2.37 \pm 0.03$, without subleading corrections.}
\label{fig:w26_2nd_der}
\end{figure}

In order to extract the exponent $\nu$, we examine the finite-size scaling of the second derivative $\partial_m^2 \Lambda$ at the critical point corresponding to $m=0$. In Fig.~\ref{fig:w26_2nd_der} we show that the derivative scales as $\partial_m^2 \Lambda \sim N_y^{2/\nu}$ with $\nu =2.37 \pm 0.03$. This is in accordance with the leading order contribution in Eq.~(\ref{eq:sec_der}) without subleading corrections. 
The value of the exponent for this specific disorder strength, $W = 2.6$,
is numerically close to that of the conjectured exponent $\nu = 7/3 \simeq 2.33$
for the Integer Quantum Hall Effect. It also agrees with
recent numerical results for disordered Dirac fermions at $E = 0$~\cite{SM_Sbierski_2021}. It
is also consistent with results for a geometrically distorted network
model~\cite{SM_Gruzberg_2017} and experiment~\cite{SM_Li_2009}. As discussed below and in the main text,
our results for the Haldane model appear to approach something closer to
$\nu\sim 5/2$ in the strong disorder regime. This is compatible with the
spread in exponents presented in Ref.~\cite{SM_Dresselhaus_2021}.

\begin{figure}[ht]
\includegraphics[width=.41\textwidth]{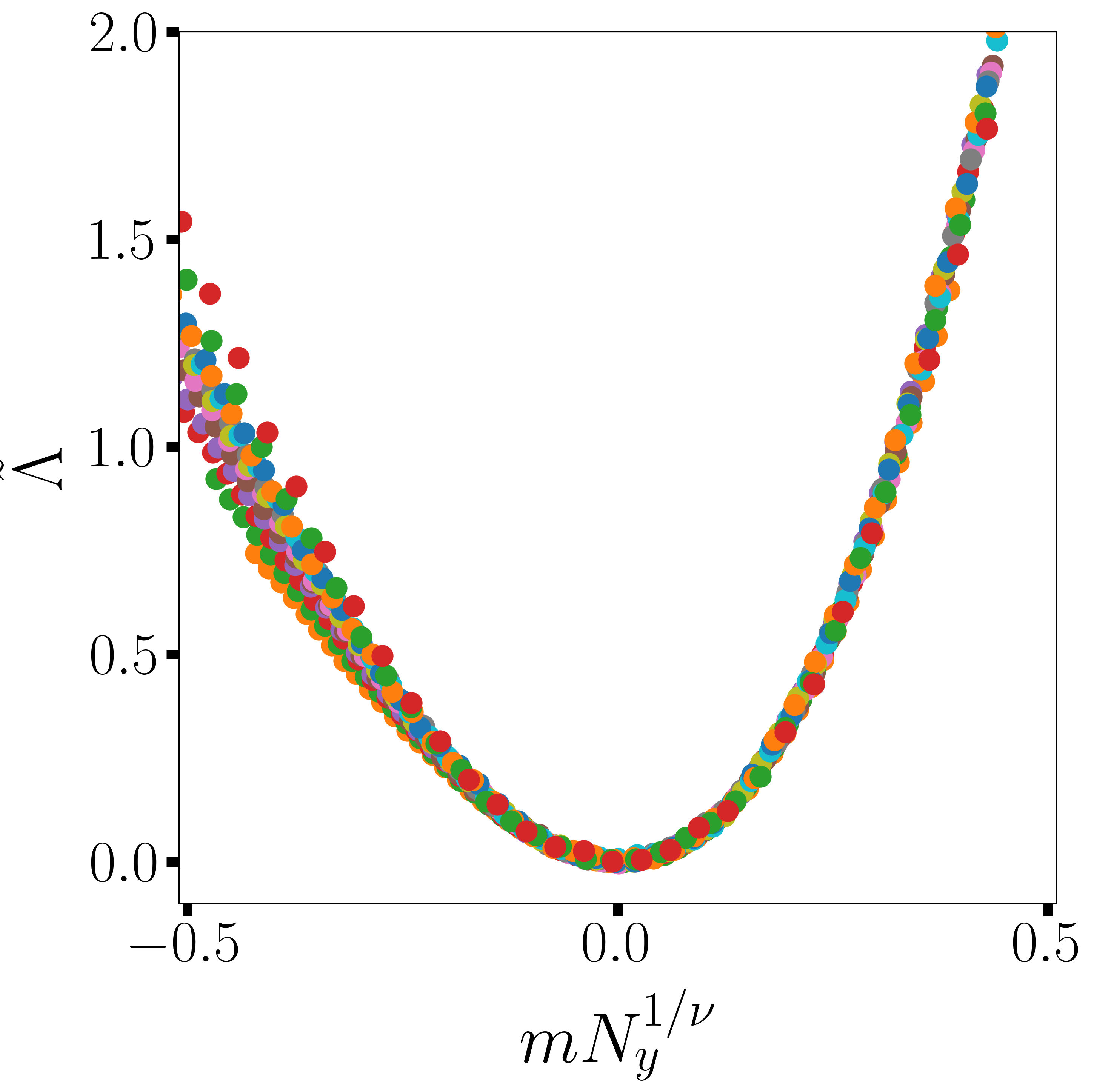}
\caption{Scaling collapse of $\dtilde{\Lambda}$ defined in Eq.~(\ref{eq:lam_cent}) for the disordered Haldane model with $W=2.6$. We set $t_1=3t_2=1$ and $\varphi=\pi/2$ and impose TBCs. In the vicinity of the critical point, the data collapse when plotted as a function of $m N_y^{1/\nu}$ with $\nu=2.37$. Away from the critical point, it can be seen that the collapse is better for $m>0$ than $m<0$.}
\label{fig:w26_collapse}
\end{figure}
\subsection{Intermediate Disorder}
\begin{figure}[ht]
\includegraphics[width=.43\textwidth]{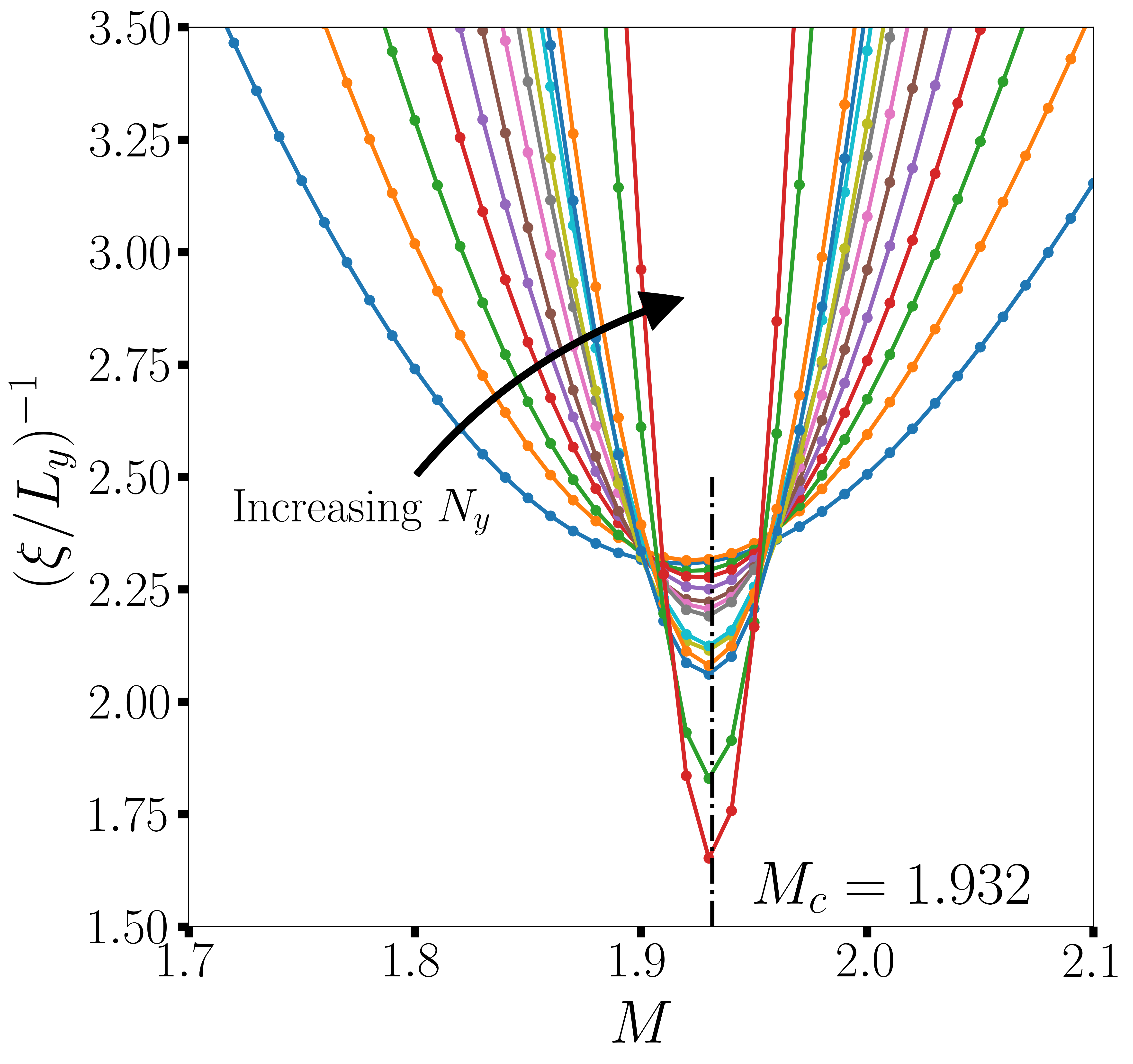}
\caption{Evolution of the dimensionless inverse gap $\Lambda$ for the disordered Haldane model with $W=1.4$ on transiting from the topological to the non-topological phase. We set $t_1=3t_2=1$, $\varphi=\pi/2$ and impose TBCs. The inverse correlation length $\xi^{-1}$ is extracted from $N_x = 10^6$ transfer matrix multiplications in Eq.~(\ref{eq:TM_def}). The location of the minimum of $\Lambda$ drifts with increasing system size $N_y$ towards the critical point at $M_c = 1.932\pm 0.003$; see Fig.~\ref{fig:W14_Xss}. The vertical displacement is strong and downwards as illustrated in Fig.~\ref{fig:W14_Yss}.}
\label{fig:W14_RawData}
\end{figure}

\begin{figure}[ht]
\includegraphics[width=.47\textwidth,height=0.6\linewidth]{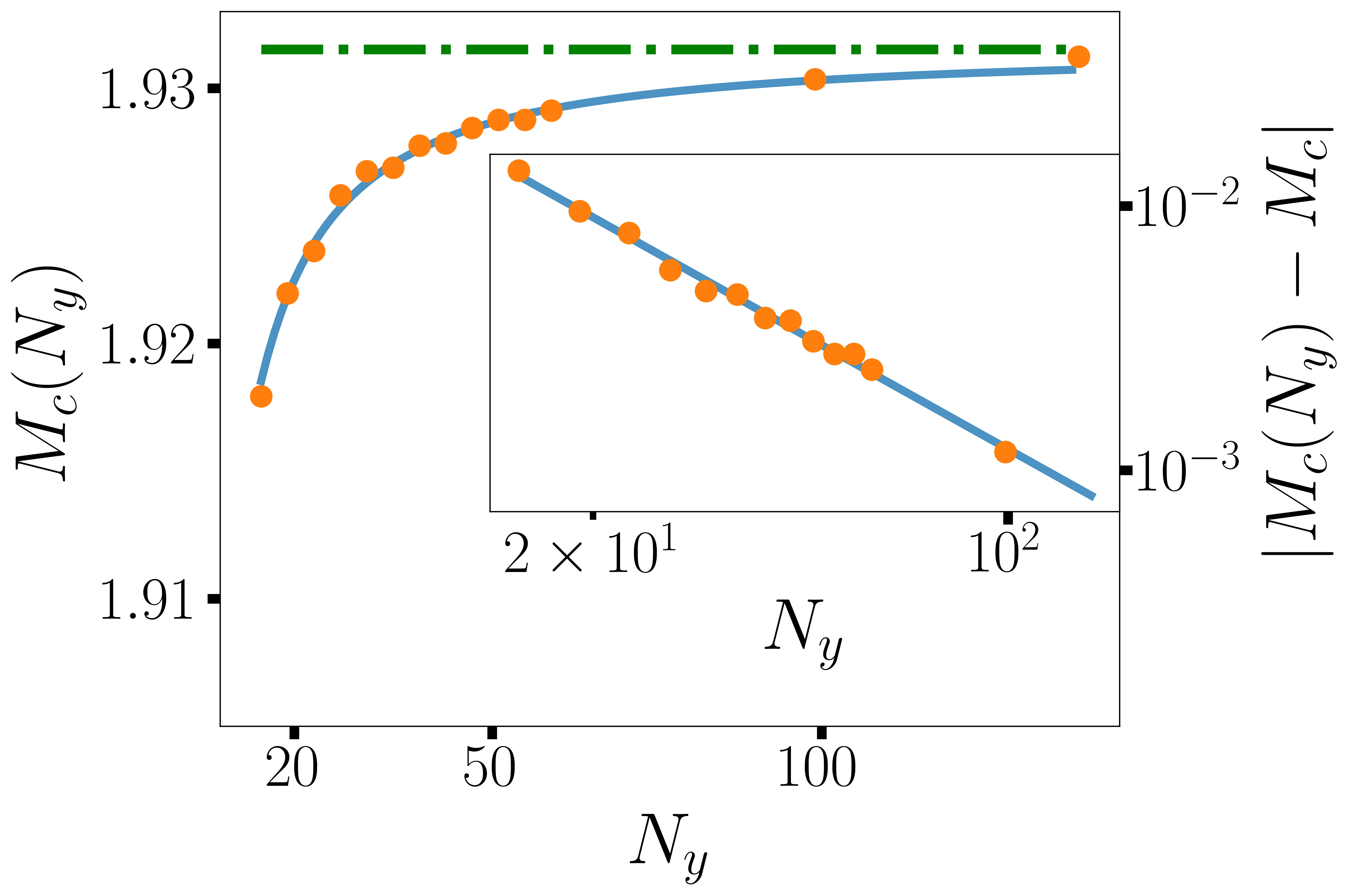}
\caption{Evolution of the finite-size critical point associated with the minimum of $\Lambda$ in Fig.~\ref{fig:W14_RawData}. The data asymptote towards a critical point at $M_c = 1.932$ (dashed line). The data is consistent with Eq.~(\ref{eq:Xdrift}) with the shift exponent $\lambda=1.30 \pm 0.05$ (blue line); see inset.}
\label{fig:W14_Xss}
\end{figure}

\begin{figure}[ht]
\includegraphics[width=.47\textwidth,height=0.6\linewidth]{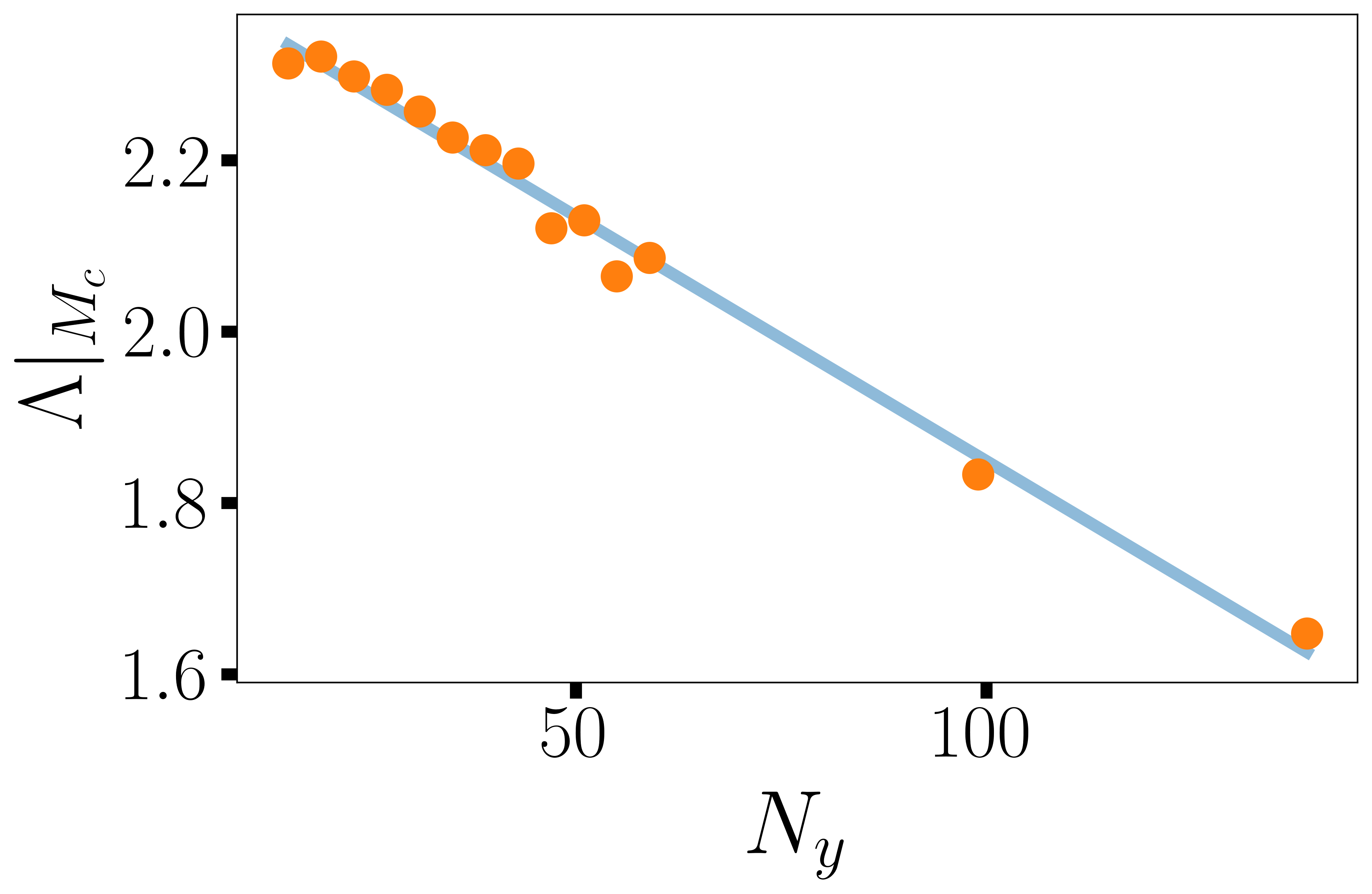}
\caption{Evolution of the dimensionless gap at the critical point $\Lambda\vert_{M_c}$ with increasing system size. For the system sizes considered, the data are consistent with the linear evolution (blue line) without a clear termination point in the thermodynamic limit.}
\label{fig:W14_Yss}
\end{figure}

Having examined the cases of weak and strong disorder, we turn our attention to the intermediate disorder regime. In Fig.~\ref{fig:W14_RawData} we plot the dimensionless gap $\Lambda = (\xi/L_y)^{-1}$ on transiting from the topological to the non-topological phase with $W=1.4$. Clear minima can be seen in the vicinity of $M_c = 1.932\pm 0.003$. In Fig.~\ref{fig:W14_Xss} we show the drift of the critical point with increasing system size $N_y$. The results yield the shift exponent $\lambda = 1.30 \pm 0.05$; see inset. In contrast, the vertical displacement in Fig.~\ref{fig:W14_RawData} shows a strong downward trend with increasing system size $N_y$. The value at the critical point $\Lambda\vert_{M_c}$ decreases linearly without saturation for the system sizes considered; see Fig.~\ref{fig:W14_Yss}.

\begin{figure}[ht]
\includegraphics[width=.41\textwidth]{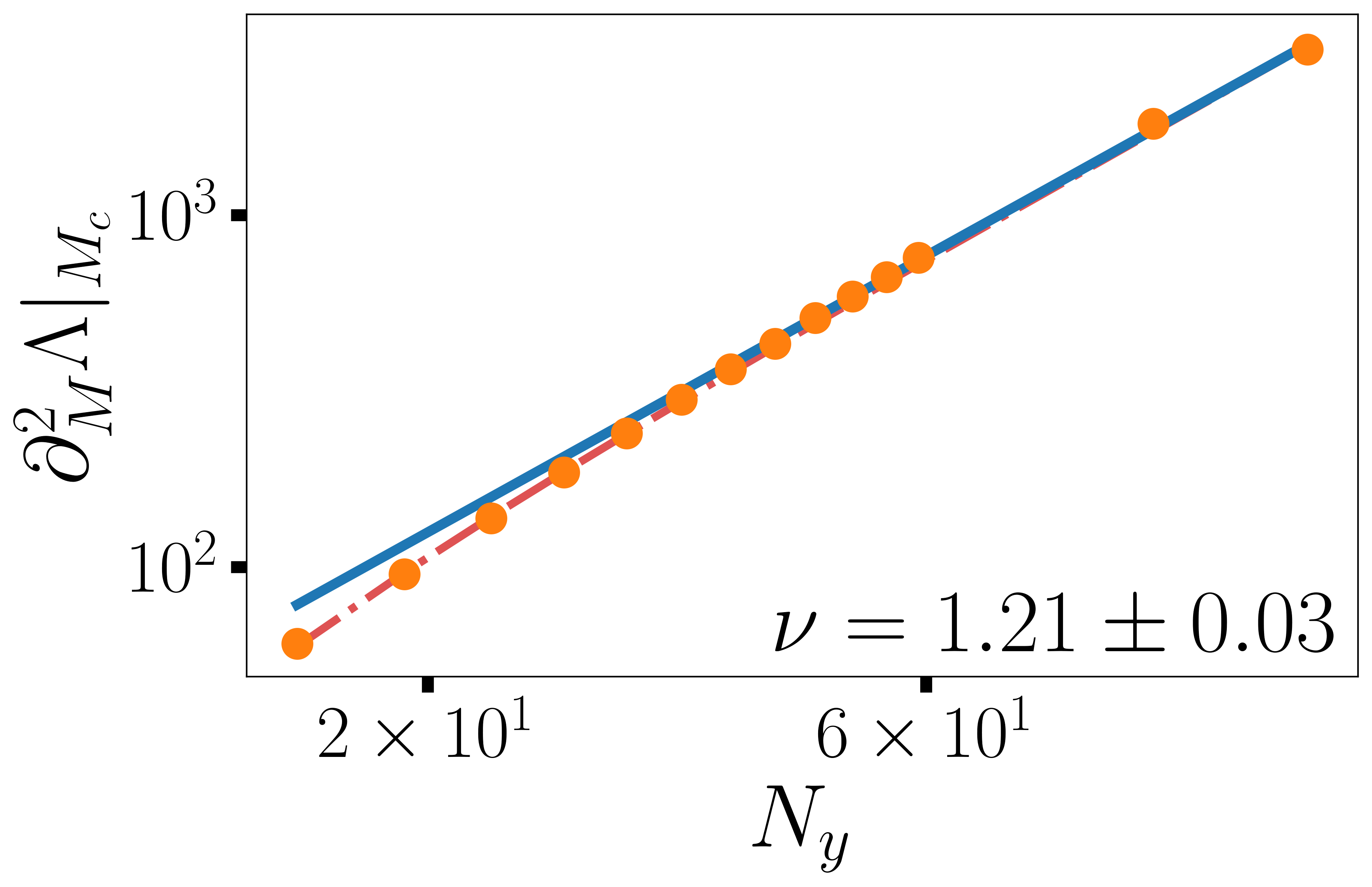}
\caption{Finite-size scaling of the second derivative $\partial_M^2 \Lambda \vert_{M_c}$ at the critical point at $M_c=1.932$ for the disordered Haldane model with $W=1.4$. We set $t_1=3t_2=1$, $\varphi=\pi/2$ and impose TBCs. For large system size $N_y>40$, the growth is well-described by the leading power-law in Eq.~(\ref{eq:sec_der}) with $\nu = 1.21 \pm 0.03$ (solid line). The extracted value $\nu=1.21 \pm 0.03$ is in agreement with that obtained via the Chern marker, $\nu = 1.23 \pm 0.05$. For smaller system sizes it is necessary to include the subleading correction in Eq.~(\ref{eq:sec_der}) (dashed line).}
\label{fig:W14_2nd_der}
\end{figure}

\begin{figure}[ht]
\includegraphics[width=.41\textwidth]{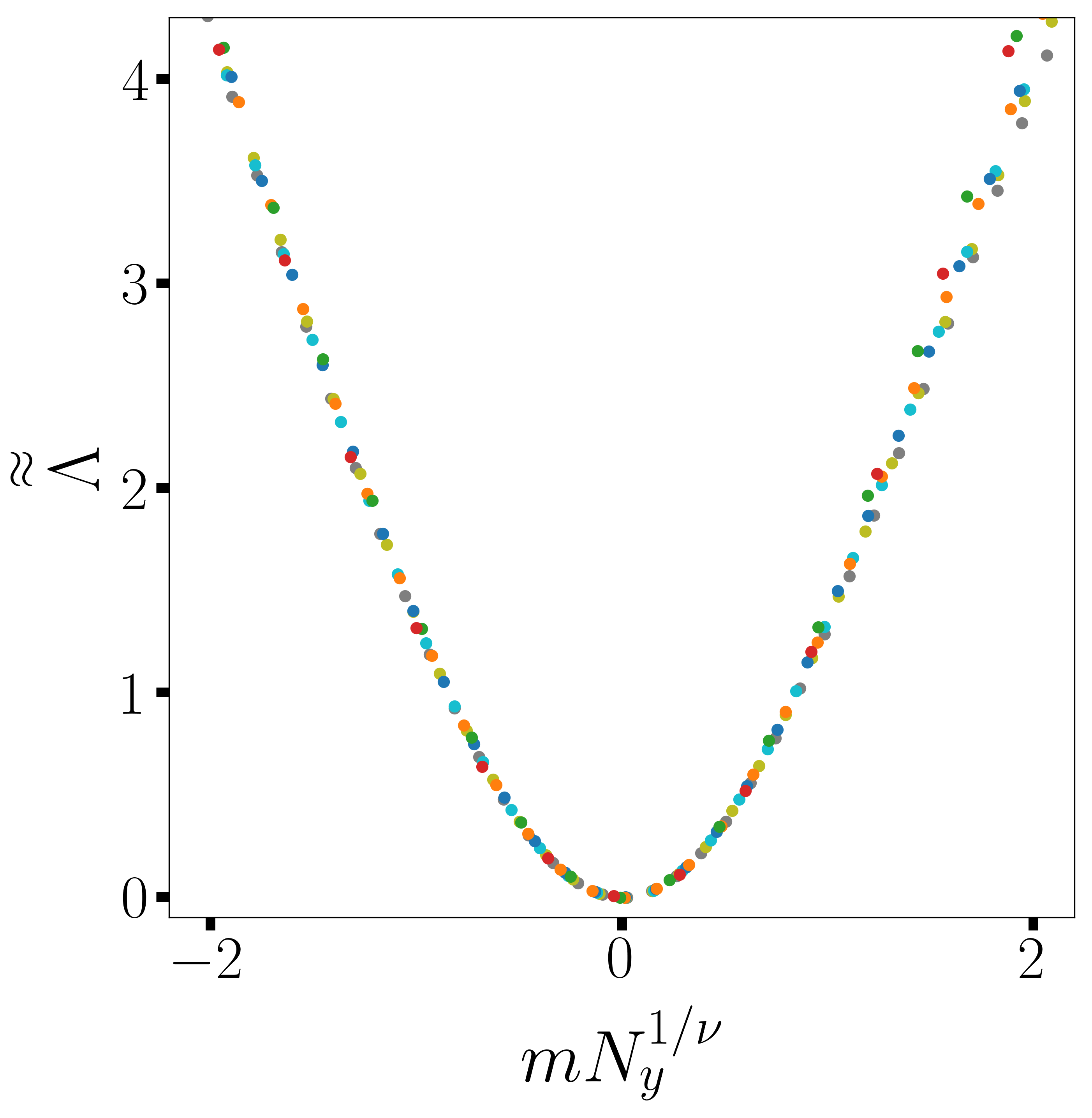}
\caption{Scaling collapse of the centred variable $\dtilde{\Lambda}$ when plotted as a function of $mN_y^{1/\nu}$ for $N_y > 40$ and $\nu=1.21$.}
\label{fig:W14_collapse}
\end{figure}

In order to extract the correlation length exponent $\nu$, we examine the finite-size scaling of the second derivative $\partial_M^2 \Lambda$ at the infinite-size critical point with $M_c=1.932$. As can be seen in Fig.~\ref{fig:W14_2nd_der} the second derivative scales as $\partial_M^2 \Lambda \vert_{M_c} \sim N_y^{2/\nu}$ with $\nu=1.21 \pm 0.03$ for the largest system sizes $N_y > 40$. For smaller system sizes it is necessary to include the subleading corrections in Eq.~\ref{eq:sec_der}. The value of the exponent is in agreement with that derived via the Chern marker, $\nu=1.23 \pm 0.05$. Further verification of this result is obtained by rescaling the horizontal axis in order to see data collapse. In Fig.~\ref{fig:W14_collapse} we plot the centred variable $\dtilde{\Lambda}$ as defined in Eq.~(\ref{eq:lam_cent}) as a function of $mN_y^{1/\nu}$ with $\nu=1.21$. It is readily seen that the data collapse onto a single curve.

\section{Variation of exponents}
\begin{figure}[ht]
    \includegraphics[width=.45\textwidth]{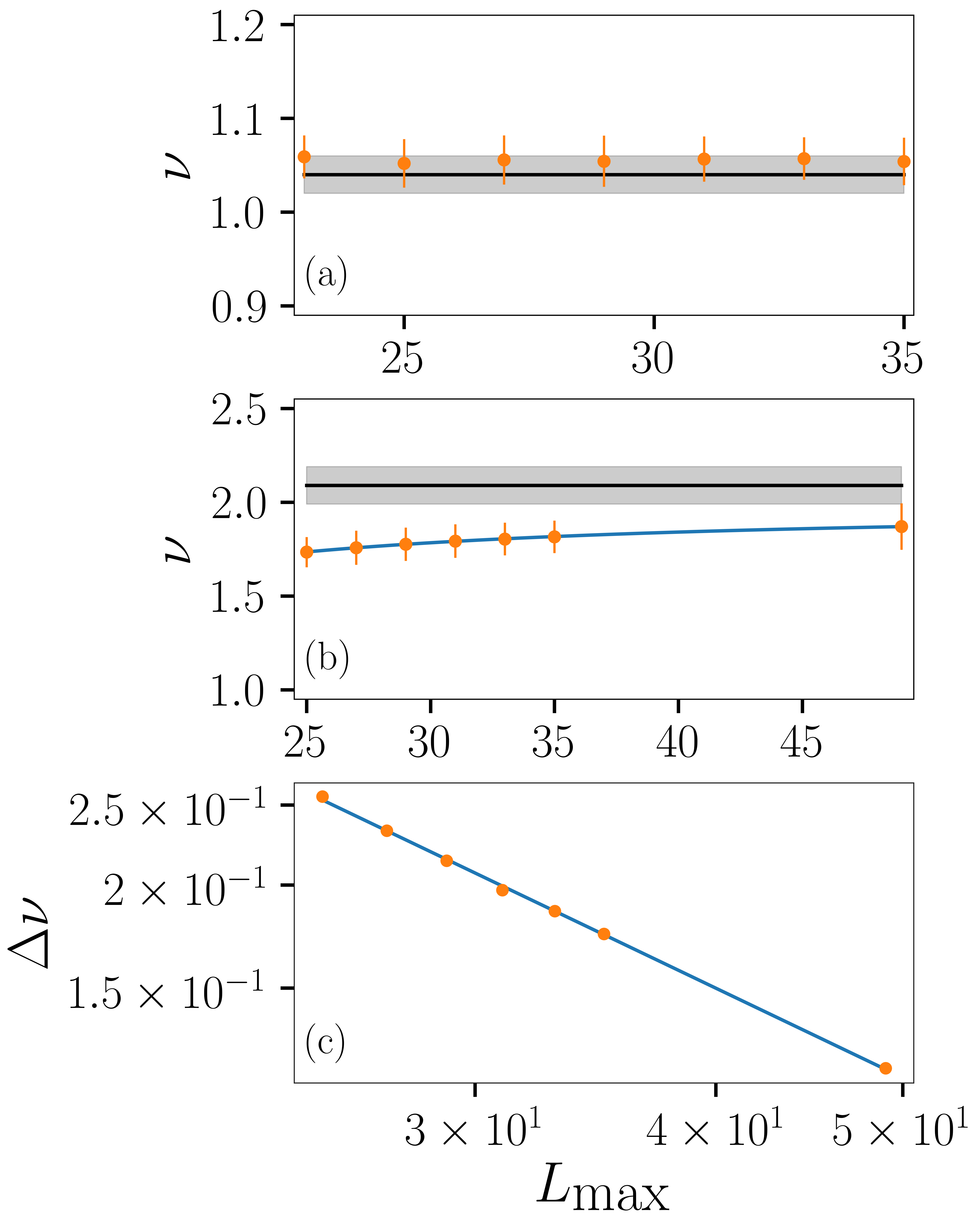}
    \caption{Evolution of the exponent $\nu$ for the mass-driven transitions extracted via the scaling of the Chern marker (orange circles) as a function of the largest system size $L_\textrm{max}$. (a) The results for $W=1$ show little dependence on $L_\textrm{max}$ and agree with the transfer matrix calculations with $\nu=1.04 \pm 0.02$ (black line and shaded area). (b) The results for $W=2$ show an upward trend towards the value obtained via transfer matrices. We obtain $\nu = 1.99 \pm 0.12$ from a na\"ive extrapolation of the Chern marker results to the thermodynamic limit, in agreement with the transfer matrix calculations. (c) Deviation $\Delta \nu$ of the exponent obtained via the Chern marker and the transfer matrix results. The deviation is compatible with a power-law approach (blue line) to the extrapolated value $\nu = 1.99$.
    }
    \label{fig:Stability}
\end{figure}

\begin{figure}[ht]
    \includegraphics[width=.4\textwidth]{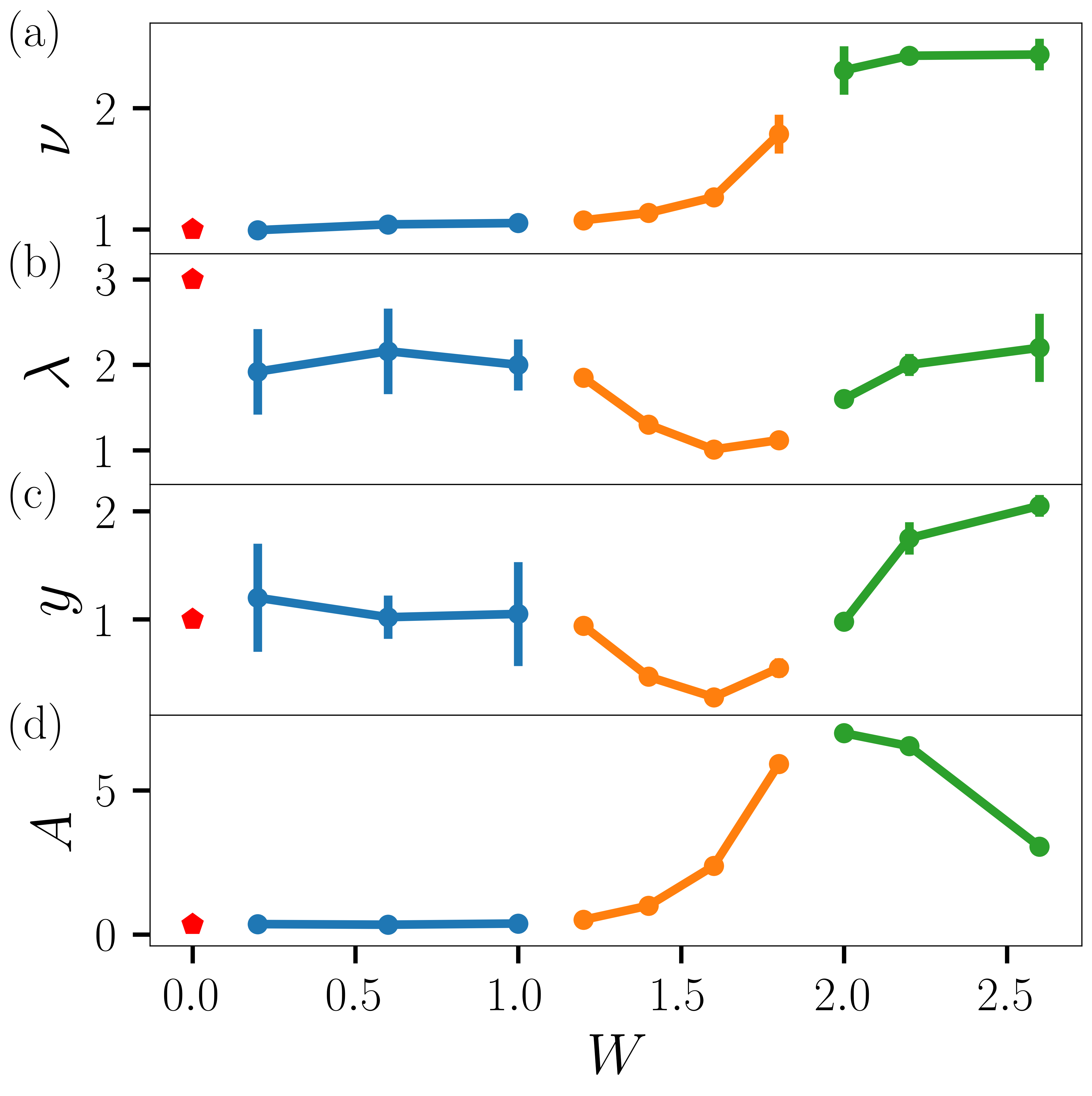}
    \caption{Evolution of the exponents $\nu$, $\lambda$ and $y$ and the amplitude $A$ from Eq.~(\ref{eq:2nd_Der_AW}) with disorder strength $W$. The correlation length exponent $\nu$ interpolates between the clean system $\nu=1$ and $\nu \sim 5/2$, a value close to what is associated with the plateau transitions in the Integer Quantum Hall Effect \cite{SM_Li_2009, SM_Gruzberg_2017, SM_Sbierski_2021}. The exponent $y$ of the irrelevant field interpolates between $y=1$ in the clean case and $y=2$ in the strong disorder regime. The red stars indicate the values obtained for the clean Haldane model.}
    \label{fig:exponents_evolution}
\end{figure}

Having provided evidence that the critical exponents $\lambda$, $y$ and $\nu$ vary with the disorder strength we plot their evolution in Figs.~\ref{fig:exponents_evolution}(a)-(c). As discussed in the main text, the exponent $\nu$ interpolates between that of free Dirac fermions with $\nu=1$ and a value $\nu\sim 5/2$ associated with the plateau transition in the IQHE. As can be seen in Fig.~7(a), of the main text the transfer matrix results are in good agreement with those obtained via the Chern marker. The departures at strong disorder are attributed to finite-size effects in the Chern marker calculations. As shown in Fig.~\ref{fig:Stability}(a), the exponent $\nu$ has little dependence on the maximum system size $L_\textrm{max}$ in the weak disorder regime. However, it exhibits a slower growth for stronger disorder; see Fig.~\ref{fig:Stability}(b). On the basis of a na\"ive extrapolation we obtain $\nu = 1.99 \pm 0.12$ for $L_\textrm{max} \rightarrow \infty$, in agreement with the transfer matrix results. The evolution towards the limiting value is well described by a power-law; see Fig.~\ref{fig:Stability}(c). This reduces the discrepancy in Fig.~7(a) of the main text.

\begin{figure}[ht]
    \includegraphics[width=.41\textwidth]{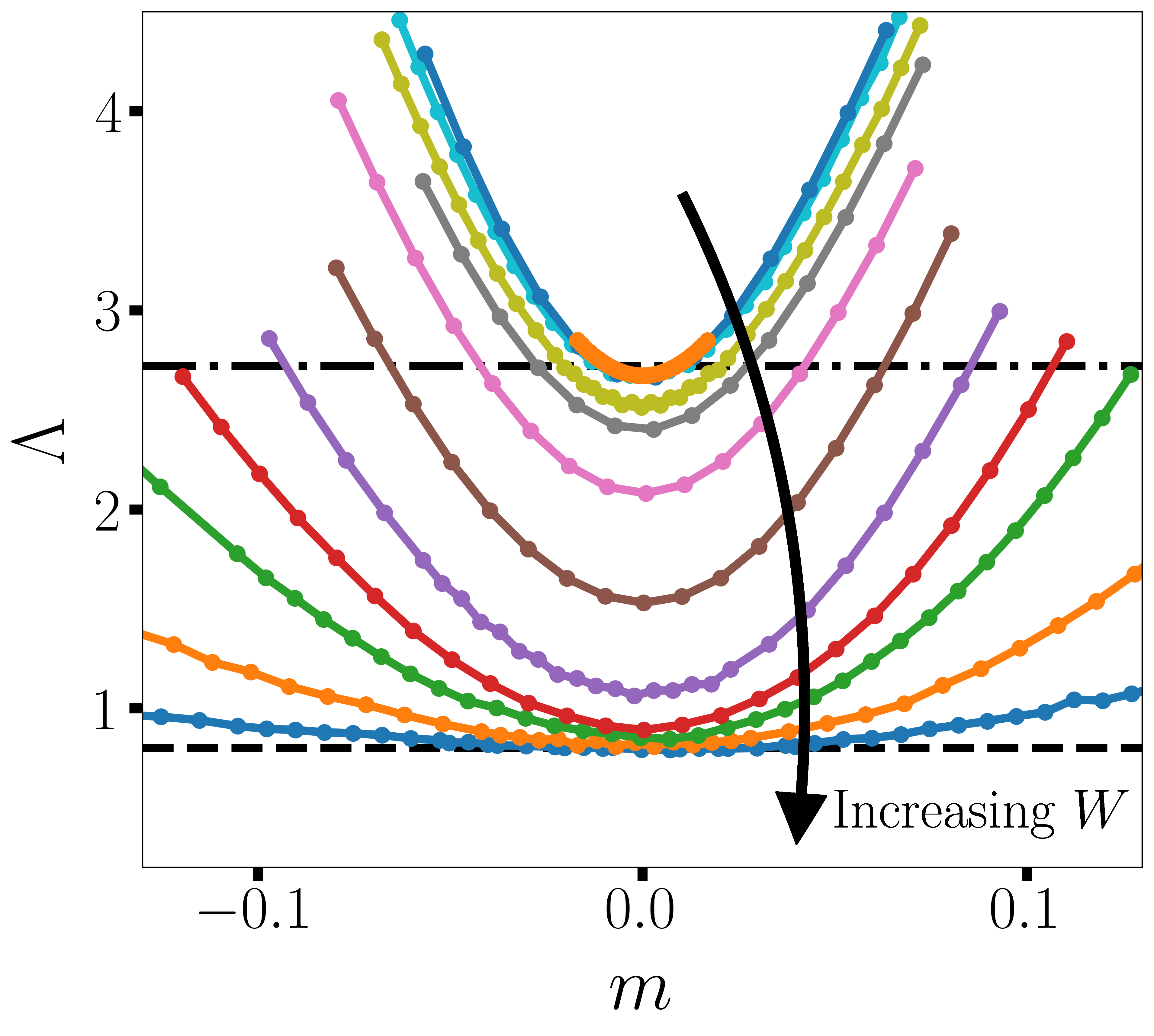}
    \caption{Variation of $\Lambda$ on transiting from the topological to the non-topological phase in the disordered Haldane model with $N_y=59$ for different disorder strengths $W$. We set $t_1=3t_2=1$ and $\varphi=\pi/2$ and impose TBCs. The minimum of $\Lambda$ interpolates that of free Dirac fermion where $\Lambda_0 = 3\pi/(2\sqrt{3})$ (dash-dotted line) and the strong disorder regime where $\Lambda_0 \sim 0.8$ (dashed line). The curvature in the vicinity of the critical point decreases with increasing $W$. This leads to a variation of the correlation length exponent $\nu$ with the disorder strength.}
    \label{fig:W_evolution}
\end{figure}

\begin{figure}[ht]
\includegraphics[width=.41\textwidth]{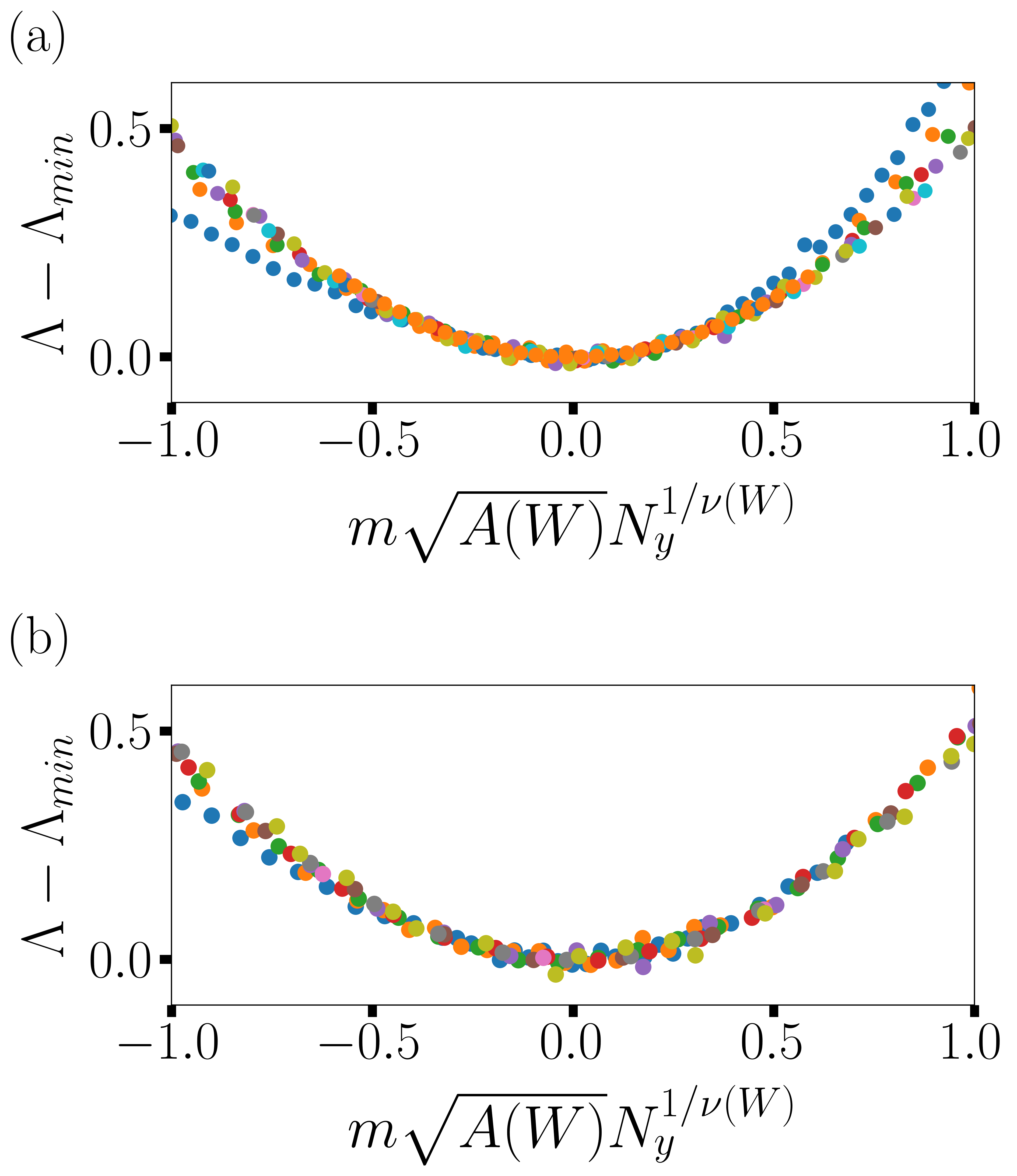}
\caption{(a) Scaling collapse of the data shown in Fig.~\ref{fig:W_evolution} for $N_y=59$ when plotted as a function of $m \sqrt{A(W)} N_y^{1/\nu}$, where $\nu$ and $A$ vary with the disorder strength $W$. (b) Analogous results for $N_y=139$.}
\label{fig:W_evolution_collapse}
\end{figure}

In tandem with the variation of $\nu$, we also observe a variation in $\lambda$ and $y$; see Figs~\ref{fig:exponents_evolution}(b) and (c). In view of the variation of the exponents with the disorder strength, it's instructive to look for scaling collapse as a function of $W$. In Fig.~\ref{fig:W_evolution} we show the variation of $\Lambda$ on transiting from the topological to the non-topological phase for different values of $W$ and a fixed system size with $N_y=59$. It can be seen that the curvature in the vicinity of the critical point decreases with increasing disorder strength.
For large system sizes $N_y$, the curvature in the vicinity of the critical point is given by Eq.~(\ref{eq:sec_der}):
\begin{equation}
    \partial^2_m \Lambda \simeq A(W) N_y^{2/\nu(W)},
    \label{eq:2nd_Der_AW}
\end{equation}
where the coefficient $A(W)$ is independent of $N_y$. Within the parabolic approximation for $\Lambda$ one obtains
\begin{equation}
    \Lambda \sim \Lambda_{min} + (m \sqrt{A(W)} N_y^{1/\nu(W)} )^2/2,
    \label{eq:Lam_dtil_quadr}
\end{equation}
where $\Lambda_{min}$ is the value of $\Lambda$ at the minimum. Using Eq.~(\ref{eq:Lam_dtil_quadr}) we can replot the data in Fig.~\ref{fig:W_evolution} as a function of the rescaled variable $m \sqrt{A(W)} N_y^{1/\nu(W)}$, where $\nu$ and $A$ are functions of disorder strength. As shown in Fig.~\ref{fig:W_evolution_collapse}(a) it can be seen that the data collapse in the vicinity of the critical point. Similar behavior is also observed for $N_y=139$ as shown in Fig.~\ref{fig:W_evolution_collapse}(b). The scaling collapse for two distinct system sizes acts as a useful cross-check on the evolution of $\nu$ with disorder strength. 

\section{Disorder-driven Transition}

\begin{figure}[ht]
\includegraphics[width=.43\textwidth]{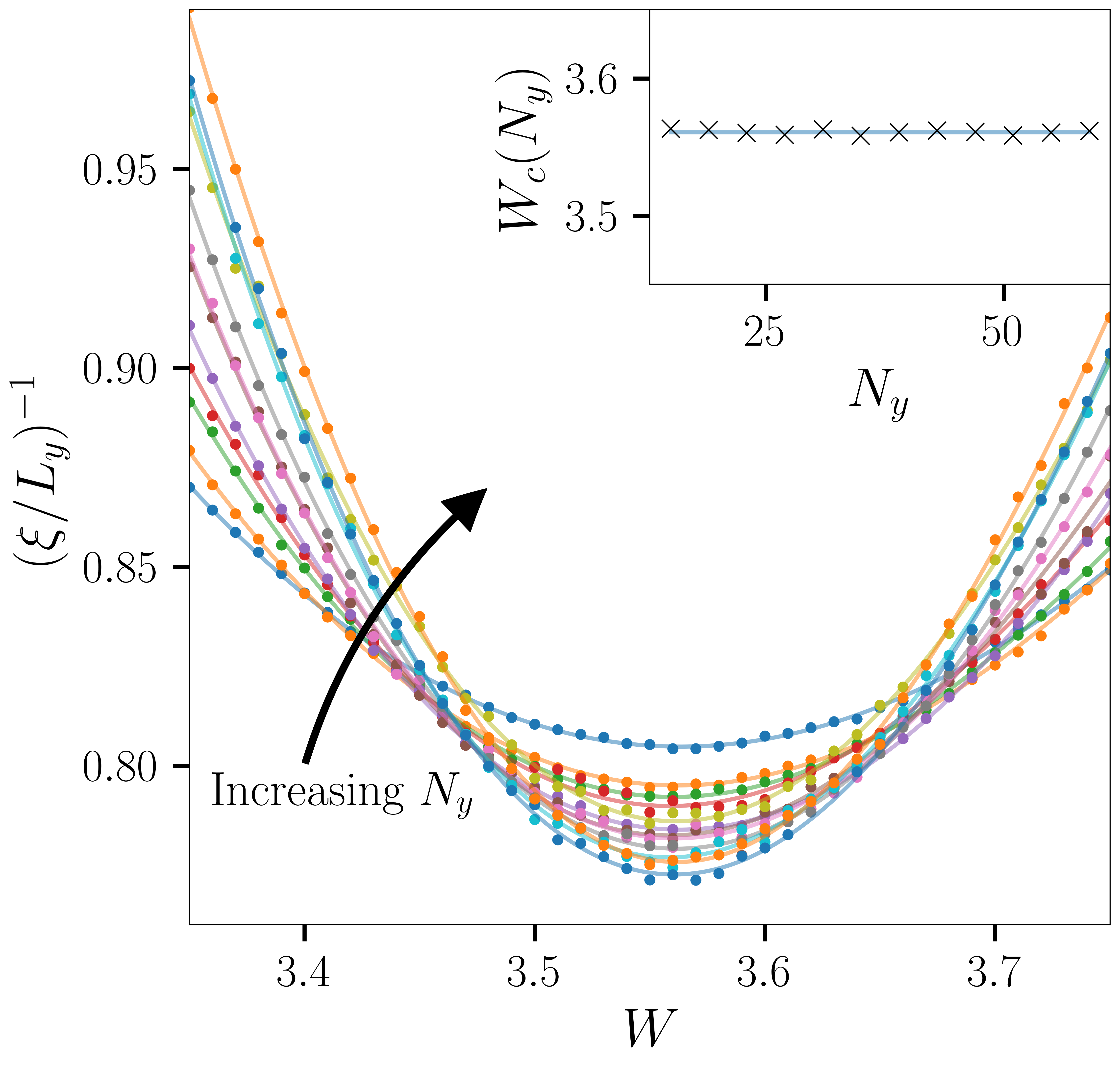}
\caption{Evolution of the dimensionless inverse gap $\Lambda$ for the disordered Haldane model with $M=0$ on transiting from the topological to the non-topological phase with increasing disorder strength. We set $t_1=3t_2=1$ and $\varphi=\pi/2$ and impose TBCs. We consider system sizes $N_y= 19,23,27,..., 59$ and use $N_x=10^7$ transfer matrix multiplications in Eq.~(\ref{eq:TM_def}). The data show a clear minimum in the vicinity of the critical disorder strength $W_c=3.56 \pm 0.01$. Inset: The location of the minimum of $\Lambda$ shows negligible drift as a function of $N_y$.}
\label{fig:DD_raw_data}
\end{figure}

Having examined the $M$-driven transitions at fixed disorder strength, we now turn our attention to the disorder-driven transitions at fixed $M$. In Fig.~\ref{fig:DD_raw_data}, we plot the dimensionless gap $\Lambda = (\xi/L_y)^{-1}$ as a function of the disorder strength on transiting from the topological to the non-topological phase with $M=0$. The data show a clear minimum in the vicinity of a critical disorder strength at $W_c = 3.56\pm 0.01$. In contrast to the $M$-driven transitions, we do not see a drift of the minimum with increasing system size; see the inset of Fig.~\ref{fig:DD_raw_data}.

\begin{figure}[ht]
\includegraphics[width=.43\textwidth]{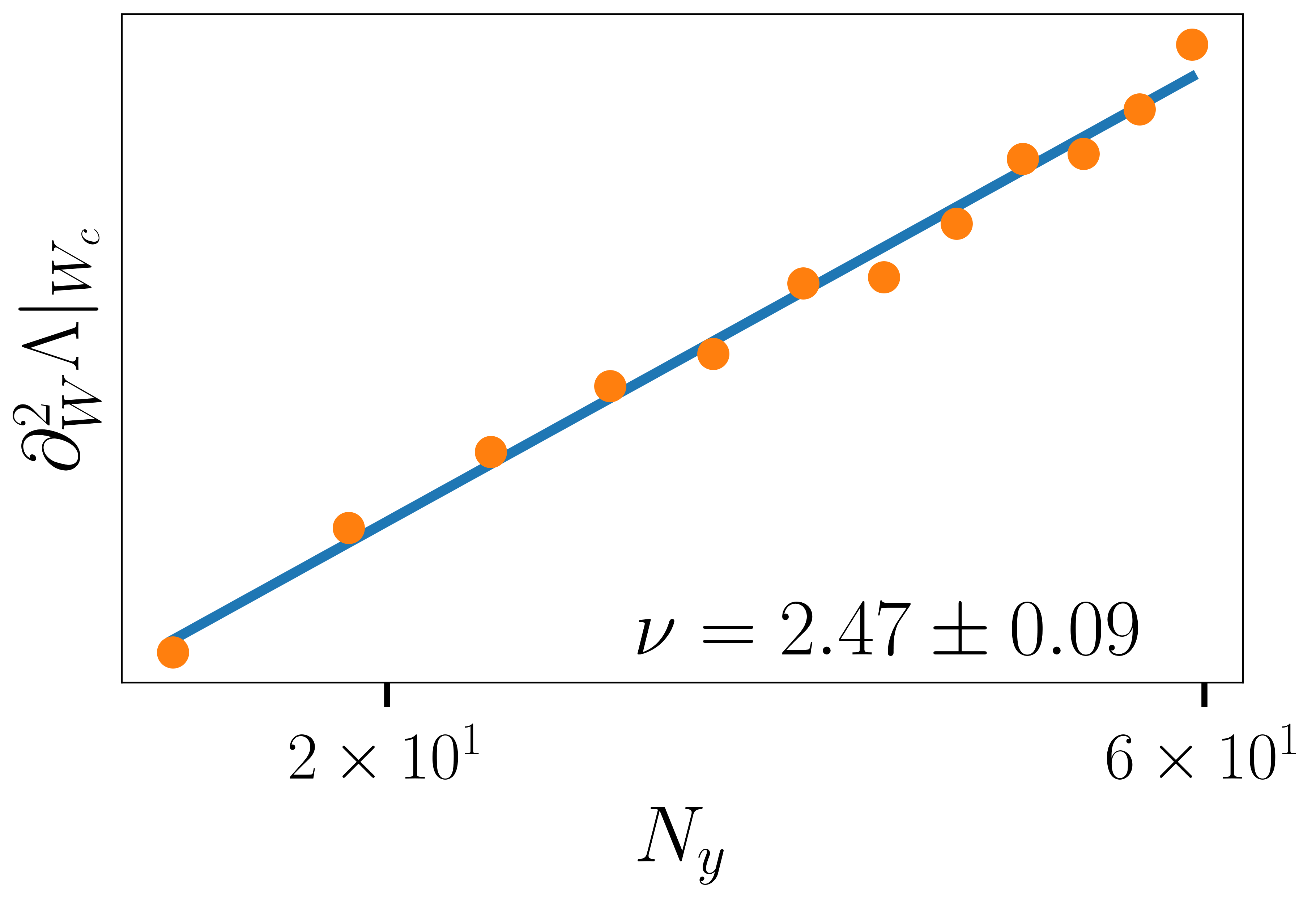}
\caption{Finite-size scaling of the second derivative $\partial_W^2 \Lambda \vert_{W_c}$ at the critical point with $W_c = 3.56$ for the disordered Haldane model. We set $t_1=3t_2=1$,$\varphi=\pi/2$ and $M=0$ with TBCs. The divergence of the second derivative is compatible with a power-law $\partial_W^2 \Lambda \vert_{W_c} \sim N_y^{2/\nu}$ with $\nu=2.47 \pm 0.09$. The result is in agreement with that obtained via the Chern marker.}
\label{fig:DD_2nd_der}
\end{figure}

In order to extract the critical exponent $\nu$, we consider the finite-size scaling of the second derivative $\partial_W^2 \Lambda$ evaluated at the critical point $W_c=3.56$; see Fig.~\ref{fig:DD_2nd_der}. It can be seen that the derivative scales as $\partial_W^2 \Lambda \vert_{W_c} \sim N_y^{2/\nu}$ with $\nu =2.47 \pm 0.09$. This is in accordance with the leading contribution in Eq.~(\ref{eq:sec_der}) without subleading corrections. The value of the exponent is in agreement with that obtained via the Chern marker, $\nu=2.42 \pm 0.11$.

\begin{figure}[ht]
\includegraphics[width=.41\textwidth]{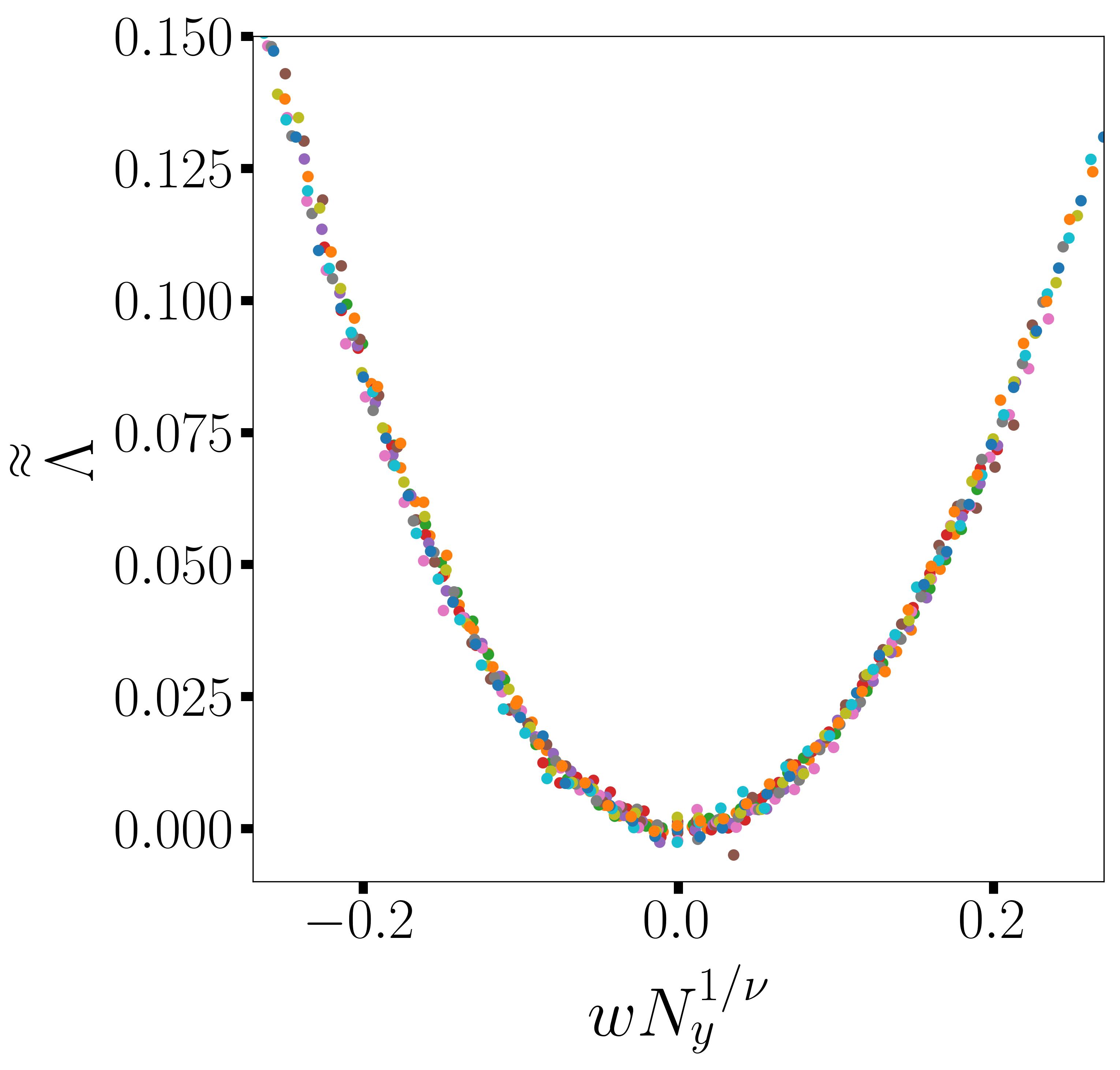}
\caption{Scaling collapse of the data shown in Fig.~\ref{fig:DD_raw_data} when plotted in terms of the centred variable $\dtilde{\Lambda}$ defined in Eq.~(\ref{eq:lam_cent}) and $wN_y^{1/\nu}$, where $w=(W-W_c)/W_c$ and $\nu=2.47$.}
\label{fig:DD_collapse}
\end{figure}

\noindent
We may further verify the extracted value $\nu=2.47$ by rescaling the data in Fig.~\ref{fig:DD_raw_data}. In Fig.~\ref{fig:DD_collapse} we plot the centred variable $\dtilde{\Lambda}$ as defined by Eq.~(\ref{eq:lam_cent}) as a function of $wN_y^{1/\nu}$, where $w=(W-W_c)/W_c$ is the reduced disorder strength. The data collapse onto a single curve for $\nu = 2.47$.

\begin{figure}[ht]
\includegraphics[width=.49\textwidth]{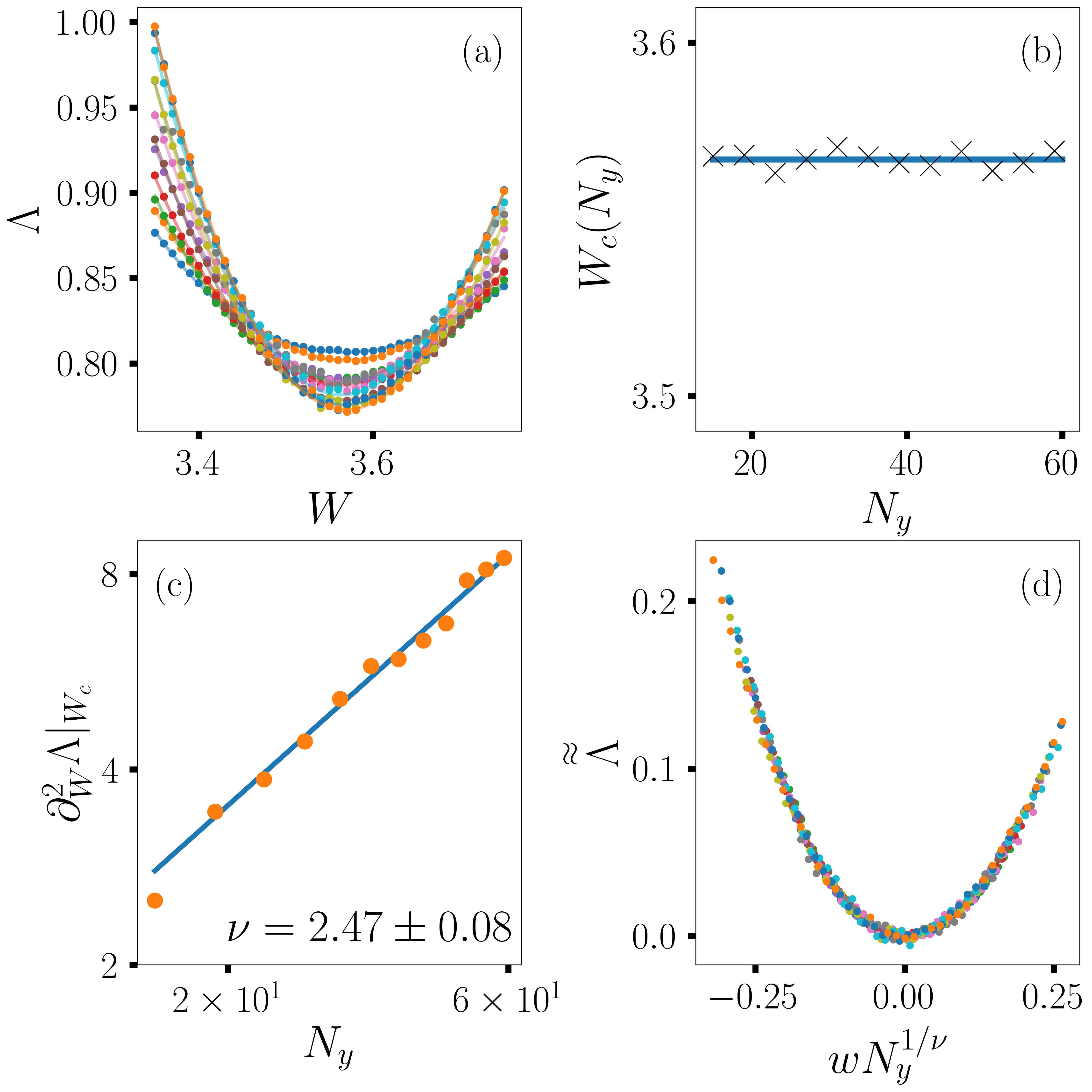}
\caption{(a) Evolution of the dimensionless gap $\Lambda$ on transiting from the topological to the non-topological phase as a function of the disorder strength, with $M=1$ fixed. We set $t_1=3t_2=1$, $\varphi=\pi/2$ with TBCs. (b) The location of the minimum of $\Lambda$ shows a negligible drift with increasing system size. The average corresponds to $W_c = 3.57$ as indicated by the blue line. (c) Finite-size scaling of the second derivative $\partial_W^2 \Lambda \vert_{W_c} \sim N_y^{2/\nu}$ at the critical point. We extract $\nu = 2.47\pm 0.08$ using the system sizes $N_y \geq 19$ (blue line). (d) Scaling collapse of the data shown in (a) when plotted in terms of the centred variable $\dtilde{\Lambda}$ and $wN_y^{1/\nu}$ with $\nu = 2.47$.}
\label{fig:M1_All}
\end{figure}

Having established results for the disorder-driven transition with $M=0$ we turn our attention to the case with $M=1$. In Fig.~\ref{fig:M1_All}(a) we plot the evolution of $\Lambda$ across this quantum phase transition. As found for $M=0$, the minimum exhibits negligible drift with increasing system size; see Fig.~\ref{fig:M1_All}(b). The location of the critical point at $W_c = 3.57 \pm 0.01$ coincides with that for $M=0$. This is consistent with the vertical phase boundary shown in Fig.~3 from the main text. The scaling of the second derivative $\partial_W^2 \Lambda \vert_{W_c} \sim N_y^{2/\nu}$ at the critical point yields $\nu = 2.47\pm 0.08$, in conformity of the result for $M=0$; see Fig.~\ref{fig:M1_All}(c). In Fig.~\ref{fig:M1_All}(d) we replot the data in terms of the centred variable $\dtilde{\Lambda}$ and $wN_y^{1/\nu}$. The data collapse onto a single curve with $\nu = 2.47$ further confirming the result. This suggest that the disorder-driven transitions across the vertical boundary in Fig.~3 of the main text are in the same universality class.

\section{Fluctuations}
Having established the scaling properties o of the disorder averaged Chern marker $\bar{c}$ and its relation to the transfer matrix approach, we now examine the fluctuations of the Chern marker $(\delta c)^2 = \overline{(c - \bar{c})^2}$.

\subsection{Mass-driven Transitions}
\begin{figure}[ht]
  \centering  
   \includegraphics[width=.49\textwidth]{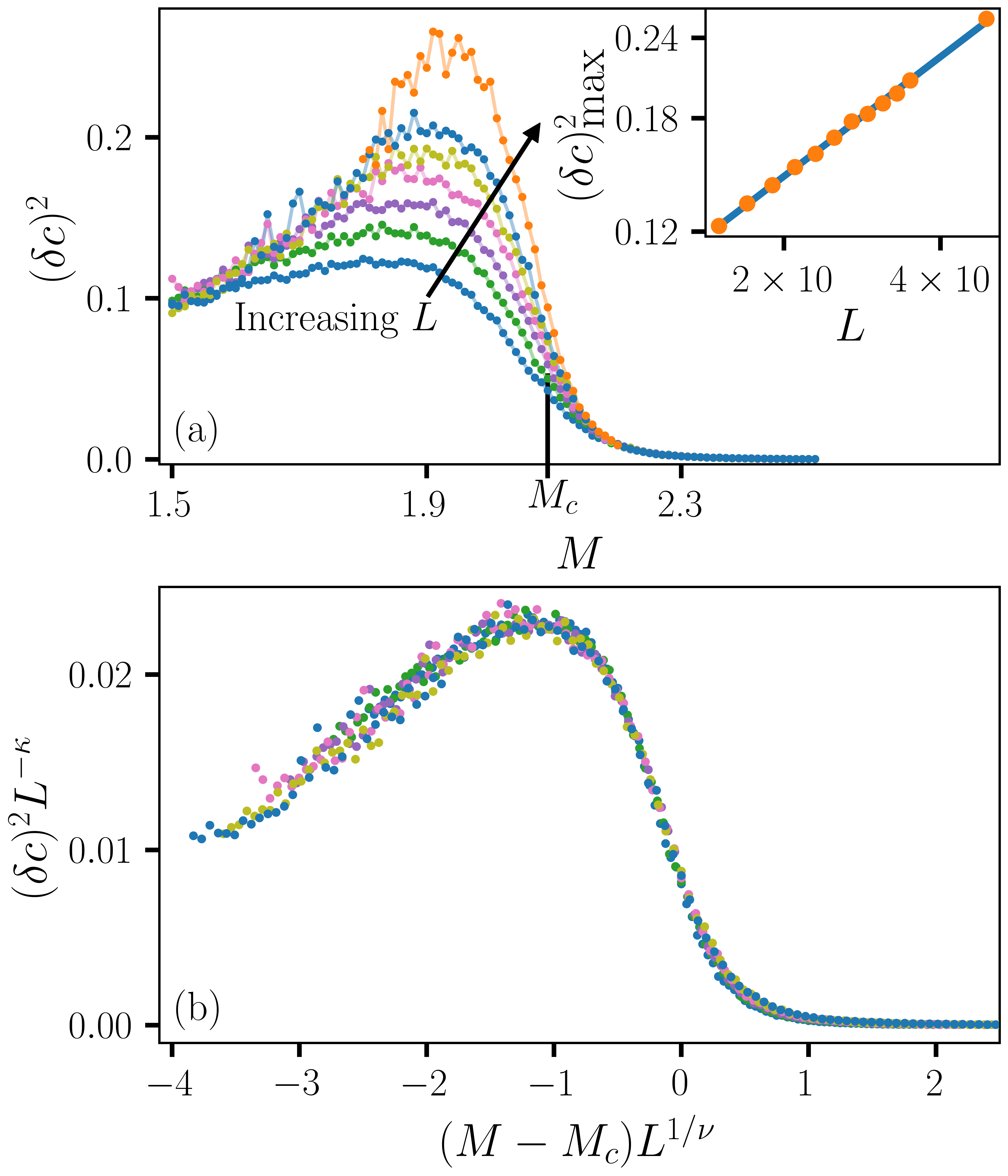}
   \caption{(a) Evolution of the Chern marker fluctuations $(\delta c)^2$ on transiting from the topological to the non-topological phase with $W=1.8$ held fixed. We set $t_1=3t_2=1$, $\varphi = \pi/2$ and average over $10^4$ disorder realisations with $L=15,19,...,35$ and $L=49$. The data show a maximum in the vicinity of the critical point at $M_c\sim 2.09$. Inset: The value of $(\delta c)^2$ at the peak exhibits a power-law scaling $(\delta c)^2_\text{max}\sim L^{\kappa}$ with $\kappa=0.62 \pm 0.02$ (blue line) (b) Scaling collapse of the data shown in panel (a) with  $M_c = 2.09$ and $\nu = 1.7$ obtained from the scaling of the Chern marker.}
   \label{fig:Fluct_W18}
 \end{figure}

In Fig.~\ref{fig:Fluct_W18}(a) we plot the evolution of $(\delta c)^2$ on transiting from the topological to the non-topological phase, with $W=1.8$ held fixed. The data show a peak on approaching the critical point at $M_c \sim 2.09$. The maximum of $(\delta c)^2$ exhibits a power-law scaling with increasing system size, $(\delta c)^2_\text{max} \sim L^\kappa$ with $\kappa = 0.62 \pm 0.02$; see inset of Fig.~\ref{fig:Fluct_W18}(a). Assuming the scaling form $(\delta c)^2 \sim L^\kappa g(\xi/L) \sim L^\kappa \tilde{g}((M-M_c)L^{1/\nu})$ with $M_c = 2.09$ and $\nu = 1.7$, the data collapse in the vicinity of the transition; see Fig.~\ref{fig:Fluct_W18}(b). The result for $\kappa$ differs from that at $W=1$ where $\kappa = 0.36 \pm 0.02$; see Fig.~6 in the main text. 
Repeating the same analysis for different disorder strengths we find $\kappa = 0.35(4), 0.35(3), 0.35(9), 0.37(6), 0.61(7)$ and $0.67(6)$ for $W = 0.2, 0.6, 1, 1.4, 1.8$ and $2.0$ respectively. The exponent interpolates between $\kappa \sim 0.35$ in the weak disorder regime and $\kappa \sim 0.65$ in the strong disorder regime as shown in Fig.~7(b) in the main text. The variation of $\kappa$ mirrors that of $\nu$.

\subsection{Disorder-driven Transition}
\begin{figure}[htb!]
  \centering 
   \includegraphics[width=.49\textwidth]{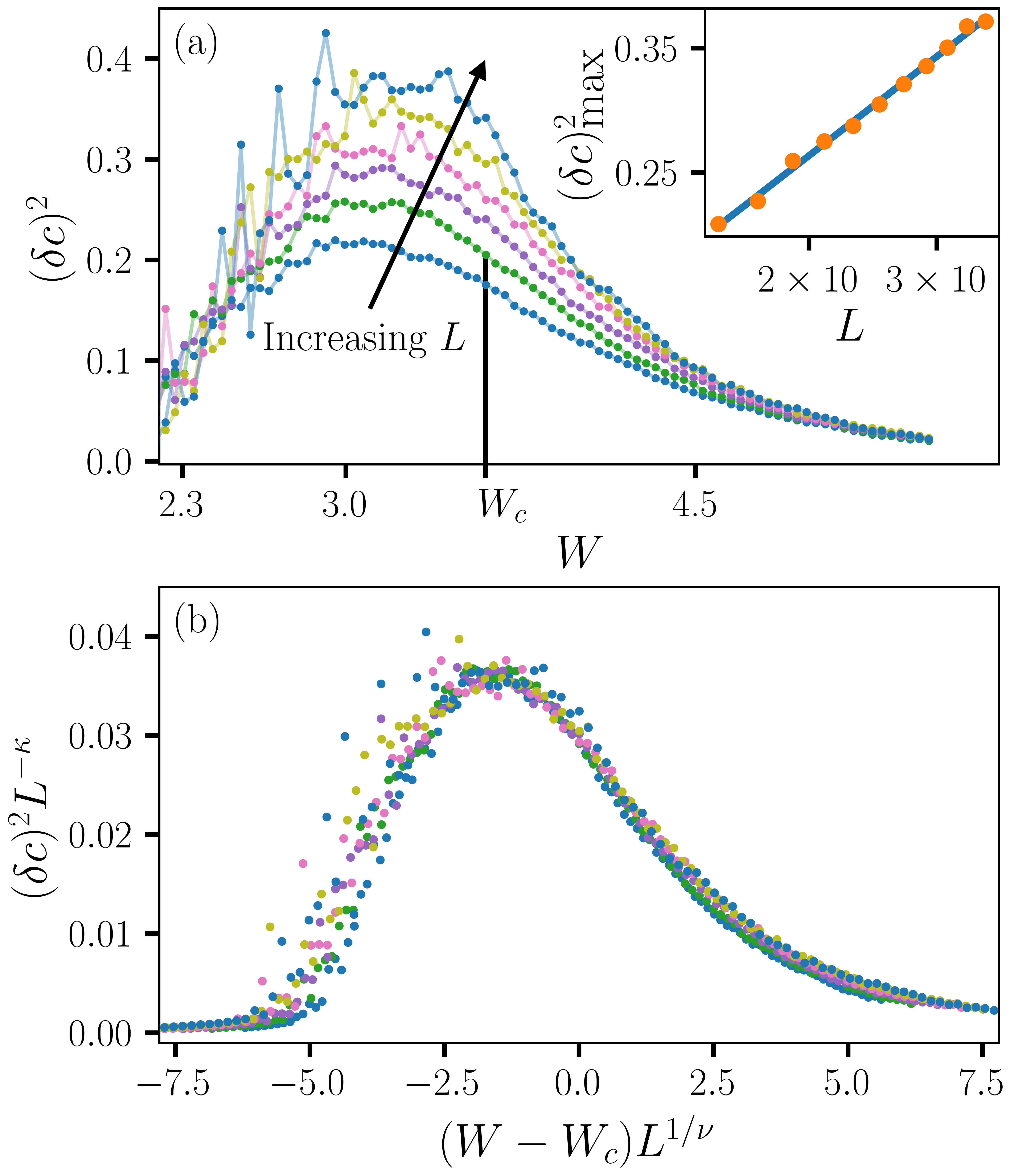}
   \caption{(a) Evolution of the Chern marker fluctuations $(\delta c)^2$ for the disorder-driven transition with $M=0$. We set $t_1=3t_2=1$, $\varphi = \pi/2$ and average over $3 \times 10^4$ disorder realisations with $L=15,19,...,35$. The data show a maximum in the vicinity of the critical point at $W_c\sim 3.6$. Inset: The value of $(\delta c)^2$ at the peak exhibits a power-law scaling $(\delta c)^2_\text{max}\sim L^{\kappa}$ with $\kappa = 0.67 \pm 0.02$. (b) Scaling collapse of the data shown in panel (a) with $W_c = 3.58$ and $\nu = 2.42$ obtained from the scaling of the Chern marker. The data collapse onto a single curve in the vicinity of the critical point.}
     \label{fig:Fluct_DD}
\end{figure}

Having established the scaling of the fluctuations of the Chern marker for the mass-driven transitions, we now examine the disorder-driven transition at $M=0$. In Fig.~\ref{fig:Fluct_DD}(a) we plot the evolution of $(\delta c)^2$ across the transition at $W_c \sim 3.6$. The data show a maximum on approaching the transition. The value of $(\delta c)^2$ at the peak grows with increasing system size and is well described by the power-law scaling $(\delta c)^2_\textrm{max} \sim L^\kappa$ with $\kappa = 0.67 \pm 0.02$; see the inset of Fig.~\ref{fig:Fluct_DD}(a). This value is close to that of the mass-driven transitions in the strong disorder regime; see Fig.~7(b) in the main text. Assuming the scaling form $(\delta c)^2 \sim L^{\kappa} f(\xi/L) \sim L^{\kappa} f((W-W_c)L^{1/\nu})$, in Fig.~\ref{fig:Fluct_DD}(b) we plot $(\delta c)^2 L^{-\kappa}$ versus $(W-W_c)L^{1/\nu}$ where $W_c = 3.58$ and $\nu = 2.42$ are obtained via the scaling of the Chern marker. The data collapse onto a single curve in the vicinity of the critical point.

\clearpage


\begin{thebibliography}{48}%
\makeatletter
\providecommand \@ifxundefined [1]{%
 \@ifx{#1\undefined}
}%
\providecommand \@ifnum [1]{%
 \ifnum #1\expandafter \@firstoftwo
 \else \expandafter \@secondoftwo
 \fi
}%
\providecommand \@ifx [1]{%
 \ifx #1\expandafter \@firstoftwo
 \else \expandafter \@secondoftwo
 \fi
}%
\providecommand \natexlab [1]{#1}%
\providecommand \enquote  [1]{``#1''}%
\providecommand \bibnamefont  [1]{#1}%
\providecommand \bibfnamefont [1]{#1}%
\providecommand \citenamefont [1]{#1}%
\providecommand \href@noop [0]{\@secondoftwo}%
\providecommand \href [0]{\begingroup \@sanitize@url \@href}%
\providecommand \@href[1]{\@@startlink{#1}\@@href}%
\providecommand \@@href[1]{\endgroup#1\@@endlink}%
\providecommand \@sanitize@url [0]{\catcode `\\12\catcode `\$12\catcode `\&12\catcode `\#12\catcode `\^12\catcode `\_12\catcode `\%12\relax}%
\providecommand \@@startlink[1]{}%
\providecommand \@@endlink[0]{}%
\providecommand \url  [0]{\begingroup\@sanitize@url \@url }%
\providecommand \@url [1]{\endgroup\@href {#1}{\urlprefix }}%
\providecommand \urlprefix  [0]{URL }%
\providecommand \Eprint [0]{\href }%
\providecommand \doibase [0]{http://dx.doi.org/}%
\providecommand \selectlanguage [0]{\@gobble}%
\providecommand \bibinfo  [0]{\@secondoftwo}%
\providecommand \bibfield  [0]{\@secondoftwo}%
\providecommand \translation [1]{[#1]}%
\providecommand \BibitemOpen [0]{}%
\providecommand \bibitemStop [0]{}%
\providecommand \bibitemNoStop [0]{.\EOS\space}%
\providecommand \EOS [0]{\spacefactor3000\relax}%
\providecommand \BibitemShut  [1]{\csname bibitem#1\endcsname}%
\let\auto@bib@innerbib\@empty
\bibitem [{\citenamefont {Ando}\ \emph {et~al.}(1975)\citenamefont {Ando}, \citenamefont {Matsumoto},\ and\ \citenamefont {Uemura}}]{Ando1975}%
  \BibitemOpen
  \bibfield  {author} {\bibinfo {author} {\bibfnamefont {T.}~\bibnamefont {Ando}}, \bibinfo {author} {\bibfnamefont {Y.}~\bibnamefont {Matsumoto}}, \ and\ \bibinfo {author} {\bibfnamefont {Y.}~\bibnamefont {Uemura}},\ }\href {\doibase 10.1143/JPSJ.39.279} {\bibfield  {journal} {\bibinfo  {journal} {J. Phys. Soc. Japan}\ }\textbf {\bibinfo {volume} {39}},\ \bibinfo {pages} {279} (\bibinfo {year} {1975})}\BibitemShut {NoStop}%
\bibitem [{\citenamefont {Klitzing}\ \emph {et~al.}(1980)\citenamefont {Klitzing}, \citenamefont {Dorda},\ and\ \citenamefont {Pepper}}]{Klitzing1980}%
  \BibitemOpen
  \bibfield  {author} {\bibinfo {author} {\bibfnamefont {K.~v.}\ \bibnamefont {Klitzing}}, \bibinfo {author} {\bibfnamefont {G.}~\bibnamefont {Dorda}}, \ and\ \bibinfo {author} {\bibfnamefont {M.}~\bibnamefont {Pepper}},\ }\href {\doibase 10.1103/PhysRevLett.45.494} {\bibfield  {journal} {\bibinfo  {journal} {Phys. Rev. Lett.}\ }\textbf {\bibinfo {volume} {45}},\ \bibinfo {pages} {494} (\bibinfo {year} {1980})}\BibitemShut {NoStop}%
\bibitem [{\citenamefont {Nayak}\ \emph {et~al.}(2008)\citenamefont {Nayak}, \citenamefont {Simon}, \citenamefont {Stern}, \citenamefont {Freedman},\ and\ \citenamefont {{Das Sarma}}}]{Nayak2008}%
  \BibitemOpen
  \bibfield  {author} {\bibinfo {author} {\bibfnamefont {C.}~\bibnamefont {Nayak}}, \bibinfo {author} {\bibfnamefont {S.~H.}\ \bibnamefont {Simon}}, \bibinfo {author} {\bibfnamefont {A.}~\bibnamefont {Stern}}, \bibinfo {author} {\bibfnamefont {M.}~\bibnamefont {Freedman}}, \ and\ \bibinfo {author} {\bibfnamefont {S.}~\bibnamefont {{Das Sarma}}},\ }\href {\doibase 10.1103/RevModPhys.80.1083} {\bibfield  {journal} {\bibinfo  {journal} {Rev. Mod. Phys.}\ }\textbf {\bibinfo {volume} {80}},\ \bibinfo {pages} {1083} (\bibinfo {year} {2008})}\BibitemShut {NoStop}%
\bibitem [{\citenamefont {Hasan}\ and\ \citenamefont {Kane}(2010)}]{Hasan2010}%
  \BibitemOpen
  \bibfield  {author} {\bibinfo {author} {\bibfnamefont {M.~Z.}\ \bibnamefont {Hasan}}\ and\ \bibinfo {author} {\bibfnamefont {C.~L.}\ \bibnamefont {Kane}},\ }\href {\doibase 10.1103/RevModPhys.82.3045} {\bibfield  {journal} {\bibinfo  {journal} {Rev. Mod. Phys.}\ }\textbf {\bibinfo {volume} {82}},\ \bibinfo {pages} {3045} (\bibinfo {year} {2010})}\BibitemShut {NoStop}%
\bibitem [{\citenamefont {Dalibard}\ \emph {et~al.}(2011)\citenamefont {Dalibard}, \citenamefont {Gerbier}, \citenamefont {Juzeliūnas},\ and\ \citenamefont {{\"{O}}hberg}}]{Dalibard2011}%
  \BibitemOpen
  \bibfield  {author} {\bibinfo {author} {\bibfnamefont {J.}~\bibnamefont {Dalibard}}, \bibinfo {author} {\bibfnamefont {F.}~\bibnamefont {Gerbier}}, \bibinfo {author} {\bibfnamefont {G.}~\bibnamefont {Juzeliūnas}}, \ and\ \bibinfo {author} {\bibfnamefont {P.}~\bibnamefont {{\"{O}}hberg}},\ }\href {\doibase 10.1103/RevModPhys.83.1523} {\bibfield  {journal} {\bibinfo  {journal} {Rev. Mod. Phys.}\ }\textbf {\bibinfo {volume} {83}},\ \bibinfo {pages} {1523} (\bibinfo {year} {2011})}\BibitemShut {NoStop}%
\bibitem [{\citenamefont {Qi}\ and\ \citenamefont {Zhang}(2011)}]{Qi2011}%
  \BibitemOpen
  \bibfield  {author} {\bibinfo {author} {\bibfnamefont {X.-L.}\ \bibnamefont {Qi}}\ and\ \bibinfo {author} {\bibfnamefont {S.-C.}\ \bibnamefont {Zhang}},\ }\href {\doibase 10.1103/RevModPhys.83.1057} {\bibfield  {journal} {\bibinfo  {journal} {Rev. Mod. Phys.}\ }\textbf {\bibinfo {volume} {83}},\ \bibinfo {pages} {1057} (\bibinfo {year} {2011})}\BibitemShut {NoStop}%
\bibitem [{\citenamefont {Carusotto}\ and\ \citenamefont {Ciuti}(2013)}]{Carusotto2013}%
  \BibitemOpen
  \bibfield  {author} {\bibinfo {author} {\bibfnamefont {I.}~\bibnamefont {Carusotto}}\ and\ \bibinfo {author} {\bibfnamefont {C.}~\bibnamefont {Ciuti}},\ }\href {\doibase 10.1103/RevModPhys.85.299} {\bibfield  {journal} {\bibinfo  {journal} {Rev. Mod. Phys.}\ }\textbf {\bibinfo {volume} {85}},\ \bibinfo {pages} {299} (\bibinfo {year} {2013})}\BibitemShut {NoStop}%
\bibitem [{\citenamefont {Lu}\ \emph {et~al.}(2014)\citenamefont {Lu}, \citenamefont {Joannopoulos},\ and\ \citenamefont {Solja{\v{c}}i{\'{c}}}}]{Lu2014}%
  \BibitemOpen
  \bibfield  {author} {\bibinfo {author} {\bibfnamefont {L.}~\bibnamefont {Lu}}, \bibinfo {author} {\bibfnamefont {J.~D.}\ \bibnamefont {Joannopoulos}}, \ and\ \bibinfo {author} {\bibfnamefont {M.}~\bibnamefont {Solja{\v{c}}i{\'{c}}}},\ }\href {\doibase 10.1038/nphoton.2014.248} {\bibfield  {journal} {\bibinfo  {journal} {Nat. Photonics}\ }\textbf {\bibinfo {volume} {8}},\ \bibinfo {pages} {821} (\bibinfo {year} {2014})}\BibitemShut {NoStop}%
\bibitem [{\citenamefont {Goldman}\ \emph {et~al.}(2014)\citenamefont {Goldman}, \citenamefont {Juzeliūnas}, \citenamefont {{\"{O}}hberg},\ and\ \citenamefont {Spielman}}]{Goldman2014}%
  \BibitemOpen
  \bibfield  {author} {\bibinfo {author} {\bibfnamefont {N.}~\bibnamefont {Goldman}}, \bibinfo {author} {\bibfnamefont {G.}~\bibnamefont {Juzeliūnas}}, \bibinfo {author} {\bibfnamefont {P.}~\bibnamefont {{\"{O}}hberg}}, \ and\ \bibinfo {author} {\bibfnamefont {I.~B.}\ \bibnamefont {Spielman}},\ }\href {\doibase 10.1088/0034-4885/77/12/126401} {\bibfield  {journal} {\bibinfo  {journal} {Rep. Prog. Phys.}\ }\textbf {\bibinfo {volume} {77}},\ \bibinfo {pages} {126401} (\bibinfo {year} {2014})}\BibitemShut {NoStop}%
\bibitem [{\citenamefont {Kim}\ \emph {et~al.}(2020)\citenamefont {Kim}, \citenamefont {Jacob},\ and\ \citenamefont {Rho}}]{Kim_2020}%
  \BibitemOpen
  \bibfield  {author} {\bibinfo {author} {\bibfnamefont {M.}~\bibnamefont {Kim}}, \bibinfo {author} {\bibfnamefont {Z.}~\bibnamefont {Jacob}}, \ and\ \bibinfo {author} {\bibfnamefont {J.}~\bibnamefont {Rho}},\ }\href {\doibase 10.1038/s41377-020-0331-y} {\bibfield  {journal} {\bibinfo  {journal} {Light Sci. Appl.}\ }\textbf {\bibinfo {volume} {9}},\ \bibinfo {pages} {130} (\bibinfo {year} {2020})}\BibitemShut {NoStop}%
\bibitem [{\citenamefont {Lv}\ \emph {et~al.}(2021)\citenamefont {Lv}, \citenamefont {Qian},\ and\ \citenamefont {Ding}}]{Lv_2021}%
  \BibitemOpen
  \bibfield  {author} {\bibinfo {author} {\bibfnamefont {B.~Q.}\ \bibnamefont {Lv}}, \bibinfo {author} {\bibfnamefont {T.}~\bibnamefont {Qian}}, \ and\ \bibinfo {author} {\bibfnamefont {H.}~\bibnamefont {Ding}},\ }\href {\doibase 10.1103/RevModPhys.93.025002} {\bibfield  {journal} {\bibinfo  {journal} {Rev. Mod. Phys.}\ }\textbf {\bibinfo {volume} {93}},\ \bibinfo {pages} {025002} (\bibinfo {year} {2021})}\BibitemShut {NoStop}%
\bibitem [{\citenamefont {Ni}\ \emph {et~al.}(2023)\citenamefont {Ni}, \citenamefont {Yves}, \citenamefont {Krasnok},\ and\ \citenamefont {Alù}}]{Ni_2023}%
  \BibitemOpen
  \bibfield  {author} {\bibinfo {author} {\bibfnamefont {X.}~\bibnamefont {Ni}}, \bibinfo {author} {\bibfnamefont {S.}~\bibnamefont {Yves}}, \bibinfo {author} {\bibfnamefont {A.}~\bibnamefont {Krasnok}}, \ and\ \bibinfo {author} {\bibfnamefont {A.}~\bibnamefont {Alù}},\ }\href {\doibase 10.1021/acs.chemrev.2c00800} {\bibfield  {journal} {\bibinfo  {journal} {Chem. Rev.}\ }\textbf {\bibinfo {volume} {123}},\ \bibinfo {pages} {7585} (\bibinfo {year} {2023})}\BibitemShut {NoStop}%
\bibitem [{\citenamefont {{Khmelnitskii}}(1983)}]{Khmelnitskii_1983}%
  \BibitemOpen
  \bibfield  {author} {\bibinfo {author} {\bibfnamefont {D.~E.}\ \bibnamefont {{Khmelnitskii}}},\ }\href@noop {} {\bibfield  {journal} {\bibinfo  {journal} {JETP Lett.}\ }\textbf {\bibinfo {volume} {38}},\ \bibinfo {pages} {454} (\bibinfo {year} {1983})}\BibitemShut {NoStop}%
\bibitem [{\citenamefont {Pruisken}(1988)}]{Pruisken_1988}%
  \BibitemOpen
  \bibfield  {author} {\bibinfo {author} {\bibfnamefont {A.~M.~M.}\ \bibnamefont {Pruisken}},\ }\href {\doibase 10.1103/PhysRevLett.61.1297} {\bibfield  {journal} {\bibinfo  {journal} {Phys. Rev. Lett.}\ }\textbf {\bibinfo {volume} {61}},\ \bibinfo {pages} {1297} (\bibinfo {year} {1988})}\BibitemShut {NoStop}%
\bibitem [{\citenamefont {Huckestein}\ and\ \citenamefont {Kramer}(1990)}]{Kramer1990}%
  \BibitemOpen
  \bibfield  {author} {\bibinfo {author} {\bibfnamefont {B.}~\bibnamefont {Huckestein}}\ and\ \bibinfo {author} {\bibfnamefont {B.}~\bibnamefont {Kramer}},\ }\href {\doibase 10.1103/PhysRevLett.64.1437} {\bibfield  {journal} {\bibinfo  {journal} {Phys. Rev. Lett.}\ }\textbf {\bibinfo {volume} {64}},\ \bibinfo {pages} {1437} (\bibinfo {year} {1990})}\BibitemShut {NoStop}%
\bibitem [{\citenamefont {Lütken}\ and\ \citenamefont {Ross}(2007)}]{Lutken_2007}%
  \BibitemOpen
  \bibfield  {author} {\bibinfo {author} {\bibfnamefont {C.}~\bibnamefont {Lütken}}\ and\ \bibinfo {author} {\bibfnamefont {G.}~\bibnamefont {Ross}},\ }\href {\doibase https://doi.org/10.1016/j.physletb.2007.08.022} {\bibfield  {journal} {\bibinfo  {journal} {Phys. Lett. B}\ }\textbf {\bibinfo {volume} {653}},\ \bibinfo {pages} {363} (\bibinfo {year} {2007})}\BibitemShut {NoStop}%
\bibitem [{\citenamefont {Obuse}\ \emph {et~al.}(2012)\citenamefont {Obuse}, \citenamefont {Gruzberg},\ and\ \citenamefont {Evers}}]{Obuse_2012}%
  \BibitemOpen
  \bibfield  {author} {\bibinfo {author} {\bibfnamefont {H.}~\bibnamefont {Obuse}}, \bibinfo {author} {\bibfnamefont {I.~A.}\ \bibnamefont {Gruzberg}}, \ and\ \bibinfo {author} {\bibfnamefont {F.}~\bibnamefont {Evers}},\ }\href {\doibase 10.1103/PhysRevLett.109.206804} {\bibfield  {journal} {\bibinfo  {journal} {Phys. Rev. Lett.}\ }\textbf {\bibinfo {volume} {109}},\ \bibinfo {pages} {206804} (\bibinfo {year} {2012})}\BibitemShut {NoStop}%
\bibitem [{\citenamefont {Dresselhaus}\ \emph {et~al.}(2022{\natexlab{a}})\citenamefont {Dresselhaus}, \citenamefont {Sbierski},\ and\ \citenamefont {Gruzberg}}]{Dresselhaus2022}%
  \BibitemOpen
  \bibfield  {author} {\bibinfo {author} {\bibfnamefont {E.~J.}\ \bibnamefont {Dresselhaus}}, \bibinfo {author} {\bibfnamefont {B.}~\bibnamefont {Sbierski}}, \ and\ \bibinfo {author} {\bibfnamefont {I.~A.}\ \bibnamefont {Gruzberg}},\ }\href {\doibase 10.1103/PhysRevLett.129.026801} {\bibfield  {journal} {\bibinfo  {journal} {Phys. Rev. Lett.}\ }\textbf {\bibinfo {volume} {129}},\ \bibinfo {pages} {026801} (\bibinfo {year} {2022}{\natexlab{a}})}\BibitemShut {NoStop}%
\bibitem [{\citenamefont {Slevin}\ and\ \citenamefont {Ohtsuki}(2023)}]{Slevin_2023}%
  \BibitemOpen
  \bibfield  {author} {\bibinfo {author} {\bibfnamefont {K.}~\bibnamefont {Slevin}}\ and\ \bibinfo {author} {\bibfnamefont {T.}~\bibnamefont {Ohtsuki}},\ }\href {\doibase https://doi.org/10.1002/pssr.202300080} {\bibfield  {journal} {\bibinfo  {journal} {Phys. Status Solidi Rapid Res. Lett.}\ }\textbf {\bibinfo {volume} {\!}},\ \bibinfo {pages} {2300080} (\bibinfo {year} {2023})}\BibitemShut {NoStop}%
\bibitem [{\citenamefont {Chalker}\ and\ \citenamefont {Coddington}(1988)}]{Chalker1988}%
  \BibitemOpen
  \bibfield  {author} {\bibinfo {author} {\bibfnamefont {J.~T.}\ \bibnamefont {Chalker}}\ and\ \bibinfo {author} {\bibfnamefont {P.~D.}\ \bibnamefont {Coddington}},\ }\href {\doibase 10.1088/0022-3719/21/14/008} {\bibfield  {journal} {\bibinfo  {journal} {J. Phys. C Solid State Phys.}\ }\textbf {\bibinfo {volume} {21}},\ \bibinfo {pages} {2665} (\bibinfo {year} {1988})}\BibitemShut {NoStop}%
\bibitem [{\citenamefont {Fulga}\ \emph {et~al.}(2011)\citenamefont {Fulga}, \citenamefont {Hassler}, \citenamefont {Akhmerov},\ and\ \citenamefont {Beenakker}}]{Fulga_2011}%
  \BibitemOpen
  \bibfield  {author} {\bibinfo {author} {\bibfnamefont {I.~C.}\ \bibnamefont {Fulga}}, \bibinfo {author} {\bibfnamefont {F.}~\bibnamefont {Hassler}}, \bibinfo {author} {\bibfnamefont {A.~R.}\ \bibnamefont {Akhmerov}}, \ and\ \bibinfo {author} {\bibfnamefont {C.~W.~J.}\ \bibnamefont {Beenakker}},\ }\href@noop {} {\bibfield  {journal} {\bibinfo  {journal} {Phys. Rev. B}\ }\textbf {\bibinfo {volume} {84}} (\bibinfo {year} {2011})}\BibitemShut {NoStop}%
\bibitem [{\citenamefont {Gruzberg}\ \emph {et~al.}(2017)\citenamefont {Gruzberg}, \citenamefont {Kl\"umper}, \citenamefont {Nuding},\ and\ \citenamefont {Sedrakyan}}]{Gruzberg2017}%
  \BibitemOpen
  \bibfield  {author} {\bibinfo {author} {\bibfnamefont {I.~A.}\ \bibnamefont {Gruzberg}}, \bibinfo {author} {\bibfnamefont {A.}~\bibnamefont {Kl\"umper}}, \bibinfo {author} {\bibfnamefont {W.}~\bibnamefont {Nuding}}, \ and\ \bibinfo {author} {\bibfnamefont {A.}~\bibnamefont {Sedrakyan}},\ }\href {\doibase 10.1103/PhysRevB.95.125414} {\bibfield  {journal} {\bibinfo  {journal} {Phys. Rev. B}\ }\textbf {\bibinfo {volume} {95}},\ \bibinfo {pages} {125414} (\bibinfo {year} {2017})}\BibitemShut {NoStop}%
\bibitem [{\citenamefont {Zirnbauer}(2019)}]{Zirnbauer_2019}%
  \BibitemOpen
  \bibfield  {author} {\bibinfo {author} {\bibfnamefont {M.~R.}\ \bibnamefont {Zirnbauer}},\ }\href {\doibase https://doi.org/10.1016/j.nuclphysb.2019.02.017} {\bibfield  {journal} {\bibinfo  {journal} {Nucl. Phys. B}\ }\textbf {\bibinfo {volume} {941}},\ \bibinfo {pages} {458} (\bibinfo {year} {2019})}\BibitemShut {NoStop}%
\bibitem [{\citenamefont {Zirnbauer}(2021)}]{Zirnbauer_2021}%
  \BibitemOpen
  \bibfield  {author} {\bibinfo {author} {\bibfnamefont {M.~R.}\ \bibnamefont {Zirnbauer}},\ }\href {\doibase https://doi.org/10.1016/j.aop.2021.168559} {\bibfield  {journal} {\bibinfo  {journal} {Ann. Phys.}\ }\textbf {\bibinfo {volume} {431}},\ \bibinfo {pages} {168559} (\bibinfo {year} {2021})}\BibitemShut {NoStop}%
\bibitem [{\citenamefont {Ludwig}\ \emph {et~al.}(1994)\citenamefont {Ludwig}, \citenamefont {Fisher}, \citenamefont {Shankar},\ and\ \citenamefont {Grinstein}}]{Ludwig1994}%
  \BibitemOpen
  \bibfield  {author} {\bibinfo {author} {\bibfnamefont {A.~W.~W.}\ \bibnamefont {Ludwig}}, \bibinfo {author} {\bibfnamefont {M.~P.~A.}\ \bibnamefont {Fisher}}, \bibinfo {author} {\bibfnamefont {R.}~\bibnamefont {Shankar}}, \ and\ \bibinfo {author} {\bibfnamefont {G.}~\bibnamefont {Grinstein}},\ }\href {\doibase 10.1103/PhysRevB.50.7526} {\bibfield  {journal} {\bibinfo  {journal} {Phys. Rev. B}\ }\textbf {\bibinfo {volume} {50}},\ \bibinfo {pages} {7526} (\bibinfo {year} {1994})}\BibitemShut {NoStop}%
\bibitem [{\citenamefont {Ho}\ and\ \citenamefont {Chalker}(1996)}]{Chalker_1996}%
  \BibitemOpen
  \bibfield  {author} {\bibinfo {author} {\bibfnamefont {C.-M.}\ \bibnamefont {Ho}}\ and\ \bibinfo {author} {\bibfnamefont {J.~T.}\ \bibnamefont {Chalker}},\ }\href {\doibase 10.1103/PhysRevB.54.8708} {\bibfield  {journal} {\bibinfo  {journal} {Phys. Rev. B}\ }\textbf {\bibinfo {volume} {54}},\ \bibinfo {pages} {8708} (\bibinfo {year} {1996})}\BibitemShut {NoStop}%
\bibitem [{\citenamefont {Guruswamy}\ \emph {et~al.}(2000)\citenamefont {Guruswamy}, \citenamefont {LeClair},\ and\ \citenamefont {Ludwig}}]{Guruswamy_2000}%
  \BibitemOpen
  \bibfield  {author} {\bibinfo {author} {\bibfnamefont {S.}~\bibnamefont {Guruswamy}}, \bibinfo {author} {\bibfnamefont {A.}~\bibnamefont {LeClair}}, \ and\ \bibinfo {author} {\bibfnamefont {A.}~\bibnamefont {Ludwig}},\ }\href {\doibase https://doi.org/10.1016/S0550-3213(00)00245-5} {\bibfield  {journal} {\bibinfo  {journal} {Nucl. Phys. B.}\ }\textbf {\bibinfo {volume} {583}},\ \bibinfo {pages} {475} (\bibinfo {year} {2000})}\BibitemShut {NoStop}%
\bibitem [{\citenamefont {Bernard}\ and\ \citenamefont {LeClair}(2002)}]{Bernard_2002}%
  \BibitemOpen
  \bibfield  {author} {\bibinfo {author} {\bibfnamefont {D.}~\bibnamefont {Bernard}}\ and\ \bibinfo {author} {\bibfnamefont {A.}~\bibnamefont {LeClair}},\ }\href {\doibase 10.1088/0305-4470/35/11/303} {\bibfield  {journal} {\bibinfo  {journal} {J. Phys. A Math. Gen.}\ }\textbf {\bibinfo {volume} {35}},\ \bibinfo {pages} {2555} (\bibinfo {year} {2002})}\BibitemShut {NoStop}%
\bibitem [{\citenamefont {Sbierski}\ \emph {et~al.}(2021)\citenamefont {Sbierski}, \citenamefont {Dresselhaus}, \citenamefont {Moore},\ and\ \citenamefont {Gruzberg}}]{Sbierski_2021}%
  \BibitemOpen
  \bibfield  {author} {\bibinfo {author} {\bibfnamefont {B.}~\bibnamefont {Sbierski}}, \bibinfo {author} {\bibfnamefont {E.~J.}\ \bibnamefont {Dresselhaus}}, \bibinfo {author} {\bibfnamefont {J.~E.}\ \bibnamefont {Moore}}, \ and\ \bibinfo {author} {\bibfnamefont {I.~A.}\ \bibnamefont {Gruzberg}},\ }\href {\doibase 10.1103/PhysRevLett.126.076801} {\bibfield  {journal} {\bibinfo  {journal} {Phys. Rev. Lett.}\ }\textbf {\bibinfo {volume} {126}},\ \bibinfo {pages} {076801} (\bibinfo {year} {2021})}\BibitemShut {NoStop}%
\bibitem [{\citenamefont {Bianco}\ and\ \citenamefont {Resta}(2011)}]{Bianco2011}%
  \BibitemOpen
  \bibfield  {author} {\bibinfo {author} {\bibfnamefont {R.}~\bibnamefont {Bianco}}\ and\ \bibinfo {author} {\bibfnamefont {R.}~\bibnamefont {Resta}},\ }\href {\doibase 10.1103/PhysRevB.84.241106} {\bibfield  {journal} {\bibinfo  {journal} {Phys. Rev. B}\ }\textbf {\bibinfo {volume} {84}},\ \bibinfo {pages} {241106} (\bibinfo {year} {2011})}\BibitemShut {NoStop}%
\bibitem [{\citenamefont {Caio}\ \emph {et~al.}(2019)\citenamefont {Caio}, \citenamefont {Möller}, \citenamefont {Cooper},\ and\ \citenamefont {Bhaseen}}]{Caio2018b}%
  \BibitemOpen
  \bibfield  {author} {\bibinfo {author} {\bibfnamefont {M.~D.}\ \bibnamefont {Caio}}, \bibinfo {author} {\bibfnamefont {G.}~\bibnamefont {Möller}}, \bibinfo {author} {\bibfnamefont {N.~R.}\ \bibnamefont {Cooper}}, \ and\ \bibinfo {author} {\bibfnamefont {M.~J.}\ \bibnamefont {Bhaseen}},\ }\href {\doibase 10.1038/s41567-018-0390-7} {\bibfield  {journal} {\bibinfo  {journal} {Nat. Phys.}\ }\textbf {\bibinfo {volume} {15}},\ \bibinfo {pages} {257} (\bibinfo {year} {2019})}\BibitemShut {NoStop}%
\bibitem [{\citenamefont {Ulcakar}\ \emph {et~al.}(2020)\citenamefont {Ulcakar}, \citenamefont {Mravlje},\ and\ \citenamefont {Rejec}}]{Ul_akar2020}%
  \BibitemOpen
  \bibfield  {author} {\bibinfo {author} {\bibfnamefont {L.}~\bibnamefont {Ulcakar}}, \bibinfo {author} {\bibfnamefont {J.}~\bibnamefont {Mravlje}}, \ and\ \bibinfo {author} {\bibfnamefont {T.}~\bibnamefont {Rejec}},\ }\href {\doibase 10.1103/PhysRevLett.125.216601} {\bibfield  {journal} {\bibinfo  {journal} {Phys. Rev. Lett.}\ }\textbf {\bibinfo {volume} {125}},\ \bibinfo {pages} {216601} (\bibinfo {year} {2020})}\BibitemShut {NoStop}%
\bibitem [{\citenamefont {Beck}\ and\ \citenamefont {Goldstein}(2021)}]{Goldstein_2021}%
  \BibitemOpen
  \bibfield  {author} {\bibinfo {author} {\bibfnamefont {A.}~\bibnamefont {Beck}}\ and\ \bibinfo {author} {\bibfnamefont {M.}~\bibnamefont {Goldstein}},\ }\href {\doibase 10.1103/PhysRevB.103.L241401} {\bibfield  {journal} {\bibinfo  {journal} {Phys. Rev. B}\ }\textbf {\bibinfo {volume} {103}},\ \bibinfo {pages} {L241401} (\bibinfo {year} {2021})}\BibitemShut {NoStop}%
\bibitem [{\citenamefont {d'Ornellas}\ \emph {et~al.}(2022)\citenamefont {d'Ornellas}, \citenamefont {Barnett},\ and\ \citenamefont {Lee}}]{Ornellas_2022}%
  \BibitemOpen
  \bibfield  {author} {\bibinfo {author} {\bibfnamefont {P.}~\bibnamefont {d'Ornellas}}, \bibinfo {author} {\bibfnamefont {R.}~\bibnamefont {Barnett}}, \ and\ \bibinfo {author} {\bibfnamefont {D.~K.~K.}\ \bibnamefont {Lee}},\ }\href {\doibase 10.1103/PhysRevB.106.155124} {\bibfield  {journal} {\bibinfo  {journal} {Phys. Rev. B}\ }\textbf {\bibinfo {volume} {106}},\ \bibinfo {pages} {155124} (\bibinfo {year} {2022})}\BibitemShut {NoStop}%
\bibitem [{\citenamefont {Haldane}(1988)}]{Haldane1988}%
  \BibitemOpen
  \bibfield  {author} {\bibinfo {author} {\bibfnamefont {F.~D.~M.}\ \bibnamefont {Haldane}},\ }\href {\doibase 10.1103/PhysRevLett.61.2015} {\bibfield  {journal} {\bibinfo  {journal} {Phys. Rev. Lett.}\ }\textbf {\bibinfo {volume} {61}},\ \bibinfo {pages} {2015} (\bibinfo {year} {1988})}\BibitemShut {NoStop}%
\bibitem [{\citenamefont {Jotzu}\ \emph {et~al.}(2014)\citenamefont {Jotzu}, \citenamefont {Messer}, \citenamefont {Desbuquois}, \citenamefont {Lebrat}, \citenamefont {Uehlinger}, \citenamefont {Greif},\ and\ \citenamefont {Esslinger}}]{Esslinger2014}%
  \BibitemOpen
  \bibfield  {author} {\bibinfo {author} {\bibfnamefont {G.}~\bibnamefont {Jotzu}}, \bibinfo {author} {\bibfnamefont {M.}~\bibnamefont {Messer}}, \bibinfo {author} {\bibfnamefont {R.}~\bibnamefont {Desbuquois}}, \bibinfo {author} {\bibfnamefont {M.}~\bibnamefont {Lebrat}}, \bibinfo {author} {\bibfnamefont {T.}~\bibnamefont {Uehlinger}}, \bibinfo {author} {\bibfnamefont {D.}~\bibnamefont {Greif}}, \ and\ \bibinfo {author} {\bibfnamefont {T.}~\bibnamefont {Esslinger}},\ }\href {\doibase 10.1038/nature13915} {\bibfield  {journal} {\bibinfo  {journal} {Nature}\ }\textbf {\bibinfo {volume} {515}},\ \bibinfo {pages} {237} (\bibinfo {year} {2014})}\BibitemShut {NoStop}%
\bibitem [{\citenamefont {Kane}\ and\ \citenamefont {Mele}(2005{\natexlab{a}})}]{Kane2005}%
  \BibitemOpen
  \bibfield  {author} {\bibinfo {author} {\bibfnamefont {C.~L.}\ \bibnamefont {Kane}}\ and\ \bibinfo {author} {\bibfnamefont {E.~J.}\ \bibnamefont {Mele}},\ }\href {\doibase 10.1103/PhysRevLett.95.226801} {\bibfield  {journal} {\bibinfo  {journal} {Phys. Rev. Lett.}\ }\textbf {\bibinfo {volume} {95}},\ \bibinfo {pages} {226801} (\bibinfo {year} {2005}{\natexlab{a}})}\BibitemShut {NoStop}%
\bibitem [{\citenamefont {Kane}\ and\ \citenamefont {Mele}(2005{\natexlab{b}})}]{Kane2005a}%
  \BibitemOpen
  \bibfield  {author} {\bibinfo {author} {\bibfnamefont {C.~L.}\ \bibnamefont {Kane}}\ and\ \bibinfo {author} {\bibfnamefont {E.~J.}\ \bibnamefont {Mele}},\ }\href {\doibase 10.1103/PhysRevLett.95.146802} {\bibfield  {journal} {\bibinfo  {journal} {Phys. Rev. Lett.}\ }\textbf {\bibinfo {volume} {95}},\ \bibinfo {pages} {146802} (\bibinfo {year} {2005}{\natexlab{b}})}\BibitemShut {NoStop}%
\bibitem [{\citenamefont {Sriluckshmy}\ \emph {et~al.}(2018)\citenamefont {Sriluckshmy}, \citenamefont {Saha},\ and\ \citenamefont {Moessner}}]{Sriluckshmy2018}%
  \BibitemOpen
  \bibfield  {author} {\bibinfo {author} {\bibfnamefont {P.~V.}\ \bibnamefont {Sriluckshmy}}, \bibinfo {author} {\bibfnamefont {K.}~\bibnamefont {Saha}}, \ and\ \bibinfo {author} {\bibfnamefont {R.}~\bibnamefont {Moessner}},\ }\href {\doibase 10.1103/PhysRevB.97.024204} {\bibfield  {journal} {\bibinfo  {journal} {Phys. Rev. B}\ }\textbf {\bibinfo {volume} {97}},\ \bibinfo {pages} {024204} (\bibinfo {year} {2018})}\BibitemShut {NoStop}%
\bibitem [{\citenamefont {Chern}(1946)}]{Chern1946}%
  \BibitemOpen
  \bibfield  {author} {\bibinfo {author} {\bibfnamefont {S.-S.}\ \bibnamefont {Chern}},\ }\href {http://www.jstor.org/stable/1969037} {\bibfield  {journal} {\bibinfo  {journal} {Ann. Math. Second Ser.}\ }\textbf {\bibinfo {volume} {47}},\ \bibinfo {pages} {85} (\bibinfo {year} {1946})}\BibitemShut {NoStop}%
\bibitem [{\citenamefont {Berry}(1984)}]{Berry1984}%
  \BibitemOpen
  \bibfield  {author} {\bibinfo {author} {\bibfnamefont {M.~V.}\ \bibnamefont {Berry}},\ }\href {http://goo.gl/n2qaJN} {\bibfield  {journal} {\bibinfo  {journal} {Proc. R. Soc. A Math. Phys. Eng. Sci.}\ }\textbf {\bibinfo {volume} {392}},\ \bibinfo {pages} {45} (\bibinfo {year} {1984})}\BibitemShut {NoStop}%
\bibitem [{\citenamefont {Ceresoli}\ and\ \citenamefont {Resta}(2007)}]{Ceresoli2007}%
  \BibitemOpen
  \bibfield  {author} {\bibinfo {author} {\bibfnamefont {D.}~\bibnamefont {Ceresoli}}\ and\ \bibinfo {author} {\bibfnamefont {R.}~\bibnamefont {Resta}},\ }\href {\doibase 10.1103/PhysRevB.76.012405} {\bibfield  {journal} {\bibinfo  {journal} {Phys. Rev. B}\ }\textbf {\bibinfo {volume} {76}},\ \bibinfo {pages} {012405} (\bibinfo {year} {2007})}\BibitemShut {NoStop}%
\bibitem [{\citenamefont {Varjas}\ \emph {et~al.}(2020)\citenamefont {Varjas}, \citenamefont {Fruchart}, \citenamefont {Akhmerov},\ and\ \citenamefont {Perez-Piskunow}}]{Varjas_2020}%
  \BibitemOpen
  \bibfield  {author} {\bibinfo {author} {\bibfnamefont {D.}~\bibnamefont {Varjas}}, \bibinfo {author} {\bibfnamefont {M.}~\bibnamefont {Fruchart}}, \bibinfo {author} {\bibfnamefont {A.~R.}\ \bibnamefont {Akhmerov}}, \ and\ \bibinfo {author} {\bibfnamefont {P.~M.}\ \bibnamefont {Perez-Piskunow}},\ }\href {\doibase 10.1103/PhysRevResearch.2.013229} {\bibfield  {journal} {\bibinfo  {journal} {Phys. Rev. Res.}\ }\textbf {\bibinfo {volume} {2}},\ \bibinfo {pages} {013229} (\bibinfo {year} {2020})}\BibitemShut {NoStop}%
\bibitem [{\citenamefont {Dresselhaus}\ \emph {et~al.}(2022{\natexlab{b}})\citenamefont {Dresselhaus}, \citenamefont {Sbierski},\ and\ \citenamefont {Gruzberg}}]{Dresselhaus_2022}%
  \BibitemOpen
  \bibfield  {author} {\bibinfo {author} {\bibfnamefont {E.~J.}\ \bibnamefont {Dresselhaus}}, \bibinfo {author} {\bibfnamefont {B.}~\bibnamefont {Sbierski}}, \ and\ \bibinfo {author} {\bibfnamefont {I.~A.}\ \bibnamefont {Gruzberg}},\ }\href {\doibase 10.1103/PhysRevLett.129.026801} {\bibfield  {journal} {\bibinfo  {journal} {Phys. Rev. Lett.}\ }\textbf {\bibinfo {volume} {129}},\ \bibinfo {pages} {026801} (\bibinfo {year} {2022}{\natexlab{b}})}\BibitemShut {NoStop}%
\bibitem [{\citenamefont {Dresselhaus}\ \emph {et~al.}(2021)\citenamefont {Dresselhaus}, \citenamefont {Sbierski},\ and\ \citenamefont {Gruzberg}}]{Dresselhaus_2021}%
  \BibitemOpen
  \bibfield  {author} {\bibinfo {author} {\bibfnamefont {E.~J.}\ \bibnamefont {Dresselhaus}}, \bibinfo {author} {\bibfnamefont {B.}~\bibnamefont {Sbierski}}, \ and\ \bibinfo {author} {\bibfnamefont {I.~A.}\ \bibnamefont {Gruzberg}},\ }\href {\doibase https://doi.org/10.1016/j.aop.2021.168676} {\bibfield  {journal} {\bibinfo  {journal} {Ann. Phys.}\ }\textbf {\bibinfo {volume} {435}},\ \bibinfo {pages} {168676} (\bibinfo {year} {2021})}\BibitemShut {NoStop}%
\bibitem [{\citenamefont {K\"uhn}(1994)}]{Kuhn_1994}%
  \BibitemOpen
  \bibfield  {author} {\bibinfo {author} {\bibfnamefont {R.}~\bibnamefont {K\"uhn}},\ }\href {\doibase 10.1103/PhysRevLett.73.2268} {\bibfield  {journal} {\bibinfo  {journal} {Phys. Rev. Lett.}\ }\textbf {\bibinfo {volume} {73}},\ \bibinfo {pages} {2268} (\bibinfo {year} {1994})}\BibitemShut {NoStop}%
\bibitem [{\citenamefont {Fytas}\ and\ \citenamefont {Malakis}(2010)}]{Fytas_2010}%
  \BibitemOpen
  \bibfield  {author} {\bibinfo {author} {\bibfnamefont {N.~G.}\ \bibnamefont {Fytas}}\ and\ \bibinfo {author} {\bibfnamefont {A.}~\bibnamefont {Malakis}},\ }\href {\doibase 10.1103/PhysRevE.81.041109} {\bibfield  {journal} {\bibinfo  {journal} {Phys. Rev. E}\ }\textbf {\bibinfo {volume} {81}},\ \bibinfo {pages} {041109} (\bibinfo {year} {2010})}\BibitemShut {NoStop}%
\bibitem [{\citenamefont {Schrauth}\ \emph {et~al.}(2018)\citenamefont {Schrauth}, \citenamefont {Richter},\ and\ \citenamefont {Portela}}]{Schrauth_2018}%
  \BibitemOpen
  \bibfield  {author} {\bibinfo {author} {\bibfnamefont {M.}~\bibnamefont {Schrauth}}, \bibinfo {author} {\bibfnamefont {J.~A.~J.}\ \bibnamefont {Richter}}, \ and\ \bibinfo {author} {\bibfnamefont {J.~S.~E.}\ \bibnamefont {Portela}},\ }\href {\doibase 10.1103/PhysRevE.97.022144} {\bibfield  {journal} {\bibinfo  {journal} {Phys. Rev. E}\ }\textbf {\bibinfo {volume} {97}},\ \bibinfo {pages} {022144} (\bibinfo {year} {2018})}\BibitemShut {NoStop}%
\end{thebibliography}

\begin{thebibliography}{19}%
\makeatletter
\providecommand \@ifxundefined [1]{%
 \@ifx{#1\undefined}
}%
\providecommand \@ifnum [1]{%
 \ifnum #1\expandafter \@firstoftwo
 \else \expandafter \@secondoftwo
 \fi
}%
\providecommand \@ifx [1]{%
 \ifx #1\expandafter \@firstoftwo
 \else \expandafter \@secondoftwo
 \fi
}%
\providecommand \natexlab [1]{#1}%
\providecommand \enquote  [1]{``#1''}%
\providecommand \bibnamefont  [1]{#1}%
\providecommand \bibfnamefont [1]{#1}%
\providecommand \citenamefont [1]{#1}%
\providecommand \href@noop [0]{\@secondoftwo}%
\providecommand \href [0]{\begingroup \@sanitize@url \@href}%
\providecommand \@href[1]{\@@startlink{#1}\@@href}%
\providecommand \@@href[1]{\endgroup#1\@@endlink}%
\providecommand \@sanitize@url [0]{\catcode `\\12\catcode `\$12\catcode
  `\&12\catcode `\#12\catcode `\^12\catcode `\_12\catcode `\%12\relax}%
\providecommand \@@startlink[1]{}%
\providecommand \@@endlink[0]{}%
\providecommand \url  [0]{\begingroup\@sanitize@url \@url }%
\providecommand \@url [1]{\endgroup\@href {#1}{\urlprefix }}%
\providecommand \urlprefix  [0]{URL }%
\providecommand \Eprint [0]{\href }%
\providecommand \doibase [0]{http://dx.doi.org/}%
\providecommand \selectlanguage [0]{\@gobble}%
\providecommand \bibinfo  [0]{\@secondoftwo}%
\providecommand \bibfield  [0]{\@secondoftwo}%
\providecommand \translation [1]{[#1]}%
\providecommand \BibitemOpen [0]{}%
\providecommand \bibitemStop [0]{}%
\providecommand \bibitemNoStop [0]{.\EOS\space}%
\providecommand \EOS [0]{\spacefactor3000\relax}%
\providecommand \BibitemShut  [1]{\csname bibitem#1\endcsname}%
\let\auto@bib@innerbib\@empty
\bibitem [{\citenamefont {Chern}(1946)}]{SM_Chern_1946}%
  \BibitemOpen
  \bibfield  {author} {\bibinfo {author} {\bibfnamefont {S.-S.}\ \bibnamefont
  {Chern}},\ }\href {http://www.jstor.org/stable/1969037} {\bibfield  {journal}
  {\bibinfo  {journal} {Ann. Math. Second Ser.}\ }\textbf {\bibinfo {volume}
  {47}},\ \bibinfo {pages} {85} (\bibinfo {year} {1946})}\BibitemShut {NoStop}%
\bibitem [{\citenamefont {Bianco}\ and\ \citenamefont
  {Resta}(2011)}]{SM_Resta2011}%
  \BibitemOpen
  \bibfield  {author} {\bibinfo {author} {\bibfnamefont {R.}~\bibnamefont
  {Bianco}}\ and\ \bibinfo {author} {\bibfnamefont {R.}~\bibnamefont {Resta}},\
  }\href {\doibase 10.1103/PhysRevB.84.241106} {\bibfield  {journal} {\bibinfo
  {journal} {Phys. Rev. B}\ }\textbf {\bibinfo {volume} {84}},\ \bibinfo
  {pages} {241106} (\bibinfo {year} {2011})}\BibitemShut {NoStop}%
\bibitem [{\citenamefont {Privitera}\ and\ \citenamefont
  {Santoro}(2016)}]{SM_Privitera_2016}%
  \BibitemOpen
  \bibfield  {author} {\bibinfo {author} {\bibfnamefont {L.}~\bibnamefont
  {Privitera}}\ and\ \bibinfo {author} {\bibfnamefont {G.~E.}\ \bibnamefont
  {Santoro}},\ }\href {\doibase 10.1103/PhysRevB.93.241406} {\bibfield
  {journal} {\bibinfo  {journal} {Phys. Rev. B}\ }\textbf {\bibinfo {volume}
  {93}},\ \bibinfo {pages} {241406} (\bibinfo {year} {2016})}\BibitemShut
  {NoStop}%
\bibitem [{\citenamefont {Caio}\ \emph {et~al.}(2019)\citenamefont {Caio},
  \citenamefont {Möller}, \citenamefont {Cooper},\ and\ \citenamefont
  {Bhaseen}}]{SM_Caio_2019}%
  \BibitemOpen
  \bibfield  {author} {\bibinfo {author} {\bibfnamefont {M.~D.}\ \bibnamefont
  {Caio}}, \bibinfo {author} {\bibfnamefont {G.}~\bibnamefont {Möller}},
  \bibinfo {author} {\bibfnamefont {N.~R.}\ \bibnamefont {Cooper}}, \ and\
  \bibinfo {author} {\bibfnamefont {M.~J.}\ \bibnamefont {Bhaseen}},\ }\href
  {\doibase 10.1038/s41567-018-0390-7} {\bibfield  {journal} {\bibinfo
  {journal} {Nature Physics}\ }\textbf {\bibinfo {volume} {15}},\ \bibinfo
  {pages} {257} (\bibinfo {year} {2019})}\BibitemShut {NoStop}%
\bibitem [{\citenamefont {MacKinnon}\ and\ \citenamefont
  {Kramer}(1981)}]{SM_MacKinnon_1981}%
  \BibitemOpen
  \bibfield  {author} {\bibinfo {author} {\bibfnamefont {A.}~\bibnamefont
  {MacKinnon}}\ and\ \bibinfo {author} {\bibfnamefont {B.}~\bibnamefont
  {Kramer}},\ }\href {\doibase 10.1103/PhysRevLett.47.1546} {\bibfield
  {journal} {\bibinfo  {journal} {Phys. Rev. Lett.}\ }\textbf {\bibinfo
  {volume} {47}},\ \bibinfo {pages} {1546} (\bibinfo {year}
  {1981})}\BibitemShut {NoStop}%
\bibitem [{\citenamefont {Pendry}\ \emph {et~al.}(1992)\citenamefont {Pendry},
  \citenamefont {MacKinnon},\ and\ \citenamefont {Roberts}}]{SM_Pendry_1992}%
  \BibitemOpen
  \bibfield  {author} {\bibinfo {author} {\bibfnamefont {J.~B.}\ \bibnamefont
  {Pendry}}, \bibinfo {author} {\bibfnamefont {A.}~\bibnamefont {MacKinnon}}, \
  and\ \bibinfo {author} {\bibfnamefont {P.~J.}\ \bibnamefont {Roberts}},\
  }\href {\doibase 10.1098/rspa.1992.0047} {\bibfield  {journal} {\bibinfo
  {journal} {Proc. R. Soc. Lond.}\ }\textbf {\bibinfo {volume} {437}},\
  \bibinfo {pages} {67} (\bibinfo {year} {1992})}\BibitemShut {NoStop}%
\bibitem [{\citenamefont {Dwivedi}\ and\ \citenamefont
  {Chua}(2016)}]{SM_Chua_2016}%
  \BibitemOpen
  \bibfield  {author} {\bibinfo {author} {\bibfnamefont {V.}~\bibnamefont
  {Dwivedi}}\ and\ \bibinfo {author} {\bibfnamefont {V.}~\bibnamefont {Chua}},\
  }\href {\doibase 10.1103/PhysRevB.93.134304} {\bibfield  {journal} {\bibinfo
  {journal} {Phys. Rev. B}\ }\textbf {\bibinfo {volume} {93}},\ \bibinfo
  {pages} {134304} (\bibinfo {year} {2016})}\BibitemShut {NoStop}%
\bibitem [{\citenamefont {MacKinnon}\ and\ \citenamefont
  {Kramer}(1983)}]{SM_Mackinnon_1983_ST}%
  \BibitemOpen
  \bibfield  {author} {\bibinfo {author} {\bibfnamefont {A.}~\bibnamefont
  {MacKinnon}}\ and\ \bibinfo {author} {\bibfnamefont {B.~K.}\ \bibnamefont
  {Kramer}},\ }\href@noop {} {\bibfield  {journal} {\bibinfo  {journal}
  {Z. Phys. B}\ }\textbf {\bibinfo {volume}
  {53}},\ \bibinfo {pages} {1} (\bibinfo {year} {1983})}\BibitemShut {NoStop}%
\bibitem [{\citenamefont {Zawadzki}\ \emph {et~al.}(2017)\citenamefont
  {Zawadzki}, \citenamefont {D'Amico},\ and\ \citenamefont
  {Oliveira}}]{SM_Zawadzki_2017}%
  \BibitemOpen
  \bibfield  {author} {\bibinfo {author} {\bibfnamefont {K.}~\bibnamefont
  {Zawadzki}}, \bibinfo {author} {\bibfnamefont {I.}~\bibnamefont {D'Amico}}, \
  and\ \bibinfo {author} {\bibfnamefont {L.~N.}\ \bibnamefont {Oliveira}},\
  }\href@noop {} {\bibfield  {journal} {\bibinfo  {journal} {Braz. J. Phys}\
  }\textbf {\bibinfo {volume} {47}},\ \bibinfo {pages} {488} (\bibinfo {year}
  {2017})}\BibitemShut {NoStop}%
\bibitem [{\citenamefont {Fisher}\ and\ \citenamefont
  {Barber}(1972)}]{SM_Fisher_1972}%
  \BibitemOpen
  \bibfield  {author} {\bibinfo {author} {\bibfnamefont {M.~E.}\ \bibnamefont
  {Fisher}}\ and\ \bibinfo {author} {\bibfnamefont {M.~N.}\ \bibnamefont
  {Barber}},\ }\href {\doibase 10.1103/PhysRevLett.28.1516} {\bibfield
  {journal} {\bibinfo  {journal} {Phys. Rev. Lett.}\ }\textbf {\bibinfo
  {volume} {28}},\ \bibinfo {pages} {1516} (\bibinfo {year}
  {1972})}\BibitemShut {NoStop}%
\bibitem [{\citenamefont {Cardy}(1996)}]{SM_Cardy_1996}%
  \BibitemOpen
  \bibfield  {author} {\bibinfo {author} {\bibfnamefont {J.}~\bibnamefont
  {Cardy}},\ }\href {\doibase Cambridge University Press} {\bibfield  {journal}
  {\bibinfo  {journal} {Scaling and Renormalization in Statistical Physics}\ }
  (\bibinfo {year} {1996}),\ Cambridge University Press}\BibitemShut {NoStop}%
\bibitem [{\citenamefont {Slevin}\ and\ \citenamefont
  {Ohtsuki}(1999)}]{SM_Slevin_1999}%
  \BibitemOpen
  \bibfield  {author} {\bibinfo {author} {\bibfnamefont {K.}~\bibnamefont
  {Slevin}}\ and\ \bibinfo {author} {\bibfnamefont {T.}~\bibnamefont
  {Ohtsuki}},\ }\href {\doibase 10.1103/PhysRevLett.82.382} {\bibfield
  {journal} {\bibinfo  {journal} {Phys. Rev. Lett.}\ }\textbf {\bibinfo
  {volume} {82}},\ \bibinfo {pages} {382} (\bibinfo {year} {1999})}\BibitemShut
  {NoStop}%
\bibitem [{\citenamefont {Slevin}\ and\ \citenamefont
  {Ohtsuki}(2009)}]{SM_Slevin_2009}%
  \BibitemOpen
  \bibfield  {author} {\bibinfo {author} {\bibfnamefont {K.}~\bibnamefont
  {Slevin}}\ and\ \bibinfo {author} {\bibfnamefont {T.}~\bibnamefont
  {Ohtsuki}},\ }\href {\doibase 10.1103/PhysRevB.80.041304} {\bibfield
  {journal} {\bibinfo  {journal} {Phys. Rev. B}\ }\textbf {\bibinfo {volume}
  {80}},\ \bibinfo {pages} {041304} (\bibinfo {year} {2009})}\BibitemShut
  {NoStop}%
\bibitem [{\citenamefont {Beck}\ and\ \citenamefont
  {Goldstein}(2021)}]{SM_Beck_2021}%
  \BibitemOpen
  \bibfield  {author} {\bibinfo {author} {\bibfnamefont {A.}~\bibnamefont
  {Beck}}\ and\ \bibinfo {author} {\bibfnamefont {M.}~\bibnamefont
  {Goldstein}},\ }\href {\doibase 10.1103/PhysRevB.103.L241401} {\bibfield
  {journal} {\bibinfo  {journal} {Phys. Rev. B}\ }\textbf {\bibinfo {volume}
  {103}},\ \bibinfo {pages} {L241401} (\bibinfo {year} {2021})}\BibitemShut
  {NoStop}%
\bibitem [{\citenamefont {Sbierski}\ \emph {et~al.}(2021)\citenamefont
  {Sbierski}, \citenamefont {Dresselhaus}, \citenamefont {Moore},\ and\
  \citenamefont {Gruzberg}}]{SM_Sbierski_2021}%
  \BibitemOpen
  \bibfield  {author} {\bibinfo {author} {\bibfnamefont {B.}~\bibnamefont
  {Sbierski}}, \bibinfo {author} {\bibfnamefont {E.~J.}\ \bibnamefont
  {Dresselhaus}}, \bibinfo {author} {\bibfnamefont {J.~E.}\ \bibnamefont
  {Moore}}, \ and\ \bibinfo {author} {\bibfnamefont {I.~A.}\ \bibnamefont
  {Gruzberg}},\ }\href {\doibase 10.1103/PhysRevLett.126.076801} {\bibfield
  {journal} {\bibinfo  {journal} {Phys. Rev. Lett.}\ }\textbf {\bibinfo
  {volume} {126}},\ \bibinfo {pages} {076801} (\bibinfo {year}
  {2021})}\BibitemShut {NoStop}%
\bibitem [{\citenamefont {Kramer}\ \emph {et~al.}(2010)\citenamefont {Kramer},
  \citenamefont {MacKinnon}, \citenamefont {Ohtsuki},\ and\ \citenamefont
  {Slevin}}]{SM_Kramer_2010}%
  \BibitemOpen
  \bibfield  {author} {\bibinfo {author} {\bibfnamefont {B.}~\bibnamefont
  {Kramer}}, \bibinfo {author} {\bibfnamefont {A.}~\bibnamefont {MacKinnon}},
  \bibinfo {author} {\bibfnamefont {T.}~\bibnamefont {Ohtsuki}}, \ and\
  \bibinfo {author} {\bibfnamefont {K.}~\bibnamefont {Slevin}},\ }\href
  {\doibase 10.1142/S0217979210064630} {\bibfield  {journal} {\bibinfo
  {journal} {Int. J. Mod. Phys. B}\ }\textbf {\bibinfo {volume} {24}},\
  \bibinfo {pages} {1841} (\bibinfo {year} {2010})}\BibitemShut {NoStop}%
\bibitem [{\citenamefont {Gruzberg}\ \emph {et~al.}(2017)\citenamefont
  {Gruzberg}, \citenamefont {Kl\"umper}, \citenamefont {Nuding},\ and\
  \citenamefont {Sedrakyan}}]{SM_Gruzberg_2017}%
  \BibitemOpen
  \bibfield  {author} {\bibinfo {author} {\bibfnamefont {I.~A.}\ \bibnamefont
  {Gruzberg}}, \bibinfo {author} {\bibfnamefont {A.}~\bibnamefont {Kl\"umper}},
  \bibinfo {author} {\bibfnamefont {W.}~\bibnamefont {Nuding}}, \ and\ \bibinfo
  {author} {\bibfnamefont {A.}~\bibnamefont {Sedrakyan}},\ }\href {\doibase10.1103/PhysRevB.95.125414} {\bibfield  {journal} {\bibinfo  {journal} {Phys.
  Rev. B}\ }\textbf {\bibinfo {volume} {95}},\ \bibinfo {pages} {125414}
  (\bibinfo {year} {2017})}\BibitemShut {NoStop}%
\bibitem [{\citenamefont {Li}\ \emph {et~al.}(2009)\citenamefont {Li},
  \citenamefont {Vicente}, \citenamefont {Xia}, \citenamefont {Pan},
  \citenamefont {Tsui}, \citenamefont {Pfeiffer},\ and\ \citenamefont
  {West}}]{SM_Li_2009}%
  \BibitemOpen
  \bibfield  {author} {\bibinfo {author} {\bibfnamefont {W.}~\bibnamefont
  {Li}}, \bibinfo {author} {\bibfnamefont {C.~L.}\ \bibnamefont {Vicente}},
  \bibinfo {author} {\bibfnamefont {J.~S.}\ \bibnamefont {Xia}}, \bibinfo
  {author} {\bibfnamefont {W.}~\bibnamefont {Pan}}, \bibinfo {author}
  {\bibfnamefont {D.~C.}\ \bibnamefont {Tsui}}, \bibinfo {author}
  {\bibfnamefont {L.~N.}\ \bibnamefont {Pfeiffer}}, \ and\ \bibinfo {author}
  {\bibfnamefont {K.~W.}\ \bibnamefont {West}},\ }\href {\doibase10.1103/PhysRevLett.102.216801} {\bibfield  {journal} {\bibinfo  {journal}
  {Phys. Rev. Lett.}\ }\textbf {\bibinfo {volume} {102}},\ \bibinfo {pages}
  {216801} (\bibinfo {year} {2009})}\BibitemShut {NoStop}%
\bibitem [{\citenamefont {Dresselhaus}\ \emph {et~al.}(2021)\citenamefont
  {Dresselhaus}, \citenamefont {Sbierski},\ and\ \citenamefont
  {Gruzberg}}]{SM_Dresselhaus_2021}%
  \BibitemOpen
  \bibfield  {author} {\bibinfo {author} {\bibfnamefont {E.~J.}\ \bibnamefont
  {Dresselhaus}}, \bibinfo {author} {\bibfnamefont {B.}~\bibnamefont
  {Sbierski}}, \ and\ \bibinfo {author} {\bibfnamefont {I.~A.}\ \bibnamefont
  {Gruzberg}},\ }\href {\doibase https://doi.org/10.1016/j.aop.2021.168676}
  {\bibfield  {journal} {\bibinfo  {journal} {Ann. Phys.}\ }\textbf {\bibinfo
  {volume} {435}},\ \bibinfo {pages} {168676} (\bibinfo {year}
  {2021})}\BibitemShut {NoStop}%
\end{thebibliography}
\end{document}